\documentclass[11pt]{article}

\usepackage{type1cm}        
                            
                            \usepackage{fullpage}
\usepackage{makeidx}         
\usepackage{graphicx}        
\usepackage{multicol}        
\usepackage[bottom]{footmisc}

\usepackage{newtxtext}       %
\usepackage[varvw]{newtxmath}       

\usepackage{wrapfig}

\usepackage{cite}
\usepackage{caption}
\usepackage{subcaption}
\usepackage{titlesec}
\usepackage{amsmath}
\usepackage{tabularx}
\usepackage{bm}
\usepackage{color}
\usepackage{url}
\usepackage{bigfoot}
\usepackage{multirow}
\usepackage{setspace}

\usepackage{placeins}
\usepackage{inputenc}
\usepackage{ulem}
\usepackage{upgreek}

\def\red{\textcolor{red}}

\def\##1{{\bf #1}}
\def\=#1{\underline{\underline #1}}
\def\^#1{{\rR}eve{#1}}
\def\`#1{{#1^\prime}}
\def\:#1{#1^{\prime\prime}}

\def\.{\mbox{ \tiny{$^\bullet$} }}

\def\ux{\hat{\#x}}
\def\uy{\hat{\#y}}
\def\uz{\hat{\#z}}

\def\muo{\mu_{\scriptscriptstyle 0}}
\def\epso{\eps_{\scriptscriptstyle 0}}

\def\lambdao{\lambda_{\scriptscriptstyle 0}}
\def\ko{k_{\scriptscriptstyle 0}}

\def\eps{\varepsilon}
\def\epsa{\eps_{\rm a}}
\def\epsb{\eps_{\rm b}}
\def\epsc{\eps_{\rm c}}

\def\le{\left(}
\def\ri{\right)}
\def\les{\left[}
\def\ris{\right]}
\def\lec{\left\{}
\def\ric{\right\}}

\def\deg{^\circ}

\def\aal{a_{\rm L}}
\def\aar{a_{\rm R}}
\def\bbl{r_{\rm L}}
\def\bbr{r_{\rm R}}
\def\ccl{t_{\rm L}}
\def\ccr{t_{\rm R}}

\def\rLL{r_{\rm LL}}
\def\rLR{r_{\rm LR}}
\def\rRL{r_{\rm RL}}
\def\rRR{r_{\rm RR}}
\def\tLL{t_{\rm LL}}
\def\tLR{t_{\rm LR}}
\def\tRL{t_{\rm RL}}
\def\tRR{t_{\rm RR}}

\def\RLL{R_{\rm LL}}
\def\RLR{R_{\rm LR}}
\def\RRL{R_{\rm RL}}
\def\RRR{R_{\rm RR}}
\def\TLL{T_{\rm LL}}
\def\TLR{T_{\rm LR}}
\def\TRL{T_{\rm RL}}
\def\TRR{T_{\rm RR}}

 \def\AL{A_{\rm L}}
\def\AR{A_{\rm R}}

 \def\TL{T_{\rm L}}
\def\TR{T_{\rm R}}
 \def\RL{R_{\rm L}}
\def\RR{R_{\rm R}}

\def\aas{a_{\rm s}}
\def\aap{a_{\rm p}}
\def\bbs{r_{\rm s}}
\def\bbp{r_{\rm p}}
\def\ccs{t_{\rm s}}
\def\ccp{t_{\rm p}}

\def\rss{r_{\rm ss}}
\def\rsp{r_{\rm sp}}
\def\rps{r_{\rm ps}}
\def\rpp{r_{\rm pp}}
\def\tss{t_{\rm ss}}
\def\tsp{t_{\rm sp}}
\def\tps{t_{\rm ps}}
\def\tpp{t_{\rm pp}}

\def\Rss{R_{\rm ss}}
\def\Rsp{R_{\rm sp}}
\def\Rps{R_{\rm ps}}
\def\Rpp{R_{\rm pp}}
\def\Tss{T_{\rm ss}}
\def\Tsp{T_{\rm sp}}
\def\Tps{T_{\rm ps}}
\def\Tpp{T_{\rm pp}}

\def\Rs{R_{\rm s}}
\def\Rp{R_{\rm p}}
\def\Ts{T_{\rm s}}
\def\Tp{T_{\rm p}}

\def\As{A_{\rm s}}
\def\Ap{A_{\rm p}}

\def\thetainc{\theta_{\rm inc}}

\def\sinczero{s_0^{\rm inc}}
\def\sincone{s_1^{\rm inc}}
\def\sinctwo{s_2^{\rm inc}}
\def\sincthree{s_3^{\rm inc}}

\def\bphit{\les{\underline \phi}\ris}
\def\bphitinc{\les{\underline \phi}^{\rm inc}\ris}

\def\srefzero{s_0^{\rm ref}}
\def\srefone{s_1^{\rm ref}}
\def\sreftwo{s_2^{\rm ref}}
\def\srefthree{s_3^{\rm ref}}

\def\bphitref{\les{\underline \phi}^{\rm ref}\ris}

\def\bphittra{\les{\underline \phi}^{\rm tr}\ris}

\def\bPhiref{\Phi^{\rm ref}}

\def\bPhirefR{\Phi_{\rm R}^{\rm ref}}

\def\bPhitra{\Phi^{\rm tr}}

\def\bPhitraR{\Phi_{\rm R}^{\rm tr}}

\def\tld{{LD}_{\rm tru}}
\def\ald{{LD}_{\rm app}}
\def\tcd{{CD}_{\rm tru}}
\def\acd{{CD}_{\rm app}}

\def\ORtra{OR^{\rm tr}}

\def\ORref{OR^{\rm ref}}

\def\EFinc{EF^{\rm inc}}

\def\EFtra{EF^{\rm tr}}

\def\EFref{EF^{\rm ref}}

\def\sp{\#s}
\def\pinc{\#p_+}
\def\pref{\#p_-}

\def\Ei{\#E^{\rm inc}(\#r)}

\def\Er{\#E^{\rm ref}(\#r)}

\def\Et{\#E^{\rm tr}(\#r)}

\def\Re{{\rm Re}}
\def\Im{{\rm Im}}

\makeindex

\begin{document}

\begin{center}

\title*{The Circular Bragg Phenomenon Updated}

\textbf{Akhlesh Lakhtakia}

\textit{Department of Engineering
Science and Mechanics, The Pennsylvania State University, University Park, PA, USA \\
 School of Mathematics, University of Edinburgh, Edinburgh EH9 3FD, UK}\\
{akhlesh@psu.edu}

\abstract{The circular Bragg phenomenon is the circular-polarization-state-selective reflection of light in a spectral regime called
the circular Bragg regime. In continuation of an expository review on this phenomenon published in 2014, an album of theoretical results 
is provided in this decadal update. Spectral variations of intensity-dependent and phase-dependent
observable quantities   in the reflection and transmission half-spaces are provided  in relation to the polarization state and the direction of propagation of the incident plane wave.}

\vspace{5mm}

\end{center}


\section{Introduction}\label{sec:intro} 

The circular Bragg phenomenon is the circular-polarization-state-selective reflection of plane waves in a spectral regime called
the circular Bragg regime that depends on the direction of incidence \cite{FLaop,LVaeu}. This phenomenon is displayed by structurally chiral materials (SCMs) exemplified by chiral
liquid crystals \cite{Chan,deG,Nityananda,Garoff,Abdulhalim,Parodi,Xiang} and chiral sculptured thin films 
\cite{STFbook,Erten2015,McAtee2018}. These linear materials are periodically non\-homogeneous along a fixed
axis, their constitutive dyadics rotating either clockwise or counterclockwise
at a fixed rate about that axis \cite{HBM1,HBM2}.  The reflection
is very high when left-circularly polarized (LCP) light is incident on a left-handed SCM which is
periodically non\-homogeneous along the thickness direction, provided that (i)  the direction of incidence
is not too oblique with respect to the thickness direction, (ii)
the free-space wavelength lies in the circular Bragg regime,  and (iii) the number of periods in SCM is sufficiently large; however,
when right-circularly polarized light (RCP) is incident on a left-handed SCM, the reflectance is very low in the circular
Bragg regime  \cite{FLaop,Erten2015,StJohn,PLA}.
An analogous statement holds for right-handed SCMs. The circular Bragg phenomenon is resilient against structural disorder \cite{dcs}
and the tilt of the axis of  periodicity \cite{Slant1,Slant2}.

An expository and detailed review of the literature on circular Bragg phenomenon was published in 2014 \cite{FLaop}, to which the
interested reader is referred. During the subsequent decade, several novel results---both experimental \cite{Erten2015,McAtee2018,FialloPhD,Das}
and theoretical \cite{Lakh2024josab,Lakh2024pra}---on the plane-wave response of SCMs
have emerged, which prompted me to compile this album of theoretical
numerical results  on the plane-wave response of SCMs. Of course, no novelty
can be claimed for these results, but this album is expected to guide relevant research for the next decade.

 This chapter is organized as follows. Section~\ref{bvp} provides the essentials of the boundary-value problem
 underlying the plane-wave response of an SCM of finite thickness. Section~\ref{idq} provides illustrative results on the
 spectral variations of intensity-dependent observable quantities and Sec.~\ref{pdq}
 is focused similarly on  phase-dependent observable quantities, in the reflection half-space as well as in the transmission
 half-space, in relation to the polarization state and the direction of propagation of the incident plane wave.

An $ \exp(-i \omega t)$  dependence on time $t$ is implicit, where $ \omega$ as the angular frequency and $i=\sqrt{-1}$.
With $ \epso$ and $\muo$, respectively,
denoting the permittivity and permeability  of free space,
the free-space wavenumber is denoted by $\ko 
= \omega \sqrt{\epso \muo}$,  and $\lambdao=2\pi/\ko$ is the free-space
wavelength. The Cartesian coordinate system $(x,y,z)$ is adopted.
Vectors are in boldface
 and unit vectors are additionally decorated by a caret on top. Dyadics \cite{Chen} are double underlined.
 Column vectors are underlined and enclosed in square brackets. The asterisk $(^\ast)$ denotes the complex conjugate
 and the dagger $(^\dag)$ denotes the conjugate transpose.

 \section{Boundary-value problem} \label{bvp}
 The half-space $z < 0$ is the region of incidence and reflection, while
the half-space $z > L$ is the region of transmission. The region $0<z<L$ is occupied by a SCM.

 \subsection{Relative permittivity dyadic of SCM}
The relative permittivity dyadic of the SCM  
is given by \cite{STFbook}
\begin{eqnarray}
 \nonumber
&&\=\eps_{\rm rel}(z)=
\=S_{\rm z}(h,\Omega,z) \.
\=S_{\rm y}(\chi)\.\left[\epsa\uz\uz+\epsb\ux\ux
+\epsc\uy\uy\right]
\nonumber
\\[5pt]
&&\qquad 
\.\=S_{\rm y}^{-1}(\chi) \.\=S_{\rm z}^{-1}(h,\Omega,z)\,,
\quad z\in(0,L)\,.
\label{epsDSCM}
\end{eqnarray}
 The frequency-dependent relative permittivity scalars
$\epsa$, $\epsb$, and $\epsc$ embody   local orthorhombicity \cite{Nye}.
 The tilt dyadic
\begin{equation}
\=S_{\rm y} (\chi) = \uy\uy + \left(\ux\ux+\uz\uz\right)\cos\chi
  + \left(\uz\ux-\ux\uz\right)\sin\chi
 \label{SY-def}
 \end{equation}
contains $\chi\in[{0,\pi/2}]$ as an angle of inclination with respect to the $xy$ plane.
The structural handedness of the SCM captured
by the rotation dyadic
\begin{equation}
\=S_z(h,\Omega,z)=\uz\uz +\left(\ux\ux+\uy\uy\right)\cos\left(h\frac{\pi z}{\Omega}\right)
+\left(\uy\ux-\ux\uy\right)\sin\left(h\frac{\pi z}{\Omega}\right)\,,
\label{Sz-def}
\end{equation} 
where $2\Omega$ is the structural period in the thickness direction (i.e., along the $z$ axis), whereas
$h \in\left\{-1,1\right\}$ is the structural-handedness parameter, with $h=-1$ for structural left-handedness and
 $h=1$ for structural right-handedness.

The foregoing equations apply to chiral sculptured thin films \cite{STFbook,Erten2015,McAtee2018}
and chiral smectic liquid crystals \cite{Parodi,Garoff,Abdulhalim},
with $\epsa\ne\epsb\ne\epsc$ and $\chi>0\deg$. They also apply to cholesteric liquid crystals with $\epsa\ne\epsb=\epsc$ and $\chi=0\deg$
\cite{deG,Nityananda} and heliconical cholesteric liquid crystals \cite{Xiang} with $\epsa\ne\epsb=\epsc$ and $\chi\in(0\deg,90\deg)$.

\subsection{Incident, reflected, and transmitted plane waves}
A plane
wave, propagating in the half-space
$z < 0$ at an angle $\thetainc\in[0,\pi/2)$ with respect to the $z$ axis and at an angle $\psi\in[0,2\pi)$ with respect
to the $x$ axis in the $xy$ plane, is incident on the SCM of thickness $L$.
The electric  field phasor associated
with the incident plane wave is represented as \cite{STFbook}
\begin{subequations}
\begin{eqnarray}
\label{eqEi-lin}
\Ei&=& \le \aas \sp + \aap\pinc\ri  \exp{\les i\kappa \le x\cos\psi + y\sin\psi \ri
\ris} \exp{\left(i\ko z{\cos\thetainc}\right)}\\
\nonumber
&=&
\les \frac{\le i\sp - \pinc \ri}{\sqrt{2}} \, \aal -\, \frac{\le i\sp +\pinc
\ri}{\sqrt{2}} \, \aar \ris
\exp{\les i\kappa \le x\cos\psi + y\sin\psi \ri
\ris}\\
 && \times \exp{\left(i\ko z{\cos\thetainc}\right)}\,, \qquad   z < 0\,,
 \label{eqEi-circ}
\end{eqnarray}
\end{subequations}
where
 \begin{equation}
 \left.\begin{array}{l}
\kappa =
\ko\sin{\thetainc}\\[5pt]
\sp=-\ux\sin\psi + \uy \cos\psi \\[5pt]
\#p_\pm=\mp\le \ux \cos\psi + \uy \sin\psi \ri \cos{\thetainc} 
 + \uz \sin{\thetainc}
\end{array}
\right\}
\, .
\end{equation}
The amplitudes  of the perpendicular- and parallel-polarized
components, respectively, are denoted by $\aas$ and $\aap$ in Eq.~(\ref{eqEi-lin}).
The amplitudes  of the
LCP  and  the 
 RCP components of the incident plane wave are denoted by 
$\aal$ and
$\aar$, respectively, in Eq.~(\ref{eqEi-circ}).  

The   electric   field phasor of the reflected plane wave is expressed as
\begin{subequations}
\begin{eqnarray}
\label{eqEr-lin}
\Er&=& \le \bbs\sp+\bbp\pref\ri \exp{\les i\kappa \le x\cos{\psi} + y\sin{\psi} \ri
\ris} \exp\left({-i\ko   z\cos{\thetainc}}\right) \, 
\\
\nonumber
&=&-\, \les \frac{\le i\sp - \pref \ri}{\sqrt{2}} \, {\bbl} -\, \frac{\le i\sp +
\pref \ri}{\sqrt{2}} \, {\bbr} \ris  \exp{\les i\kappa \le x\cos{\psi} + y\sin{\psi} \ri
\ris} \\
&&\times\,\exp\left({-i\ko   z\cos{\thetainc}}\right)\,,\qquad z < 0 \, ,
\label{eqEr-circ}
\end{eqnarray}
\end{subequations}
and the electric   field phasor of the transmitted plane wave is represented as
\begin{subequations}
\begin{eqnarray}
\label{eqEt-lin}
\Et&=&\le \ccs\sp +\ccp\pinc\ri    \exp{\les i\kappa \le x\cos{\psi} + y\sin{\psi} \ri
\ris}  \exp\les{i\ko   (z-L)\cos{\thetainc}}\ris
\\
\nonumber
&=& \les \frac{\le i\sp - \pinc \ri}{\sqrt{2}} \, {\ccl} -\, \frac{\le i\sp +\pinc
\ri}{\sqrt{2}} \, {\ccr} \ris  \exp{\les i\kappa \le x\cos{\psi} + y\sin{\psi} \ri
\ris}\\
\label{eqEt-circ}
& &  \times \,  \exp\les{i\ko   (z-L)\cos{\thetainc}}\ris \, , \qquad z > L \, .
\end{eqnarray}
\end{subequations}
Linear reflection amplitudes are denoted by $\bbs$ and $\bbp$  in Eq.~(\ref{eqEr-lin}), whereas the
circular reflection amplitudes are denoted by $\bbl$ and $\bbr$ in Eq.~(\ref{eqEr-circ}).
Similarly,  $\ccs$ and $\ccp$ are the linear transmission amplitudes in Eq.~(\ref{eqEt-lin}),
whereas $\ccl$ and $\ccr$ are the circular transmission amplitudes in Eq.~(\ref{eqEt-circ}).

\subsection{Reflection and transmission coefficients}
The reflection amplitudes ${\bbs}$ and ${\bbp}$ as well as the transmission
amplitudes ${\ccs}$ and ${\ccp}$ (equivalently, $\bbl$, $\bbr$, $\ccl$, and $\ccr$)
are unknown. A boundary-value problem must be solved in order to determine these amplitudes in terms
of $\aas$ and $\aap$ (equivalently, $\aal$ and $\aar$). Several numerical techniques exist to solve
this problem \cite{Dreher,Sugita, Oldano,LW95}. The most straightforward
technique requires the use of the piecewise uniform approximation of 
$\=\eps_{\rm rel}(z)$ followed by application of the 4$\times$4 transfer-matrix method
\cite{MLbook}. The interested reader is
referred to  Ref.~\citenum{STFbook} for a detailed description of this technique.

Interest generally lies in determining the reflection and transmission coefficients
entering the 2$\times$2 matrixes on the left side in each of the following four relations \cite{STFbook}:
\begin{subequations}
\begin{equation}
\label{eq13a}
\les \begin{array}{c} \bbs \\ \bbp  \end{array}\ris  =
\les \begin{array}{cc} \rss &\quad \rsp \\ \rps &\quad \rpp \end{array}\ris \,
\les \begin{array}{c} \aas \\ \aap  \end{array}\ris
\, , \end{equation}
\begin{equation}
\label{eq13b}
\les \begin{array}{c} \ccs \\ \ccp  \end{array}\ris  =
\les \begin{array}{cc} \tss &\quad \tsp \\ \tps &\quad \tpp \end{array}\ris \,
\les \begin{array}{c} \aas \\ \aap \end{array}\ris
\, ,
\end{equation}
\begin{equation}
\label{eq13c}
\les \begin{array}{c} \bbl \\ \bbr  \end{array}\ris  =
\les \begin{array}{cc} \rLL&\quad \rLR \\ \rRL &\quad \rRR \end{array}\ris \,
\les \begin{array}{c} \aal \\ \aar  \end{array}\ris
\, , 
\end{equation}
and
\begin{equation}
\label{eq13d}
\les \begin{array}{c} \ccl \\ \ccr  \end{array}\ris  =
\les \begin{array}{cc} \tLL &\quad \tLR \\ \tRL &\quad \tRR \end{array}\ris \,
\les \begin{array}{c} \aal \\ \aar  \end{array}\ris
\, , 
\end{equation}
\end{subequations}
These coefficients are doubly subscripted:
those with both subscripts identical refer to co-polarized,
while those with two different subscripts denote
cross-polarized, reflection or transmission. Clearly from Eqs. (\ref{eqEi-lin})--(\ref{eqEt-circ}),
the coefficients defined {\it via} Eqs. (\ref{eq13a}) and (\ref{eq13b})
are simply related to those defined {\it via} Eqs. (\ref{eq13c}) and (\ref{eq13d}).

\subsection{Parameters chosen for calculations}

An album of  numerical results is presented in the remainder of this chapter, with  the frequency-dependent constitutive
parameters  
\begin{equation}
\label{resonance}
\eps_{\rm a,b,c}(\lambdao) = 1+ \frac{p_{\rm a,b,c}}
{1 + (N_{\rm a,b,c}^{-1}  - i\lambdao^{-1} \lambda_{\rm a,b,c}  )^2}\,
\end{equation}
chosen to be single-resonance Lorentzian functions 
\cite{Wooten},
this choice being consistent with the requirement of causality 
\cite{Frisch,Silva,Kinsler-EJP}.
The oscillator strengths are determined by the values of $p_{\ell}$,  $\lambda_{\ell} (1 + N_{\ell}^{-2})^{-1/2} $  are the  resonance wavelengths, and $\lambda_{\ell}/N_{\ell}$ are the  resonance linewidths, $\ell\in\left\{{\rm a, b, c}\right\}$.
 Values of the parameters used for all theoretical results reported in this chapter are as follows:  $p_{\rm a} = 2.3$, $p_{\rm b} =3.0$, $p_{\rm c} =2.2 $, $\lambda_{\rm a} = \lambda_{\rm c} =260$~nm, $\lambda_{\rm b} = 270$~nm,   and
$N_{\rm a} = N_{\rm b} =N_{\rm c}=130$. Furthermore, $\chi = 37\deg$, $L=30\Omega$,  and $\Omega = 150$~nm. 

The album comprising Figs.~\ref{CircReflectance}--\ref{LinGPtrans} contains 2D plots of the theoretically 
calculated spectral variations of diverse
observable quantities in the reflection and transmission half-spaces for either
\begin{itemize}
\item $\thetainc\in [0\deg,90\deg)$ and $\psi= 0\deg$ or  
\item $\thetainc=0\deg$ and $\psi\in[ 0\deg,360\deg)$.
\end{itemize}
These plots are provided for both $h=1$ and $h=-1$ to facilitate easy comparison of the effect of structural handedness.

\section{Intensity-dependent quantities}\label{idq}

\subsection{Circular  remittances}

The square of the magnitude
of a circular reflection or transmission coefficient is the corresponding circular
reflectance or transmittance;  thus, $\RLR = \vert \rLR\vert^2\in[0,1]$ is
the circular reflectance corresponding to the circular reflection coefficient $\rLR$,
$\TLR = \vert \tLR\vert^2\in[0,1]$ is
the circular transmittance corresponding to the circular transmission coefficient $\tLR$,
and so on. 
The total circular reflectances are given by
\begin{equation}
\left. \begin{array}{l}
\RL =  \RLL+\RRL\in[0,1]\\[5pt]
\RR=\RRR+\RLR\in[0,1]
\end{array}\ric
\label{def-Rcirc}
\end{equation}
and the total circular transmittances by
\begin{equation}
\left. \begin{array}{l}
\TL =  \TLL+\TRL\in[0,1]\\[5pt]
\TR=\TRR+\TLR\in[0,1]
\end{array}\ric\,.
\label{def-Tcirc}
\end{equation}
As the principle
of conservation of energy must be satisfied by the
presented formalism, the inequalities \cite{STFbook}
\begin{equation}
R_{\ell}+T_{\ell} \leq1\,, \quad {\ell} \in\lec{\rm L,R}\ric \,.
\label{eq16}
\end{equation}
hold, with
the equalities relevant  only if the SCM is non-dissipative
at a particular frequency of interest. 

The  Bragg phenomenon was discovered as a reflection phenomenon, so that its
chief signature comprises high-reflectance spectral regimes \cite{Braggs1913,WHBragg1913,Ewald1916-2}.
The same is true of the circular
Bragg phenomenon, which has been confirmed by time-domain simulations \cite{ML2000,GL2000,GML2000}.

\begin{figure}[ht]
\begin{subfigure}{0.5\textwidth}
\centering
\includegraphics[width=7cm]{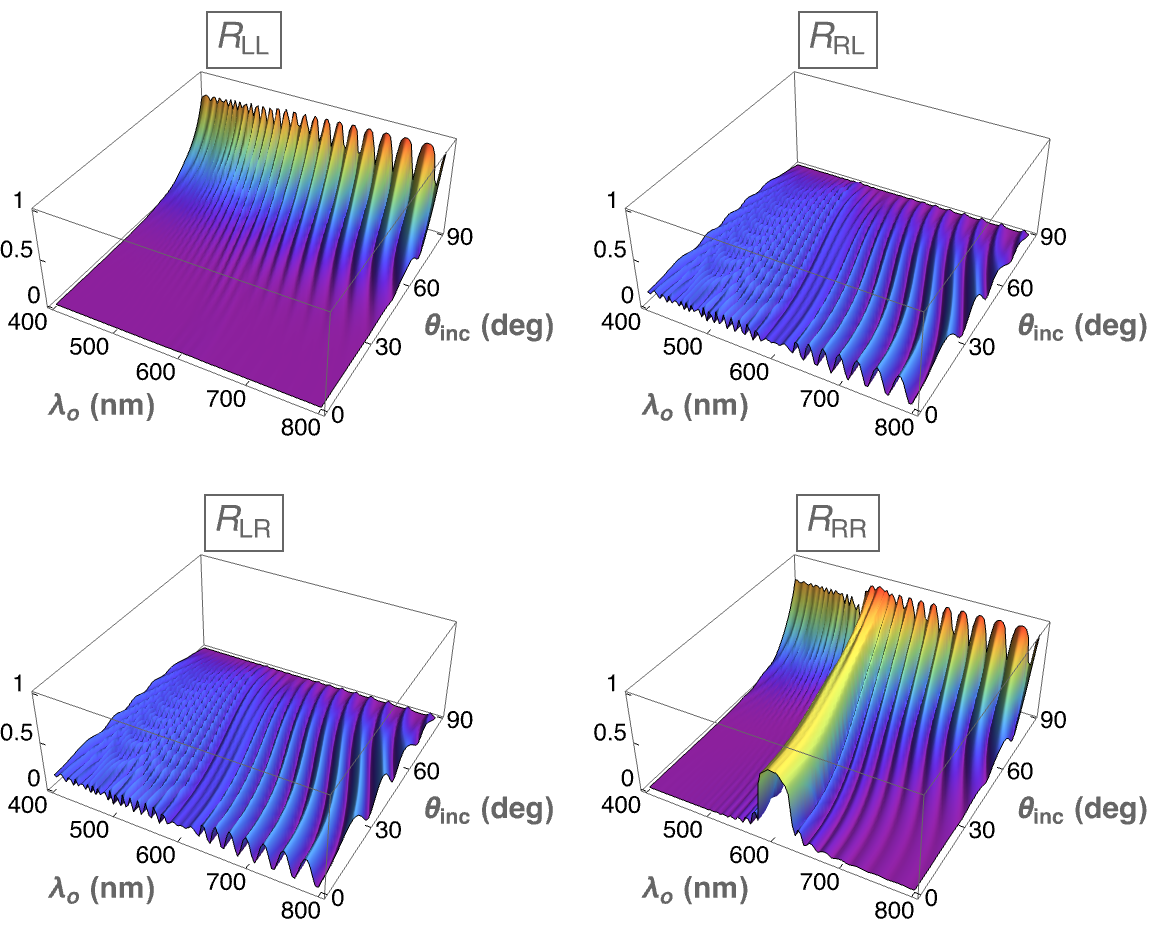} 
\hfill \vspace{0mm} 
 \caption{\label{CircReflectance-1} $h=1$, $\thetainc\in[0\deg,90\deg)$, and $\psi=0\deg$}
\end{subfigure}  \hspace{-5mm} 
\begin{subfigure}{0.5\textwidth}
\centering
\includegraphics[width=7cm]{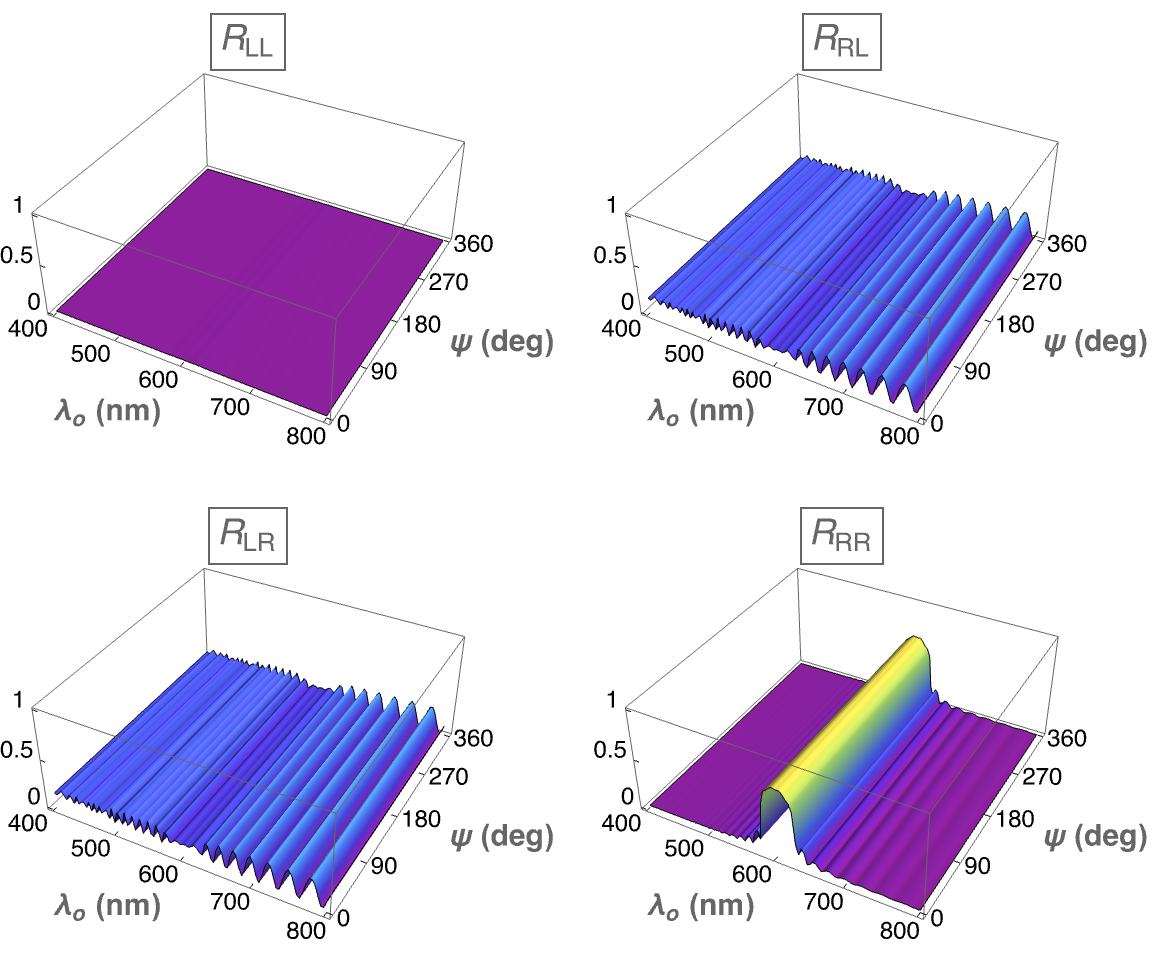} 
\hfill \vspace{0mm} 
 \caption{\label{CircReflectance-2} $h=1$, $\thetainc=0\deg$, and $\psi\in[0\deg,360\deg)$}
\end{subfigure} 
\\ ------------------------------------------------------------------------------------------------- \\
\begin{subfigure}{0.5\textwidth}
\centering
\includegraphics[width=7cm]{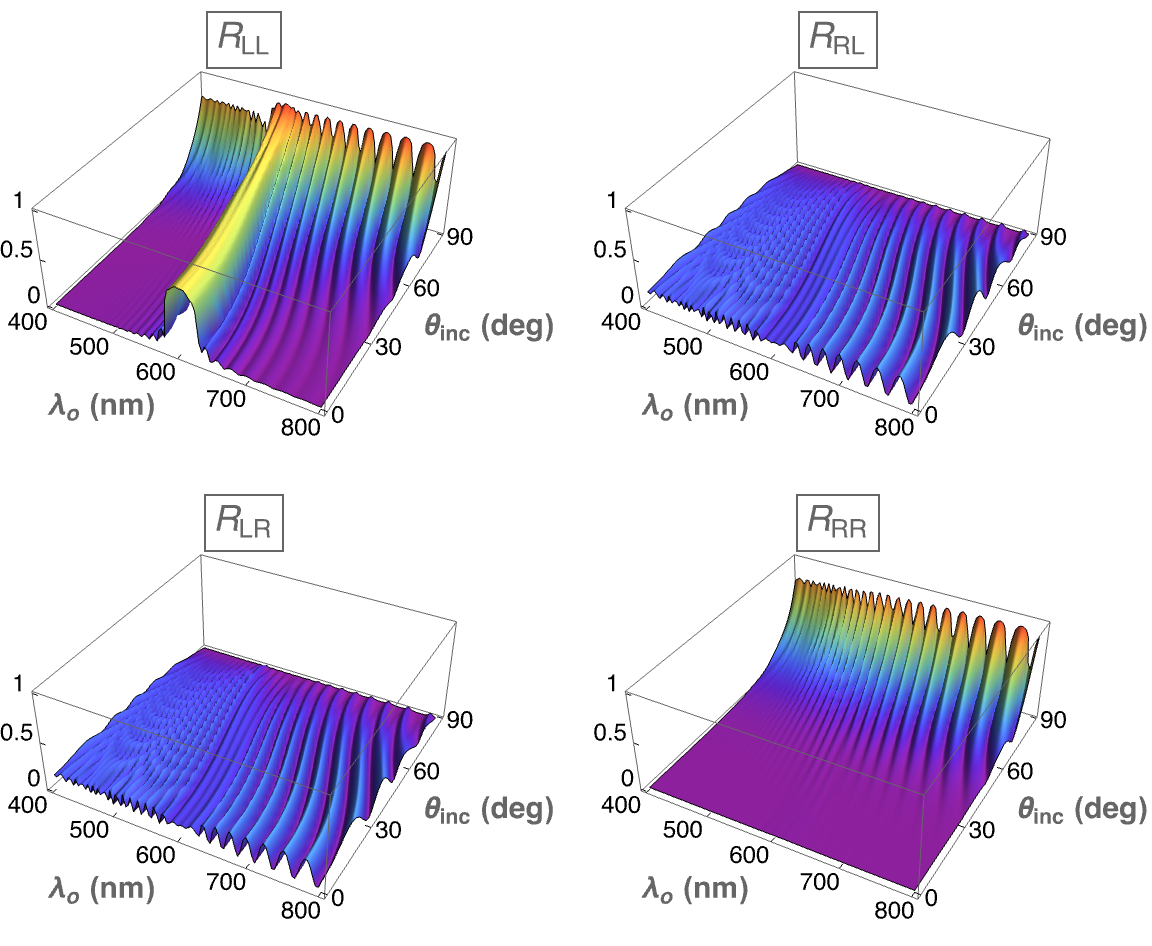} 
\hfill \vspace{0mm} 
 \caption{\label{CircReflectance-3} $h=-1$, $\thetainc\in[0\deg,90\deg)$, and $\psi=0\deg$}
\end{subfigure}  \hspace{-5mm} 
\begin{subfigure}{0.5\textwidth}
\centering
\includegraphics[width=7cm]{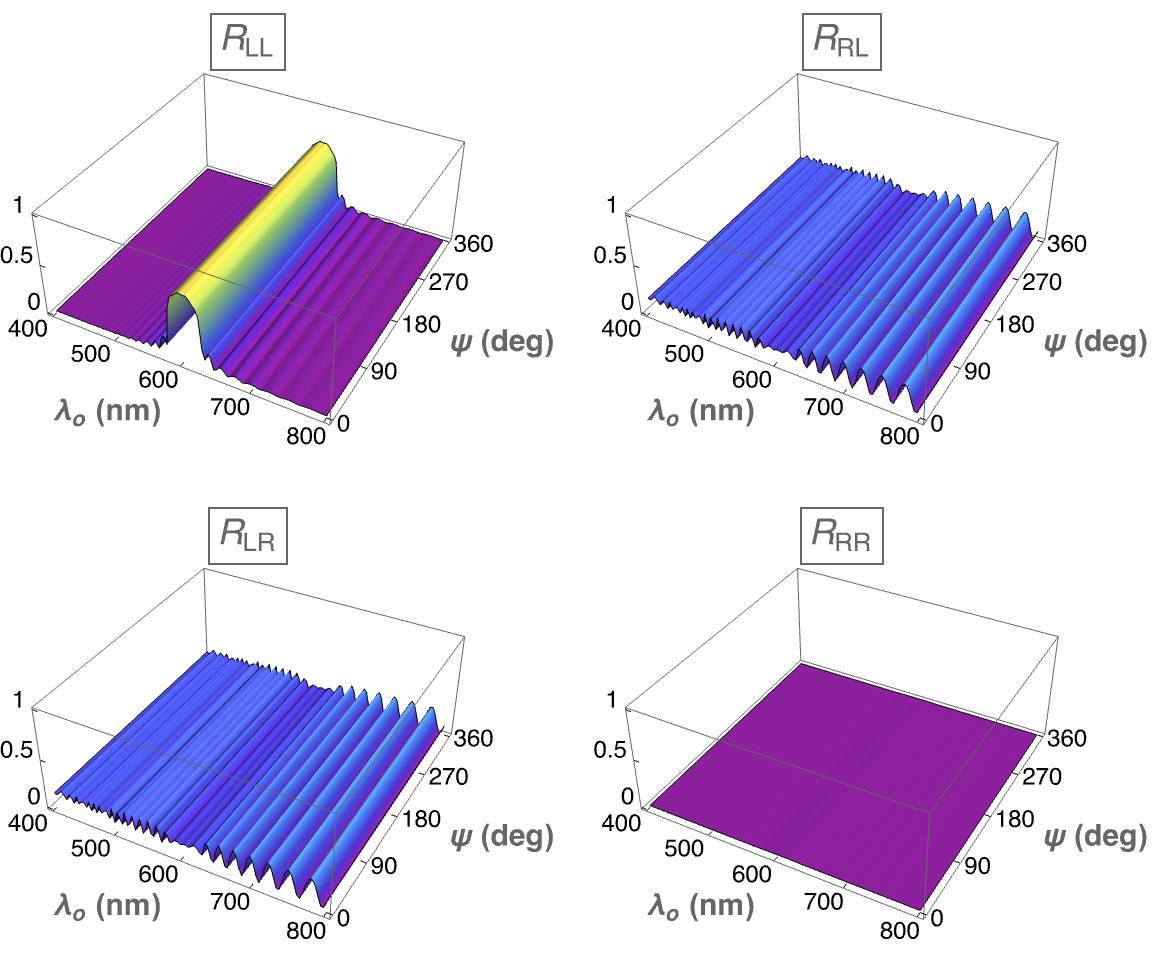} 
\hfill \vspace{0mm} 
 \caption{\label{CircReflectance-4} $h=-1$, $\thetainc=0\deg$, and $\psi\in[0\deg,360\deg)$}
\end{subfigure} 
	\caption{Spectral variations of circular reflectances $R_{\mu\nu}(\lambdao,\thetainc,\psi)$,   $\mu\in\left\{\rm L,R\right\}$ and $\nu\in\left\{\rm L,R\right\}$, 
	when
	(a,b) $h=1$ and (c,d) $h=-1$. (a,c) $\thetainc\in [0\deg,90\deg)$ and $\psi= 0\deg$;  (b,d) $\thetainc=0\deg$ and $\psi\in[ 0\deg,360\deg)$.
	}
	\label{CircReflectance}
\end{figure}

 In addition, the circular Bragg phenomenon is best manifested as
 the circular-polarization-state-selective reflection of light. Therefore it is best to begin the album
 with the spectral variations of the circular reflectances $R_{\mu\nu}(\lambdao,\thetainc,\psi)$,   $\mu\in\left\{\rm L,R\right\}$ and $\nu\in\left\{\rm L,R\right\}$. 
 These are presented in Fig.~\ref{CircReflectance} for $h=\pm1$. 
 
 Note the presence of a high-reflectance ridge in the plots
 of $\RRR$ for $h=1$ and in the plots of $\RLL$ for $h=-1$ in Fig.~\ref{CircReflectance}. For fixed $\psi$, the high-reflectance ridge curves towards
 shorter wavelengths as $\thetainc$ increases, which has been experimentally verified \cite{Erten2015,FialloPhD,McAtee2018}. For fixed $\thetainc$, the high-reflectance
 ridge is more or less invariant with respect to $\psi$. The ridge is absent in the  plots of $\RLL$ for $h=1$ and in the plots of $\RRR$ for $h=-1$;
 however, the ridge is vestigially present in the plots of both cross-polarized reflectances.
 
\begin{figure}[ht]
\begin{subfigure}{0.5\textwidth}
\centering
\includegraphics[width=7cm]{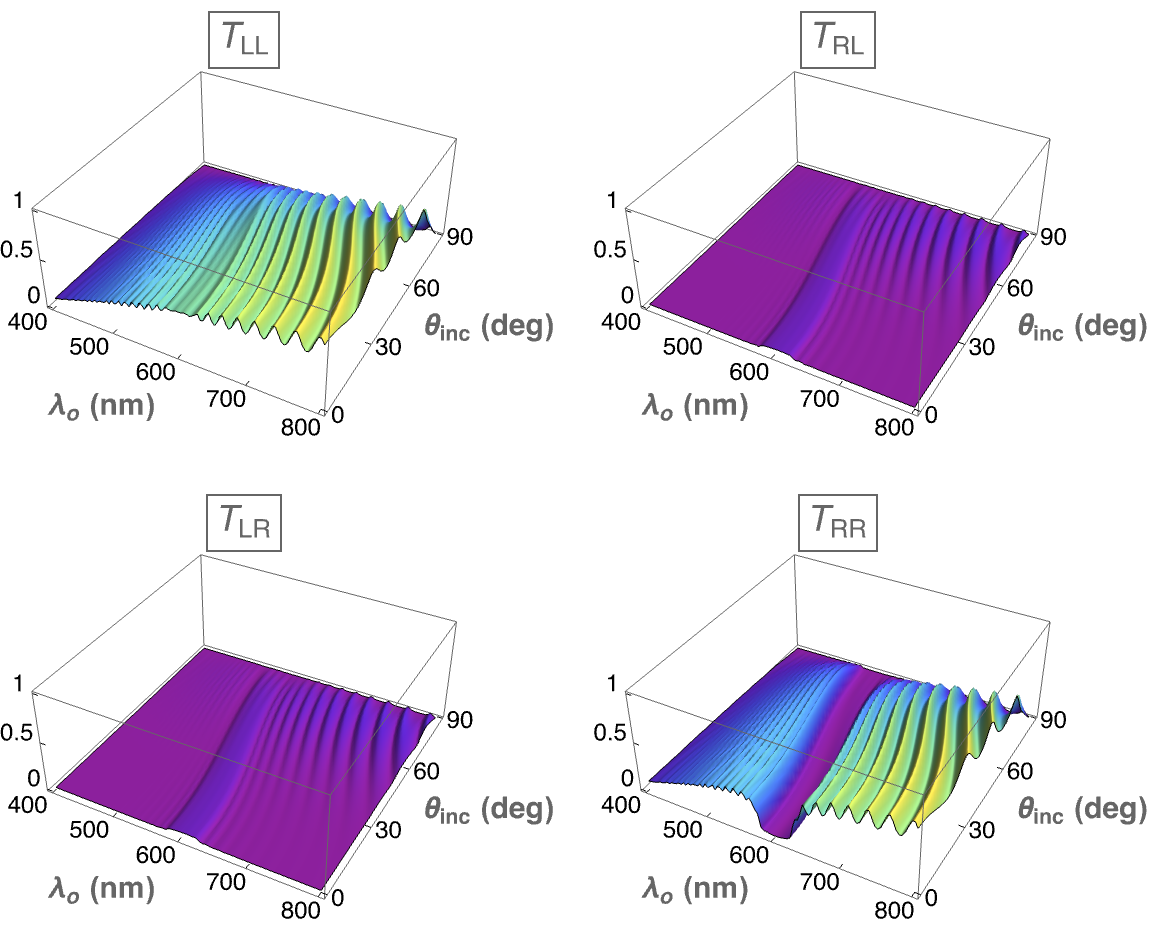} 
\hfill \vspace{0mm} 
 \caption{\label{CircTransmittance-1} $h=1$, $\thetainc\in[0\deg,90\deg)$, and $\psi=0\deg$}
\end{subfigure}  \hspace{-5mm} 
\begin{subfigure}{0.5\textwidth}
\centering
\includegraphics[width=7cm]{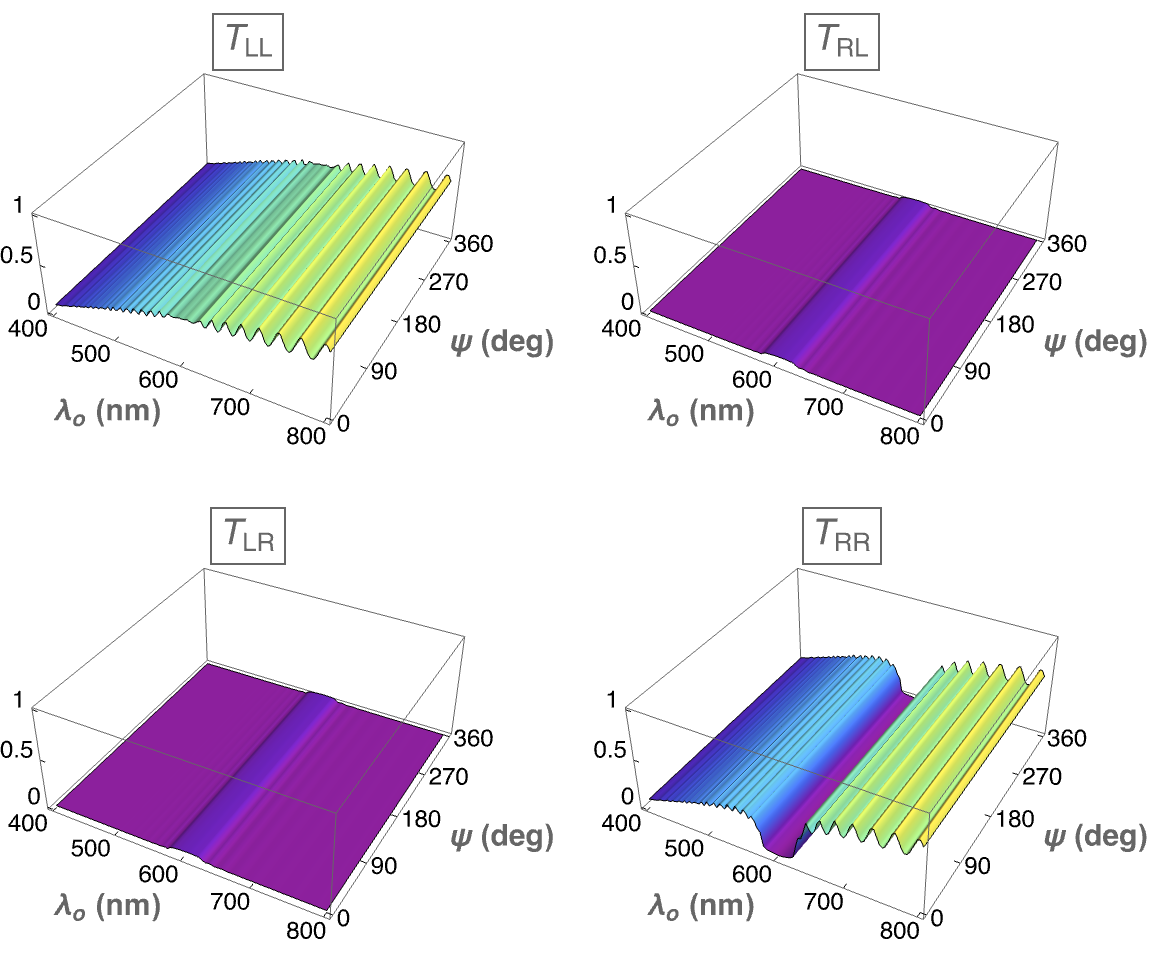} 
\hfill \vspace{0mm} 
 \caption{\label{CircTransmittance-2} $h=1$, $\thetainc=0\deg$, and $\psi\in[0\deg,360\deg)$}
\end{subfigure} 
\\ ------------------------------------------------------------------------------------------------- \\
\begin{subfigure}{0.5\textwidth}
\centering
\includegraphics[width=7cm]{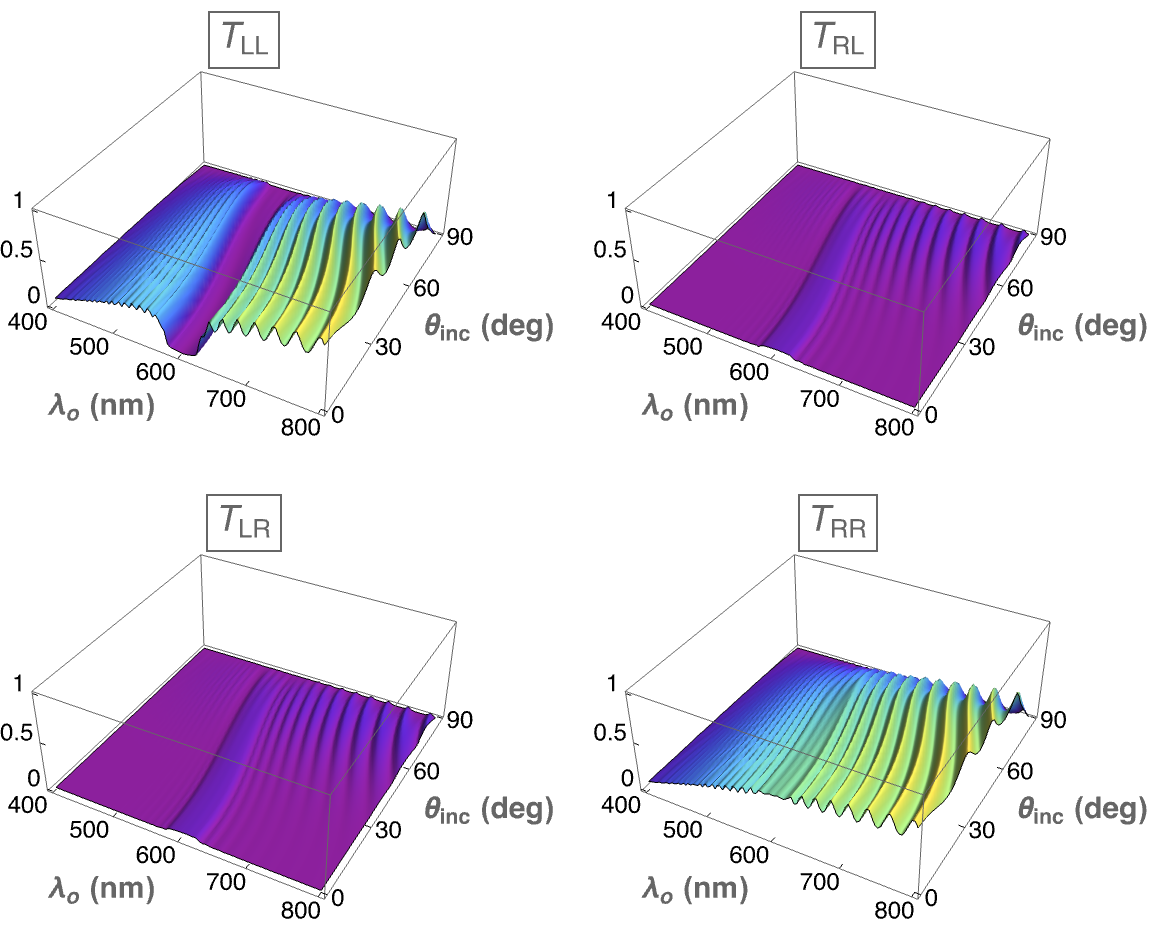} 
\hfill \vspace{0mm} 
 \caption{\label{CircTransmittance-3} $h=-1$, $\thetainc\in[0\deg,90\deg)$, and $\psi=0\deg$}
\end{subfigure}  \hspace{-5mm} 
\begin{subfigure}{0.5\textwidth}
\centering
\includegraphics[width=7cm]{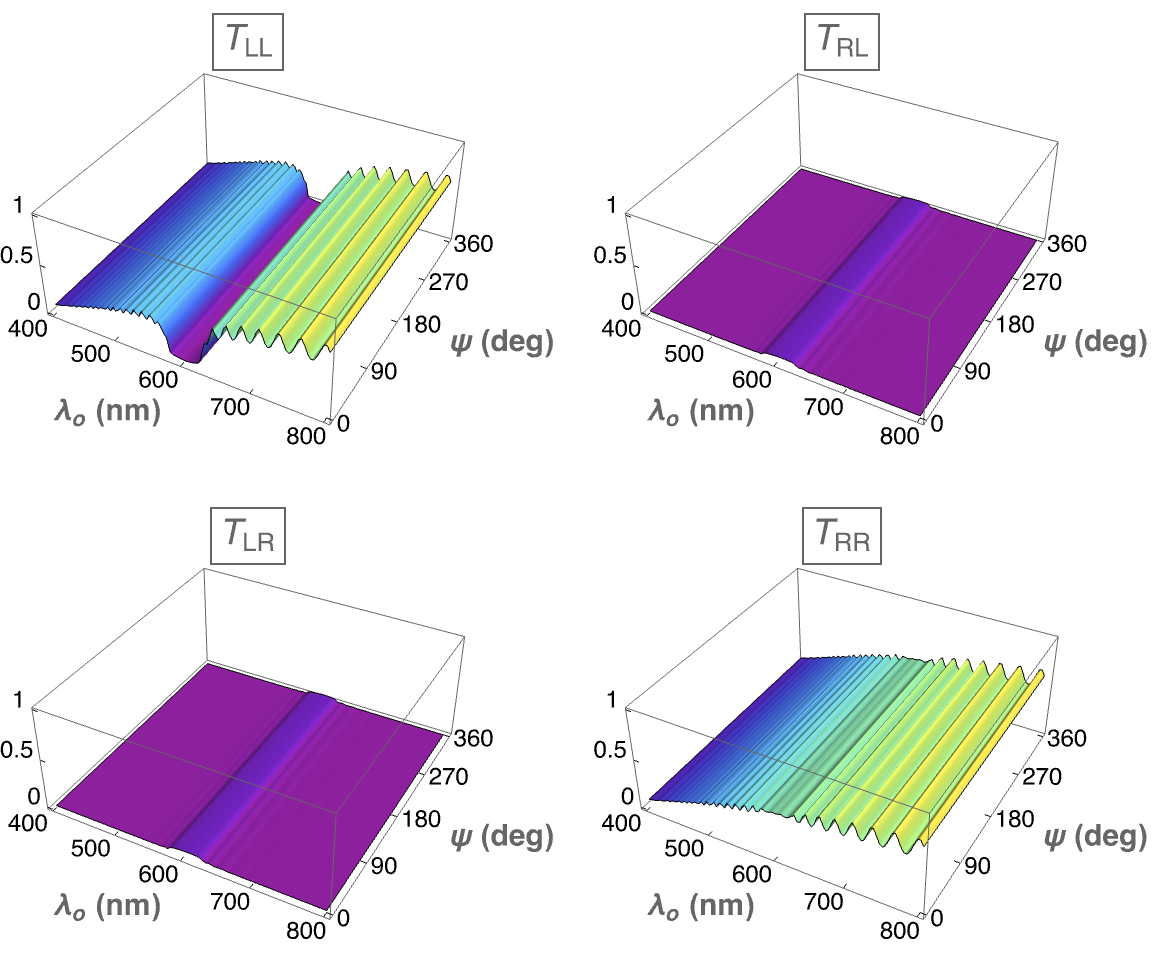} 
\hfill \vspace{0mm} 
 \caption{\label{CircTransmittance-4} $h=-1$, $\thetainc=0\deg$, and $\psi\in[0\deg,360\deg)$}
\end{subfigure} 
	\caption{Spectral variations of circular transmittances $T_{\mu\nu}(\lambdao,\thetainc,\psi)$,   $\mu\in\left\{\rm L,R\right\}$ and $\nu\in\left\{\rm L,R\right\}$, 
	when
	(a,b) $h=1$ and (c,d) $h=-1$. (a,c) $\thetainc\in [0\deg,90\deg)$ and $\psi= 0\deg$;  (b,d) $\thetainc=0\deg$ and $\psi\in[ 0\deg,360\deg)$.
	}
	\label{CircTransmittance}
\end{figure}

 The fraction of the power density of the incident plane wave that
 is not reflected is either transmitted into the half-space $z>L$ or absorbed in the SCM ($0<z<L$). Since $\Im\left(\eps_\ell\right)>0$, $\ell\in\lec{\rm a,b,c}\ric$,
 there is some absorption \cite{FialloPhD,McAtee2018}. Accordingly, in Fig.~\ref{CircTransmittance}, the circular Bragg phenomenon is manifested
 as a low-transmittance trough in the plots
 of $\TRR$ for $h=1$ and in the plots of $\TLL$ for $h=-1$, that trough being absent in  the
 plots of $\TLL$ for $h=1$ and in the plots of $\TRR$ for $h=-1$. Vestigial presence of the trough
 in the plots of $\TLR$ and $\TRL$ for $h=\pm1$ should also be noted.

\subsection{Linear remittances}
The square of the magnitude
of a linear reflection or transmission coefficient is the corresponding linear
reflectance or transmittance;  thus, $\Rsp = \vert \rsp\vert^2\in[0,1]$ is
the linear reflectance corresponding to the linear reflection coefficient $\rsp$ and
$\Tps = \vert \tps\vert^2\in[0,1]$ is
the linear transmittance corresponding to the linear transmission coefficient $\tps$,
etc. 
The total linear reflectances are given by
\begin{equation}
\left. \begin{array}{l}
\Rs =  \Rss+\Rps\in[0,1]\\[5pt]
\Rp=\Rpp+\Rsp\in[0,1]
\end{array}\ric
\label{def-Rlin}
\end{equation}
and the total linear transmittances by
\begin{equation}
\left. \begin{array}{l}
\Ts=  \Tss+\Tps\in[0,1]\\[5pt]
\Tp=\Tpp+\Tps\in[0,1]
\end{array}\ric\,.
\label{def-Tlin}
\end{equation}
The inequalities (\ref{eq16}) still hold with $\ell\in\lec{\rm s,p}\ric$
and convert to equalities
 only if the SCM is non-dissipative
at a particular frequency of interest. 

\begin{figure}[ht]
\begin{subfigure}{0.5\textwidth}
\centering
\includegraphics[width=7cm]{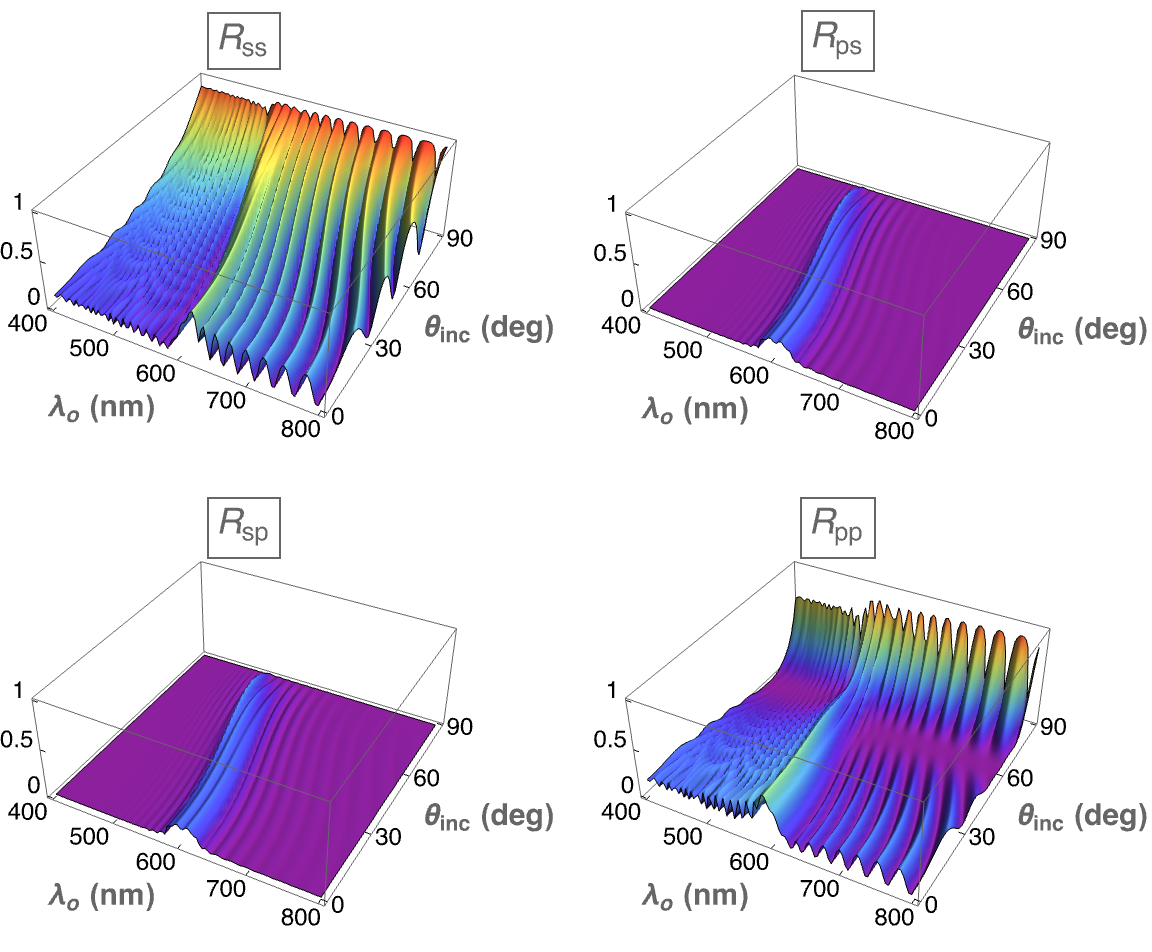} 
\hfill \vspace{0mm} 
 \caption{\label{LinReflectance-1} $h=1$, $\thetainc\in[0\deg,90\deg)$, and $\psi=0\deg$}
\end{subfigure}  \hspace{-5mm} 
\begin{subfigure}{0.5\textwidth}
\centering
\includegraphics[width=7cm]{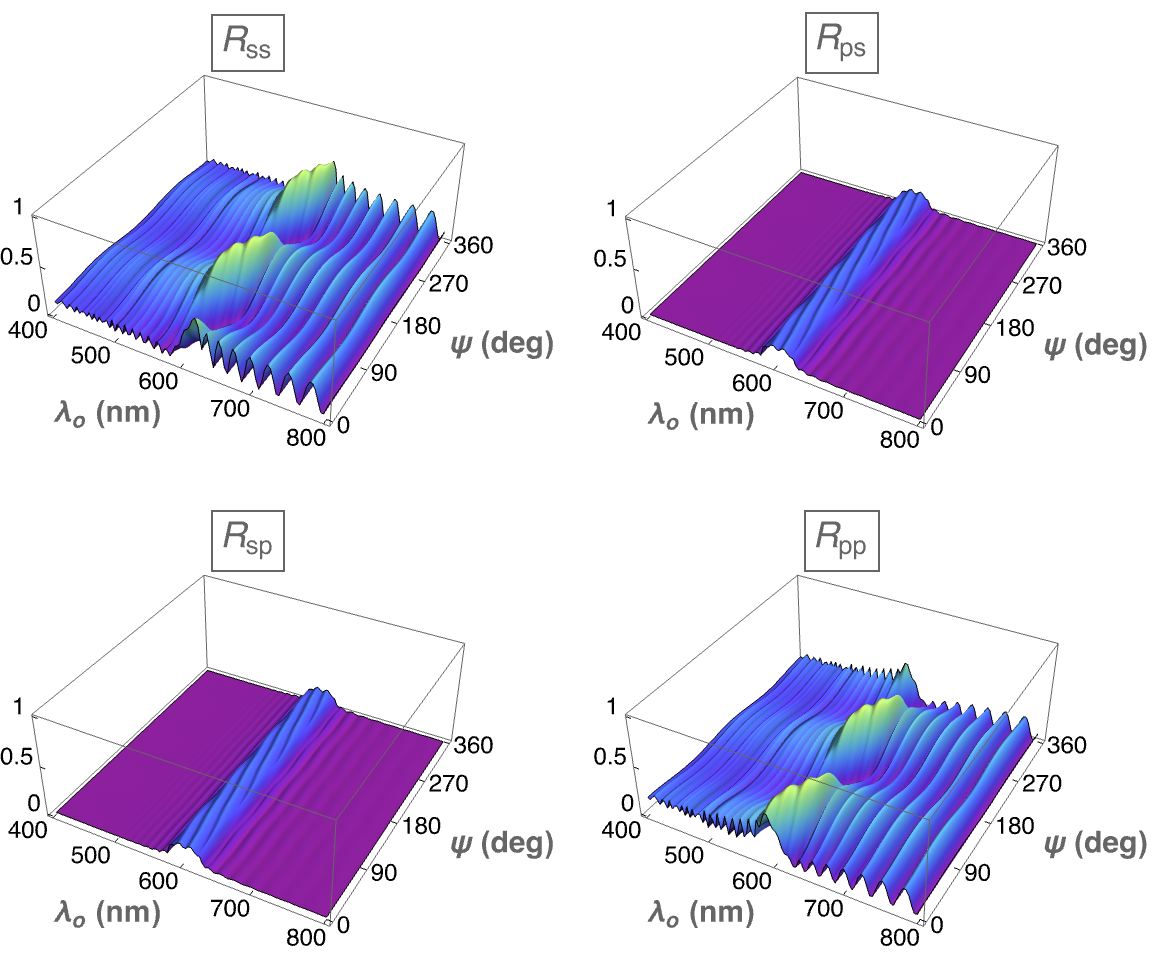} 
\hfill \vspace{0mm} 
 \caption{\label{LinReflectance-2} $h=1$, $\thetainc=0\deg$, and $\psi\in[0\deg,360\deg)$}
\end{subfigure} 
\begin{subfigure}{0.5\textwidth}
\centering
\includegraphics[width=7cm]{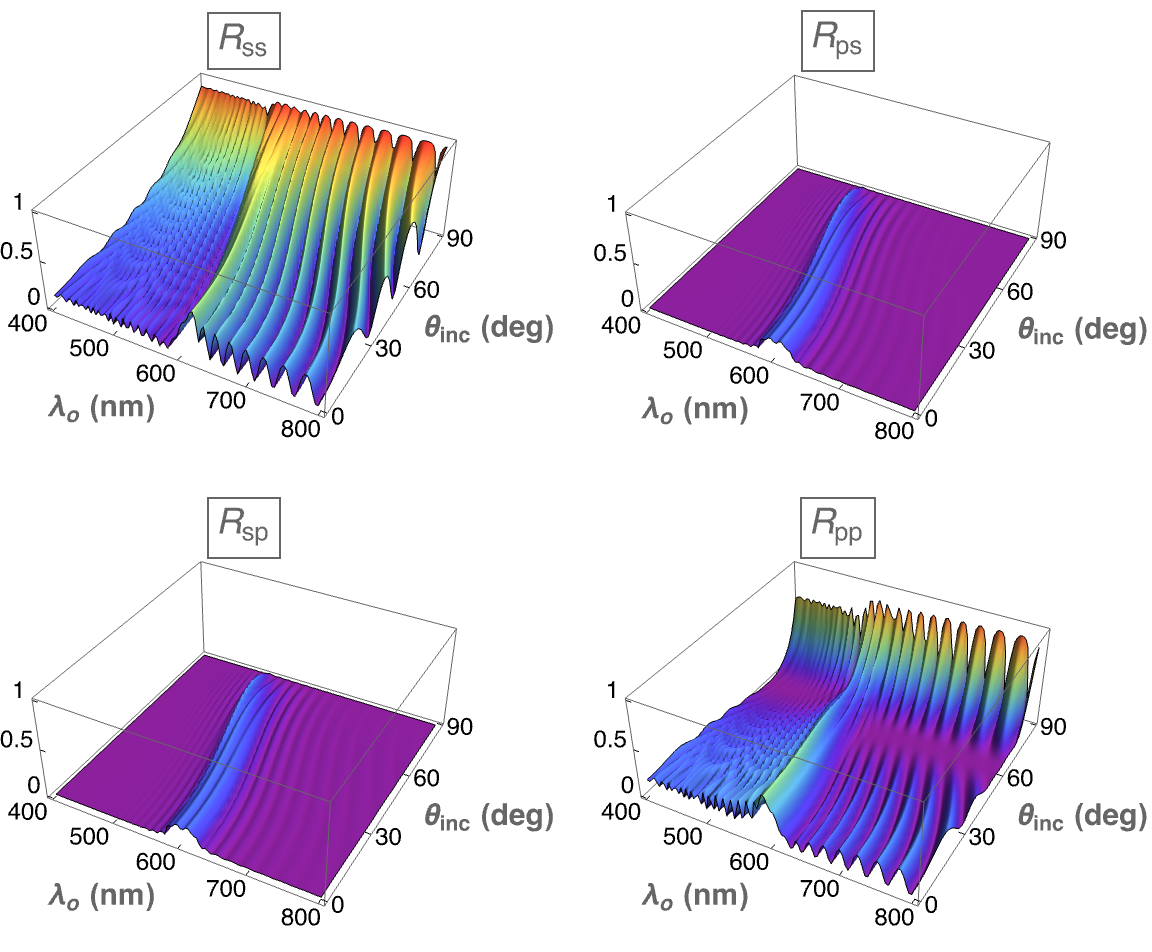} 
\hfill \vspace{0mm} 
 \caption{\label{LinReflectance-3} $h=-1$, $\thetainc\in[0\deg,90\deg)$, and $\psi=0\deg$}
\end{subfigure}  \hspace{-5mm} 
\begin{subfigure}{0.5\textwidth}
\centering
\includegraphics[width=7cm]{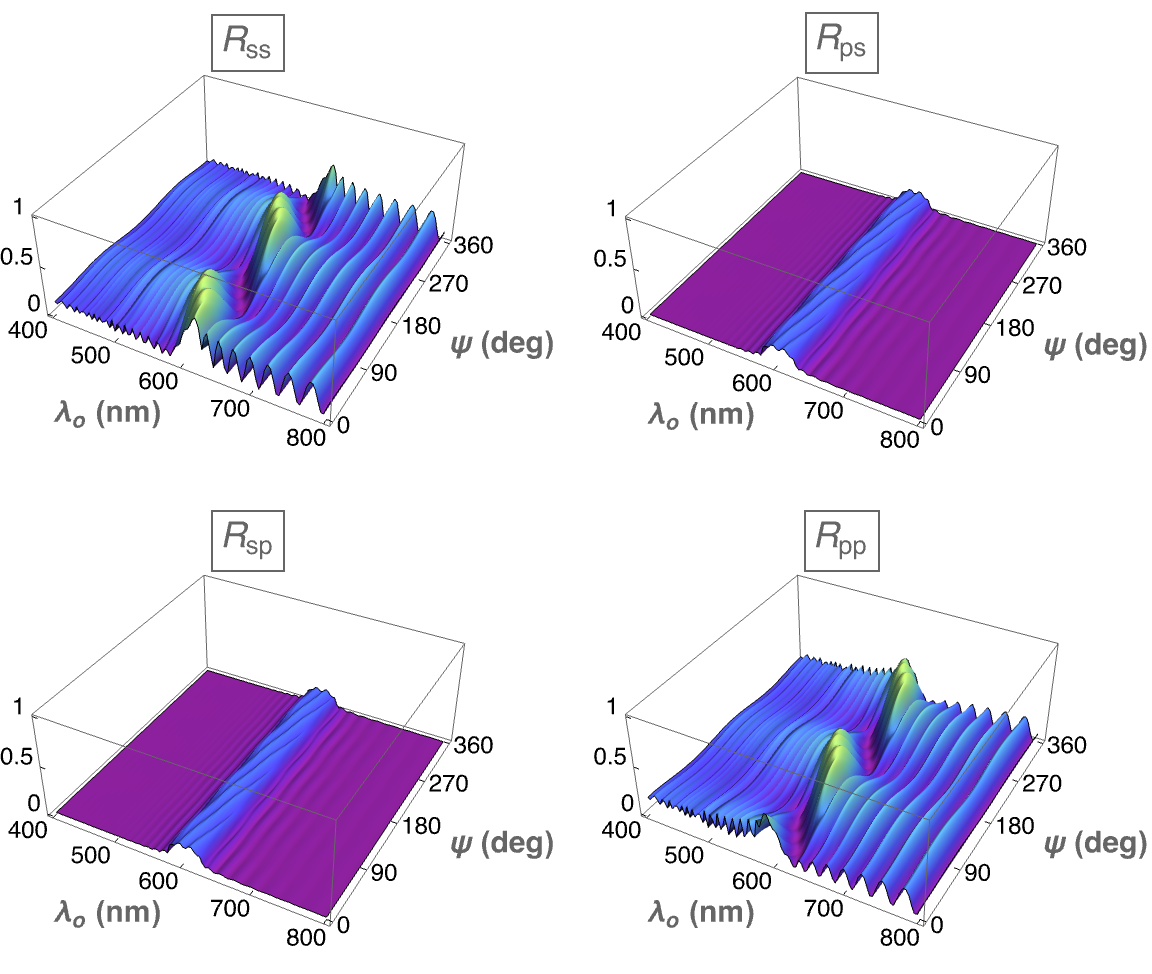} 
\hfill \vspace{0mm} 
 \caption{\label{LinReflectance-4} $h=-1$, $\thetainc=0\deg$, and $\psi\in[0\deg,360\deg)$}
\end{subfigure}
	\caption{Spectral variations of linear reflectances $R_{\mu\nu}(\lambdao,\thetainc,\psi)$,   $\mu\in\left\{\rm s,p\right\}$ and $\nu\in\left\{\rm s,p\right\}$, 
	when
	(a,b) $h=1$ and (c,d) $h=-1$. (a,c) $\thetainc\in [0\deg,90\deg)$ and $\psi= 0\deg$;  (b,d) $\thetainc=0\deg$ and $\psi\in[ 0\deg,360\deg)$.
	}
	\label{LinReflectance}
\end{figure}

Linear reflectances can be written in terms of circular reflectances \cite{STFbook}. Therefore, the circular
Bragg regime is evident in the spectral
variations of both co-polarized linear reflectances as a
medium-reflectance ridge and in the spectral variations of both cross-polarized reflectances
as a low-reflectance ridge \cite{FialloPhD,McAtee2018},  in Fig.~\ref{LinReflectance} for both $h=1$ and $h=-1$.
For fixed $\psi$, the  ridge curves towards
 shorter wavelengths as $\thetainc$ increases. For fixed $\thetainc$, the low-reflectance
 ridge in the plots of $\Rps$ and $\Rsp$ is more or less invariant with respect to $\psi$;
 but the medium-reflectance ridge  in the plots of $\Rss$ and $\Rpp$ has two periods of undulations.
 
\begin{figure}[ht]
\begin{subfigure}{0.5\textwidth}
\centering
\includegraphics[width=7cm]{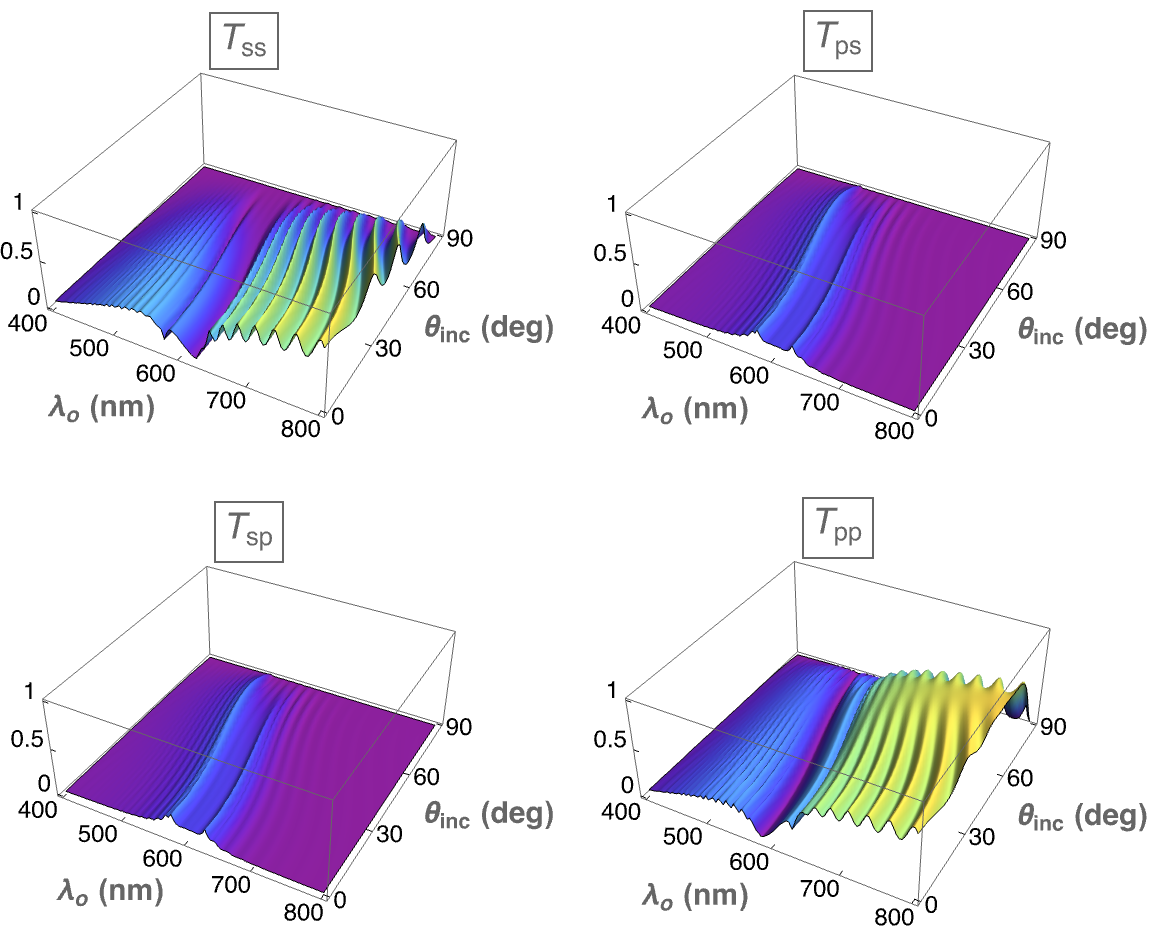} 
\hfill \vspace{0mm} 
 \caption{\label{LinTransmittance-1} $h=1$, $\thetainc\in[0\deg,90\deg)$, and $\psi=0\deg$}
\end{subfigure}  \hspace{-5mm} 
\begin{subfigure}{0.5\textwidth}
\centering
\includegraphics[width=7cm]{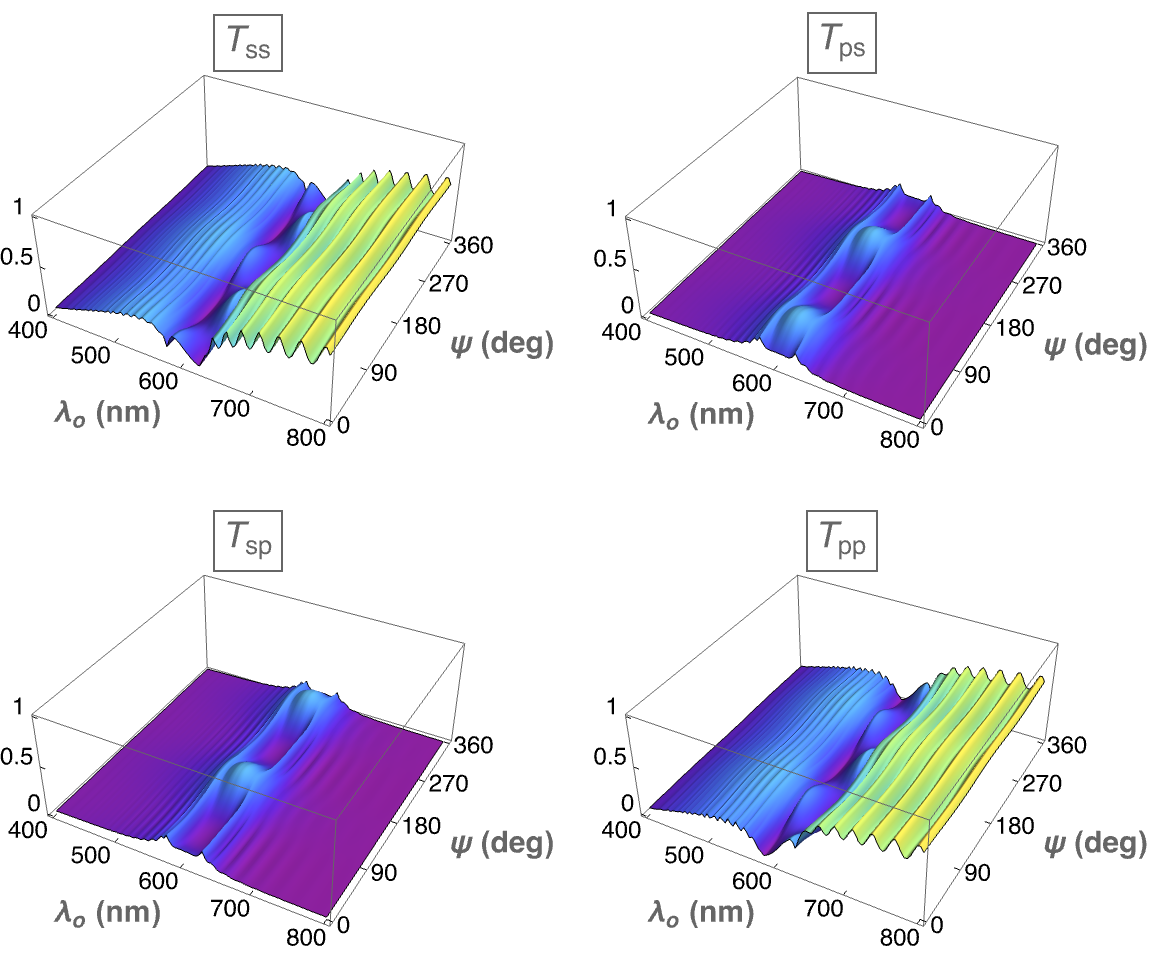} 
\hfill \vspace{0mm} 
 \caption{\label{LinTransmittance-2} $h=1$, $\thetainc=0\deg$, and $\psi\in[0\deg,360\deg)$}
\end{subfigure} 
\\
\begin{subfigure}{0.5\textwidth}
\centering
\includegraphics[width=7cm]{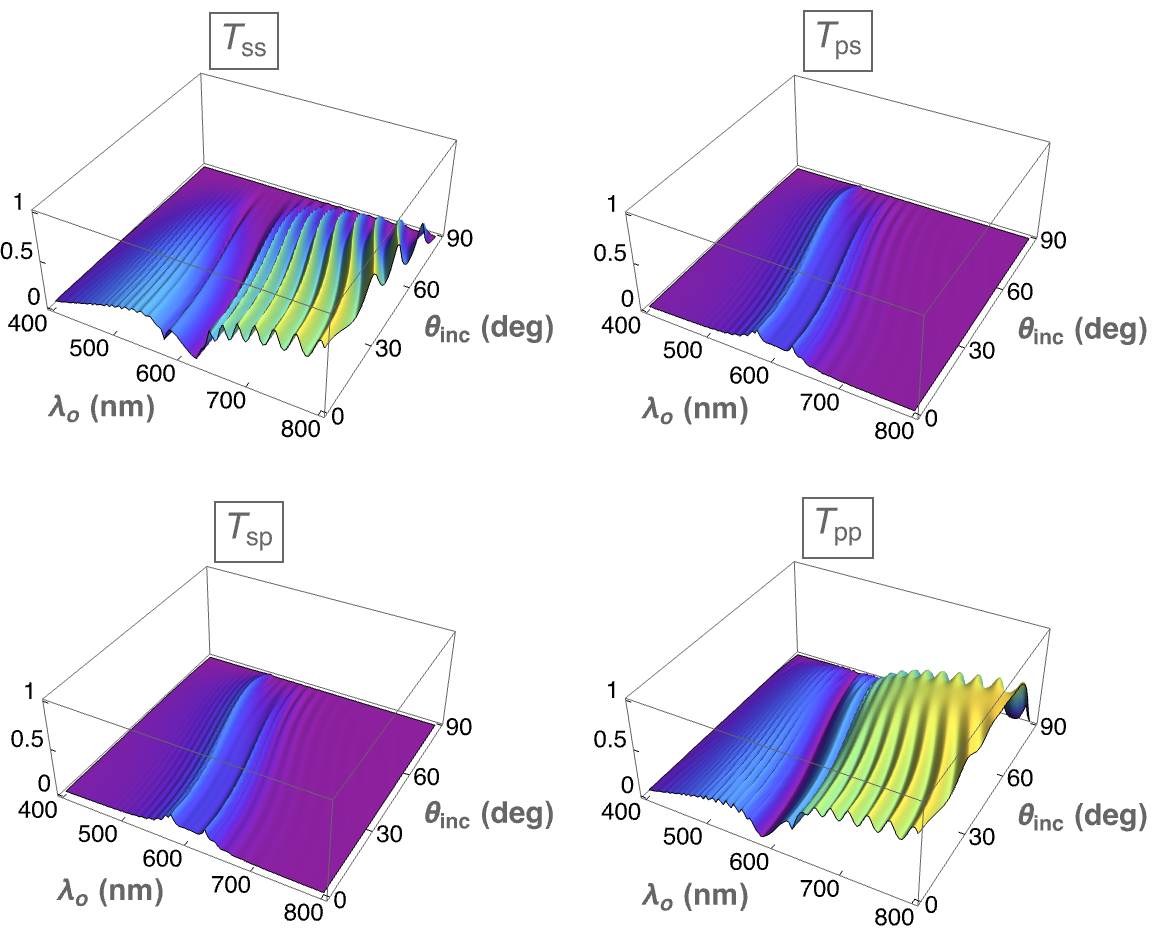} 
\hfill \vspace{0mm} 
 \caption{\label{LinTransmittance-3} $h=-1$, $\thetainc\in[0\deg,90\deg)$, and $\psi=0\deg$}
\end{subfigure}  \hspace{-5mm} 
\begin{subfigure}{0.5\textwidth}
\centering
\includegraphics[width=7cm]{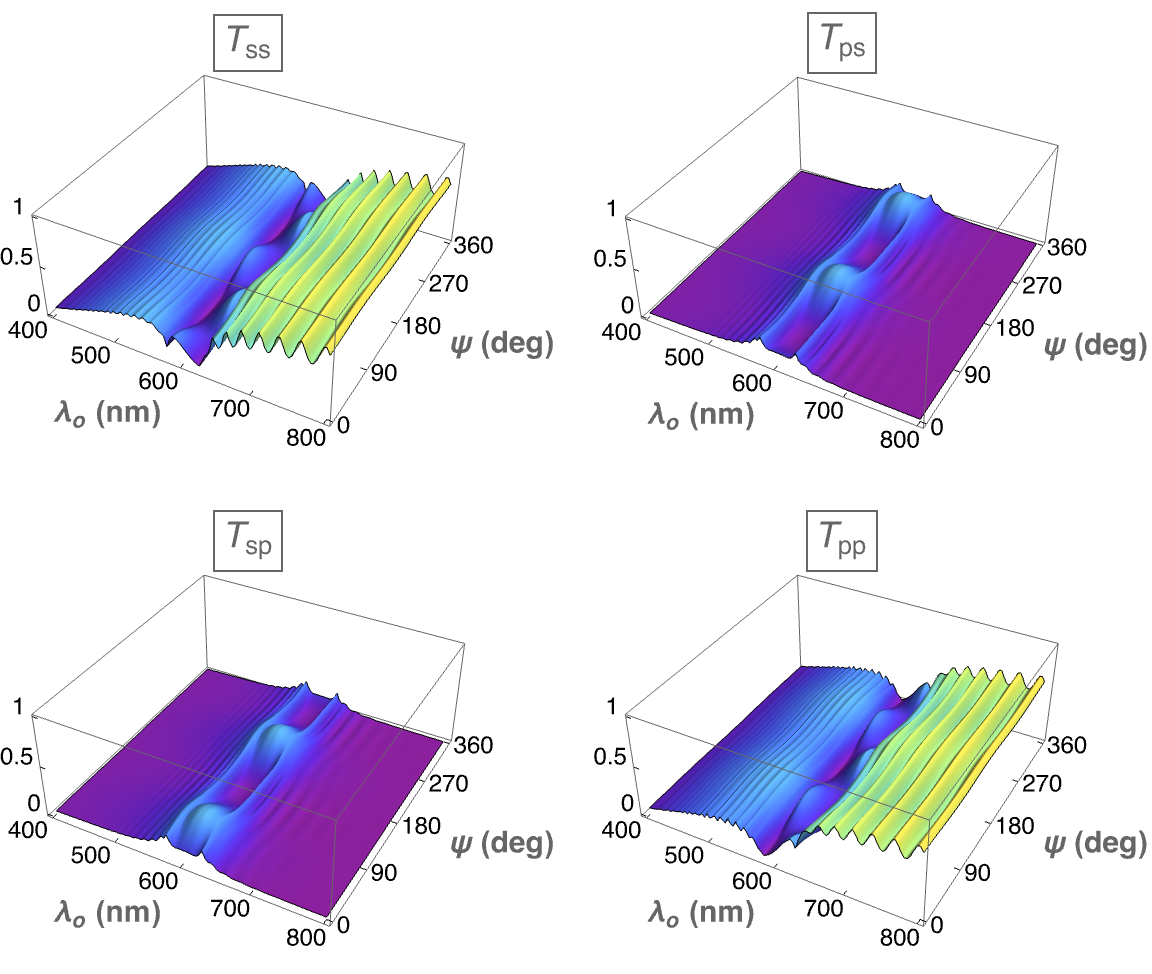} 
\hfill \vspace{0mm} 
 \caption{\label{LinTransmittance-4} $h=-1$, $\thetainc=0\deg$, and $\psi\in[0\deg,360\deg)$}
 \end{subfigure} 
	\caption{Spectral variations of linear transmittances $T_{\mu\nu}(\lambdao,\thetainc,\psi)$,   $\mu\in\left\{\rm s,p\right\}$ and $\nu\in\left\{\rm s,p\right\}$, 
	when
	(a,b) $h=1$ and (c,d) $h=-1$. (a,c) $\thetainc\in [0\deg,90\deg)$ and $\psi= 0\deg$;  (b,d) $\thetainc=0\deg$ and $\psi\in[ 0\deg,360\deg)$.
	}	\label{LinTransmittance}
\end{figure}

 Linear transmittances can be written in terms of circular transmittances \cite{STFbook,Lakh2024josab}.
The spectral
variations of both co-polarized linear transmittances exhibit a
medium-transmittance trough and both cross-polarized linear transmittances
show a low-reflectance ridge \cite{FialloPhD,McAtee2018}, in Fig.~\ref{LinTransmittance} for both $h=1$ and $h=-1$. Indicative
of the circular Bragg phenomenon, these features curve
towards shorter wavelengths as $\thetainc$ increases when $\psi$ is held fixed.
For fixed $\thetainc$ but variable $\psi$, the low-transmittance
 ridge in the plots of $\Tps$ and $\Tsp$  
 and the medium-transmittance trough  in the plots of $\Tss$ and $\Tpp$ have two periods of undulations.

\subsection{Circular and linear dichroisms}
With 
\begin{equation}
\left. \begin{array}{l}
\AL = 1- \left(\RL+\TL\right)\in[0,1]\\
[5pt]
\AR = 1- \left(\RR  +\TR \right)  \in[0,1]
\end{array}\ric
\end{equation}
as the circular absorptances,  
\begin{equation}
\tcd = \sqrt {\AR} - \sqrt{\AL}\in[-1,1]
\end{equation}
is the {\it true} circular dichroism which quantitates the circular-polarization-dependence of
absorption. The {\it apparent} circular dichroism
\begin{equation}
\acd =  \sqrt {\TR} - \sqrt{\TL}\in[-1,1]
\end{equation}
is a measure of the circular-polarization-state-dependence of transmission \cite{McAtee2018}.
Whereas $\acd$  may not equal zero for a non-dissipative
SCM, $\tcd$ must be. 

\begin{figure}[ht]
\begin{subfigure}{0.5\textwidth}
\centering
\includegraphics[width=7cm]{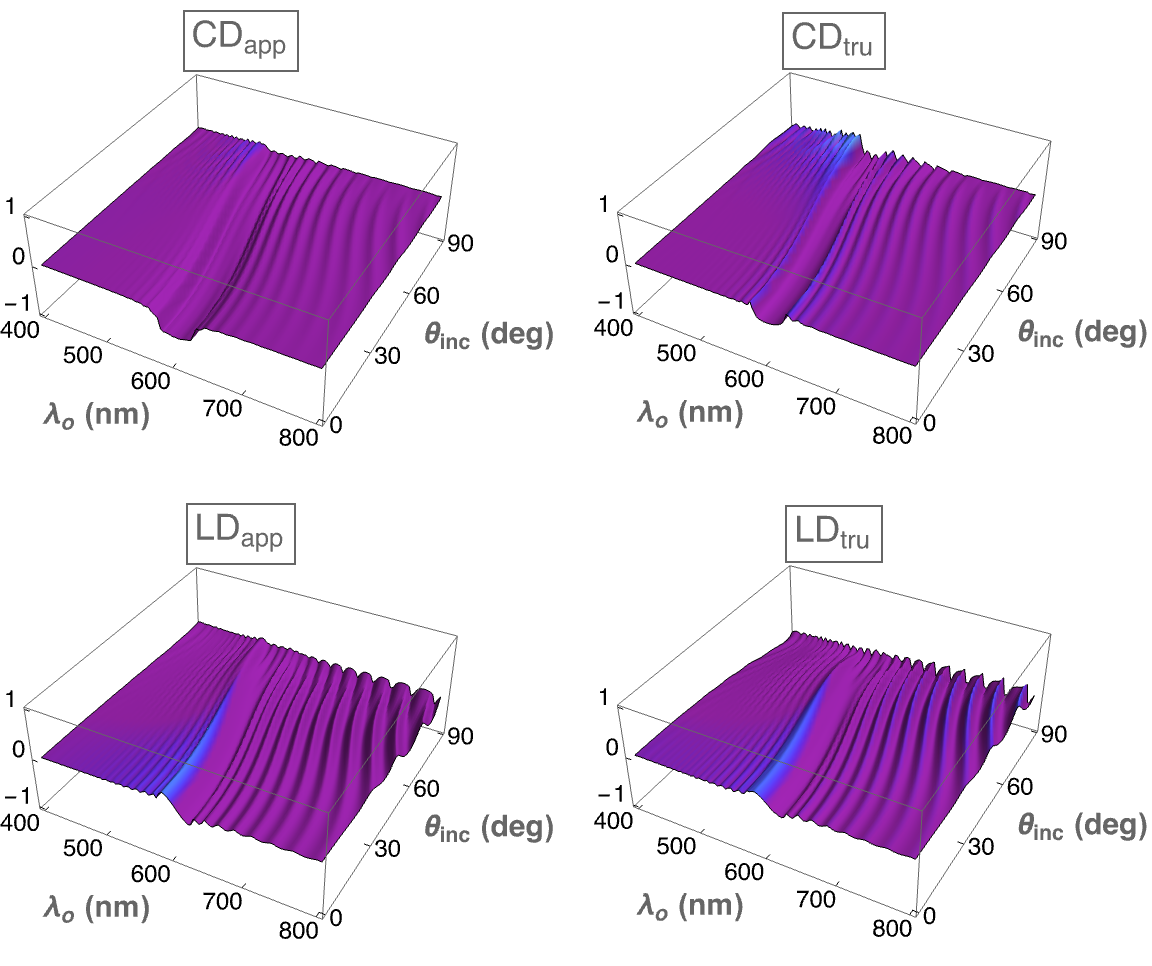} 
\hfill \vspace{0mm} 
 \caption{\label{Dichroism-1} $h=1$, $\thetainc\in[0\deg,90\deg)$, and $\psi=0\deg$}
\end{subfigure}  \hspace{-5mm} 
\begin{subfigure}{0.5\textwidth}
\centering
\includegraphics[width=7cm]{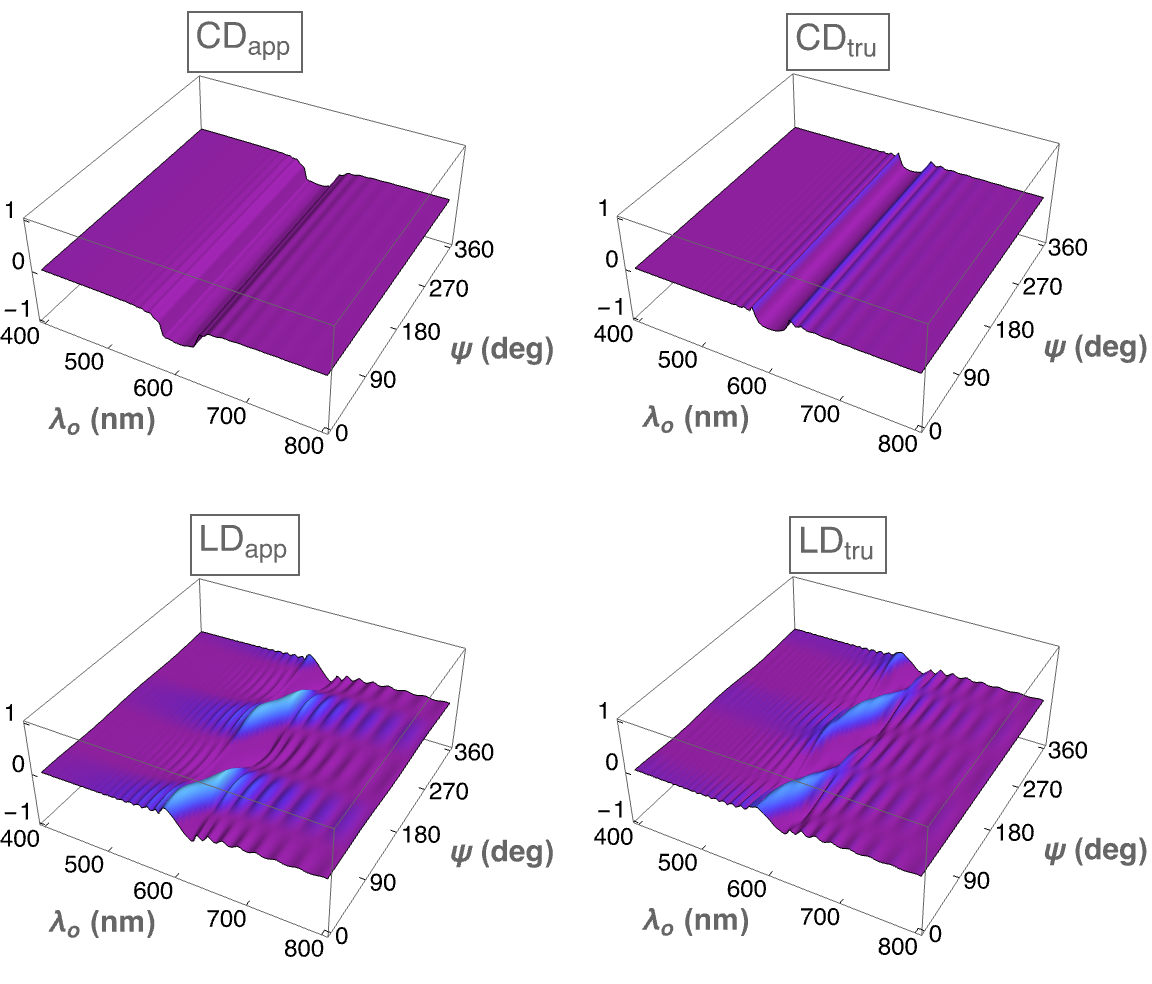} 
\hfill \vspace{0mm} 
 \caption{\label{Dichroism-2} $h=1$, $\thetainc=0\deg$, and $\psi\in[0\deg,360\deg)$}
\end{subfigure} 
\\
\begin{subfigure}{0.5\textwidth}
\centering
\includegraphics[width=7cm]{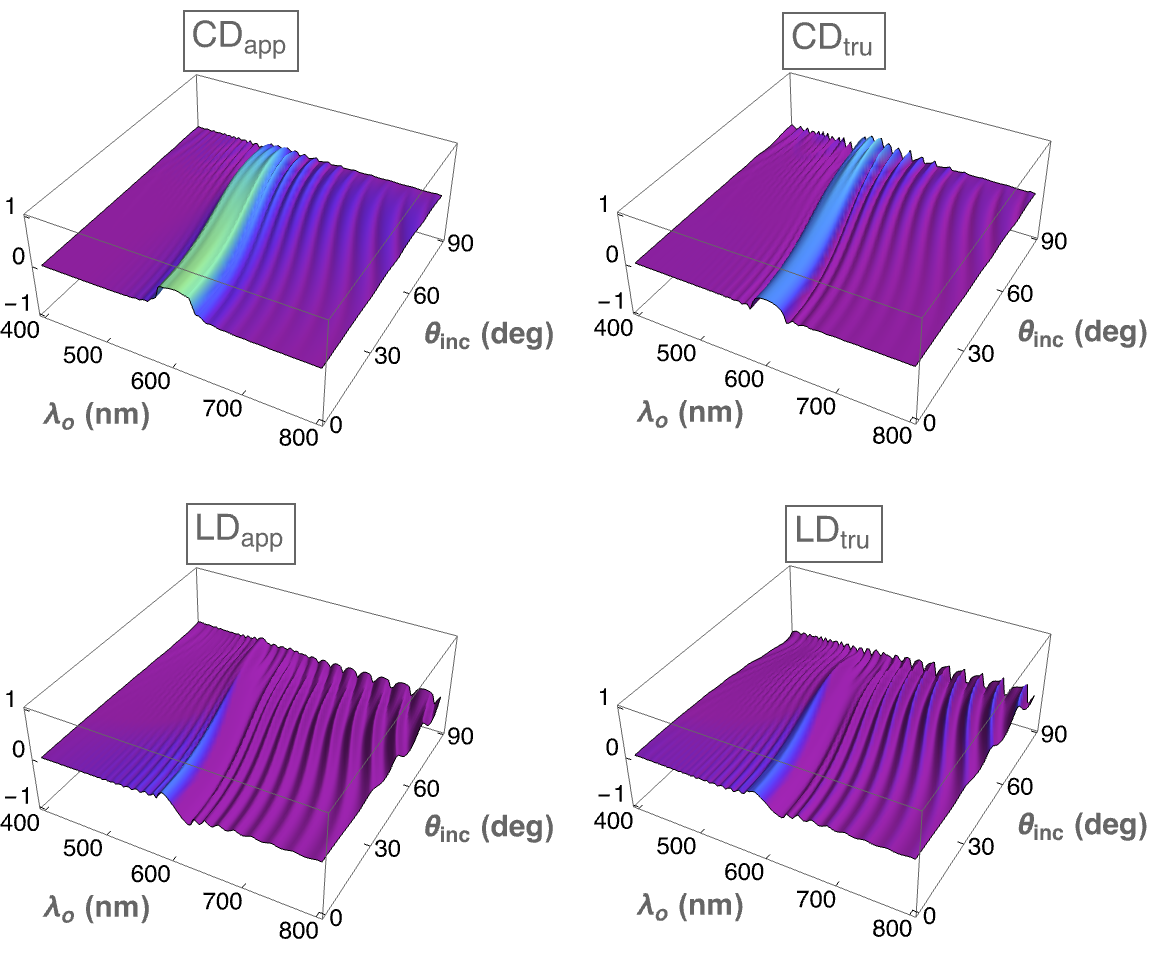} 
\hfill \vspace{0mm} 
 \caption{\label{Dichroism-3} $h=-1$, $\thetainc\in[0\deg,90\deg)$, and $\psi=0\deg$}
\end{subfigure}  \hspace{-5mm} 
\begin{subfigure}{0.5\textwidth}
\centering
\includegraphics[width=7cm]{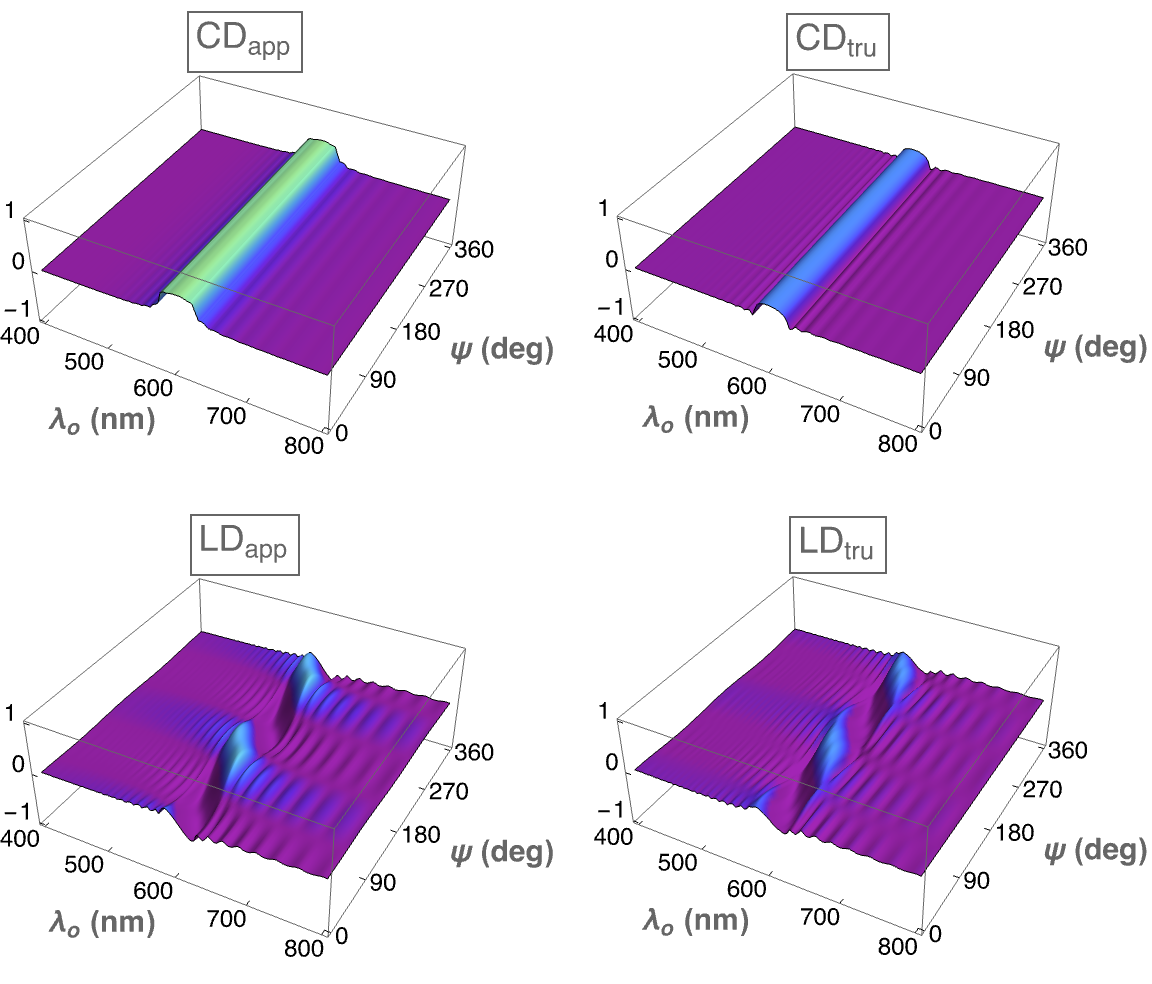} 
\hfill \vspace{0mm} 
 \caption{\label{Dichroism-4} $h=-1$, $\thetainc=0\deg$, and $\psi\in[0\deg,360\deg)$}
\end{subfigure} 
	\caption{Spectral variations of apparent circular dichroism $\acd(\lambdao,\thetainc,\psi)$, true circular dichroism $\tcd(\lambdao,\thetainc,\psi)$, apparent linear dichroism $\ald(\lambdao,\thetainc,\psi)$, 
	and true   linear dichroism $\tld(\lambdao,\thetainc,\psi)$, when
	(a,b) $h=1$ and (c,d) $h=-1$. (a,c) $\thetainc\in [0\deg,90\deg)$ and $\psi= 0\deg$;  (b,d) $\thetainc=0\deg$ and $\psi\in[ 0\deg,360\deg)$.
	}
	\label{Dichroism}
\end{figure}

Figure~\ref{Dichroism} contains plots of the spectral variations of both $\acd$ and $\tcd$
in relation to the direction of plane-wave incidence. The circular Bragg phenomenon is evident
as a trough in all plots of  $\acd$ and $\tcd$ for $h=1$, and as a ridge in    all plots of  $\acd$ and $\tcd$ for $h=-1$.
Furthermore, the quantities $h\acd$ and $h\tcd$ are invariant if the sign of $h$ is changed. These features
curve towards shorter wavelengths as $\thetainc$ increases while $\psi$ is held fixed, 
as has been experimentally verified \cite{McAtee2018}. For normal incidence
(i.e., $\thetainc=0\deg$), the effect of $\psi$ is minimal.
 
Similarly to the circular absorptances,
\begin{equation}
\left. \begin{array}{l}
\As = 1- \left(\Rss +\Ts\right)\in[0,1]\\
[5pt]
\Ap = 1- \left(\Rp +  \Tp\right) \in[0,1] 
\end{array}\ric
\end{equation}
are the linear absorptances.   The {\it true} linear dichroism is defined as \cite{McAtee2018}
\begin{equation}
\tld=\sqrt{\As}-\sqrt{\Ap}\in[-1,1]
\end{equation}
and the {\it apparent} linear dichroism as
\begin{equation}
\ald=\sqrt{\Ts}-\sqrt{\Tp}\in[-1,1]\,.
\end{equation}
Whereas
$\tld \equiv 0$ for a non-dissipative SCM, $\ald$ may not
be null valued.

Figure~\ref{Dichroism} also contains plots of the spectral variations of both $\ald$ and $\tld$
in relation to the direction of plane-wave incidence. The circular Bragg phenomenon is featured
in all plots.
For fixed $\psi$, the feature curves towards
 shorter wavelengths as $\thetainc$ increases, which has been experimentally verified \cite{McAtee2018}. For normal
 incidence, the feature
 has two undulations with increasing $\psi$, and the   replacement $h \to -h$ affects both $\ald$ and $\tld$ non-trivially.

\section{Phase-dependent quantities}\label{pdq}

\subsection{Ellipticity and optical rotation}

The most general plane wave in free space
is elliptically polarized \cite{BW}.
Signed ellipticity functions  
\begin{equation}
\left. \begin{array}{l}
\EFinc =\displaystyle{ 
- 2\, \frac{ \Im \left( {\aas} \aap^* \right)}
{ \vert {\aas}  \vert ^2
+ \vert {\aap}  \vert ^2 } 
}\\ [10pt]
\EFref =\displaystyle{ 
- 2\, \frac{ \Im \left( {\bbs} \bbp^* \right)}
{ \vert {\bbs}  \vert ^2
+ \vert {\bbp}  \vert ^2 } 
}\\ [10pt]
\EFtra = \displaystyle{
- 2\, \frac{ \Im \left( \ccs \ccp^* \right)}
{   \vert \ccs  \vert ^2
+  \vert \ccp \vert ^2  }
} \end{array} \ric \, 
\end{equation}
characterize the shapes of the vibration ellipses
of the incident, reflected, and   transmitted plane waves. 

Note that $EF^{\ell}\in[-1,1]$, $\ell \in \lec{\rm inc}, \, {\rm ref},\,  {\rm tr}\ric$.
The magnitude of $EF^{\ell}$
is the ellipticity of the plane wave labelled $\ell$. The vibration
ellipse simplifies to a circle when
 $\big \vert  EF^{\ell} \big \vert  = 1$
(circular polarization state), and it
degenerates into  a straight line when
$EF^{\ell} = 0$  (linear polarization state). 
The plane wave is
left-handed  for $EF^{\ell} > 0$ and right-handed
for $EF^{\ell} < 0$.

The major axes of the vibration ellipses of the incident and the
reflected/transmitted plane wave may not coincide, the angular offset between the two major axes
known as optical rotation.  
The
 auxiliary vectors \cite{STFbook}
\begin{equation}
\left.
\begin{array}{l}
\#{F}^{\rm inc} =  \lec 1+ \les1 - \left(\EFinc\right)^2\ris^{1/2} \ric \,
\Re \left( {\aas}\sp +{\aap}\pinc  \right) +\EFinc \,
\Im \left({\aas}\pinc -{\aap}\sp  \right) \\ [8pt]
\#{F}^{\rm ref} =  \lec 1+ \les1 - \left(\EFref\right)^2\ris^{1/2} \ric \,
\Re \left( {\bbs}\sp +{\bbp}\pref  \right) +\EFref \,
\Im \left({\bbs}\pref -{\bbp}\sp  \right) \\ [8pt]
\#{F}^{\rm tr} =  \lec 1+ \les1 - \left(\EFtra\right)^2\ris^{1/2} \ric \,
\Re \left( {\ccs}\sp +{\ccp}\pinc  \right) +\EFtra \,
\Im \left({\ccs}\pinc -{\ccp}\sp  \right)
\end{array}\ric\, 
\end{equation}
 are parallel to the major axes of the
respective vibration ellipses. Therefrom, the  angles $\tau^{\rm inc}$, $\tau^{\rm ref}$,
and $\tau^{\rm tr}$ are calculated using the following expressions:
\begin{equation}
\left.\begin{array}{ll}
\cos\tau^{\ell} = \left( {\#{F}^{\ell}\.\sp}\right)/
{\big \vert  \#{F}^{\ell} \big \vert } \,, &\quad \ell \in \lec{\rm inc}, \,  {\rm ref},\,{\rm tr}\ric
\\[5pt]
\sin\tau^{\ell} = \left( {\#{F}^{\ell}\.\pinc}\right)/
{\big \vert  \#{F}^{\ell} \big \vert } \,, &\quad \ell \in \lec{\rm inc}, \,{\rm tr}\ric
\\[5pt]
\sin\tau^{\rm ref} = \left( {\#{F}^{\rm ref}\.\pref}\right)/
{\big \vert  \#{F}^{\rm ref} \big \vert } 
\end{array}\right\} \, .
\end{equation}
The optical rotation of the reflected/transmitted plane wave
then is the angle
\begin{equation}
OR^\ell = \lec
\begin{array}{lll}
\tau^{\ell}-\tau^{\rm inc}+\pi\, ,\quad &{\rm if}\, -\pi\leq\tau^{\ell}-\tau^{\rm inc} \leq -\pi/2\,,\\[5pt]
\tau^{\ell}-\tau^{\rm inc}\, ,\quad &{\rm if}\, \big \vert \tau^{\ell}-\tau^{\rm inc}\big \vert  < \pi/2\,,\\[5pt]
\tau^{\ell}-\tau^{\rm inc}-\pi\, ,\quad &{\rm if}\, \pi/2\leq\tau^{\ell}-\tau^{\rm inc} \leq \pi\,,
\end{array}\right.\,, \quad\ell \in \lec{\rm ref}, \,{\rm tr}\ric\,.
\end{equation}
 
The ellipticity function of the reflected/transmitted plane wave is denoted by $EF^\ell_{\rm s}$
and $EF^\ell_{\rm p}$, respectively, and
the
optical rotation of the reflected/transmitted plane wave is denoted by $OR^\ell_{\rm s}$ and $OR^\ell_{\rm p}$, respectively,
for   incident perpendicular-polarized and parallel-polarized plane waves, $\ell \in \lec{\rm ref}, \,{\rm tr}\ric$.

\begin{figure}[ht]
\begin{subfigure}{0.5\textwidth}
\centering
\includegraphics[width=7cm]{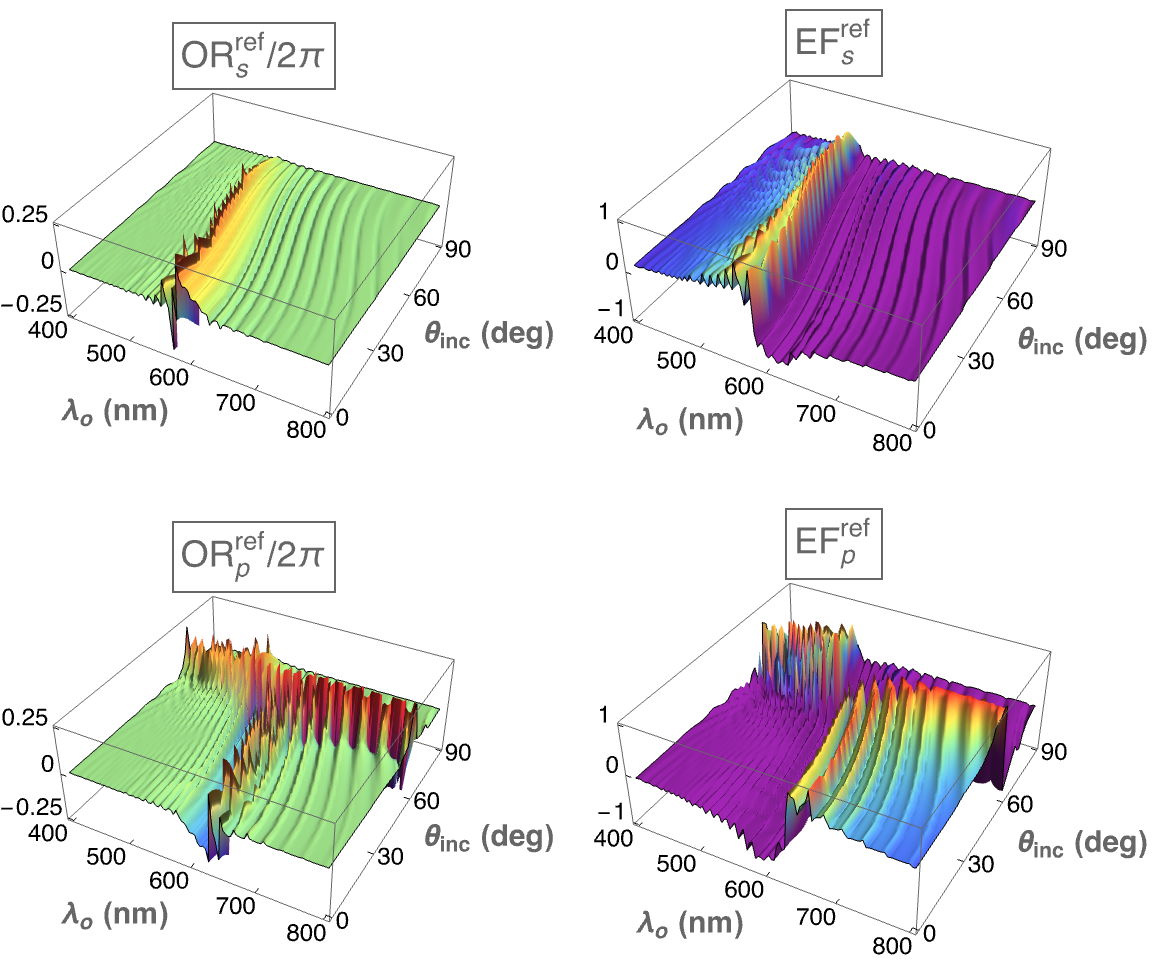} 
\hfill \vspace{0mm} 
 \caption{\label{OREFref-1} $h=1$, $\thetainc\in[0\deg,90\deg)$, and $\psi=0\deg$}
\end{subfigure}  \hspace{-5mm} 
\begin{subfigure}{0.5\textwidth}
\centering
\includegraphics[width=7cm]{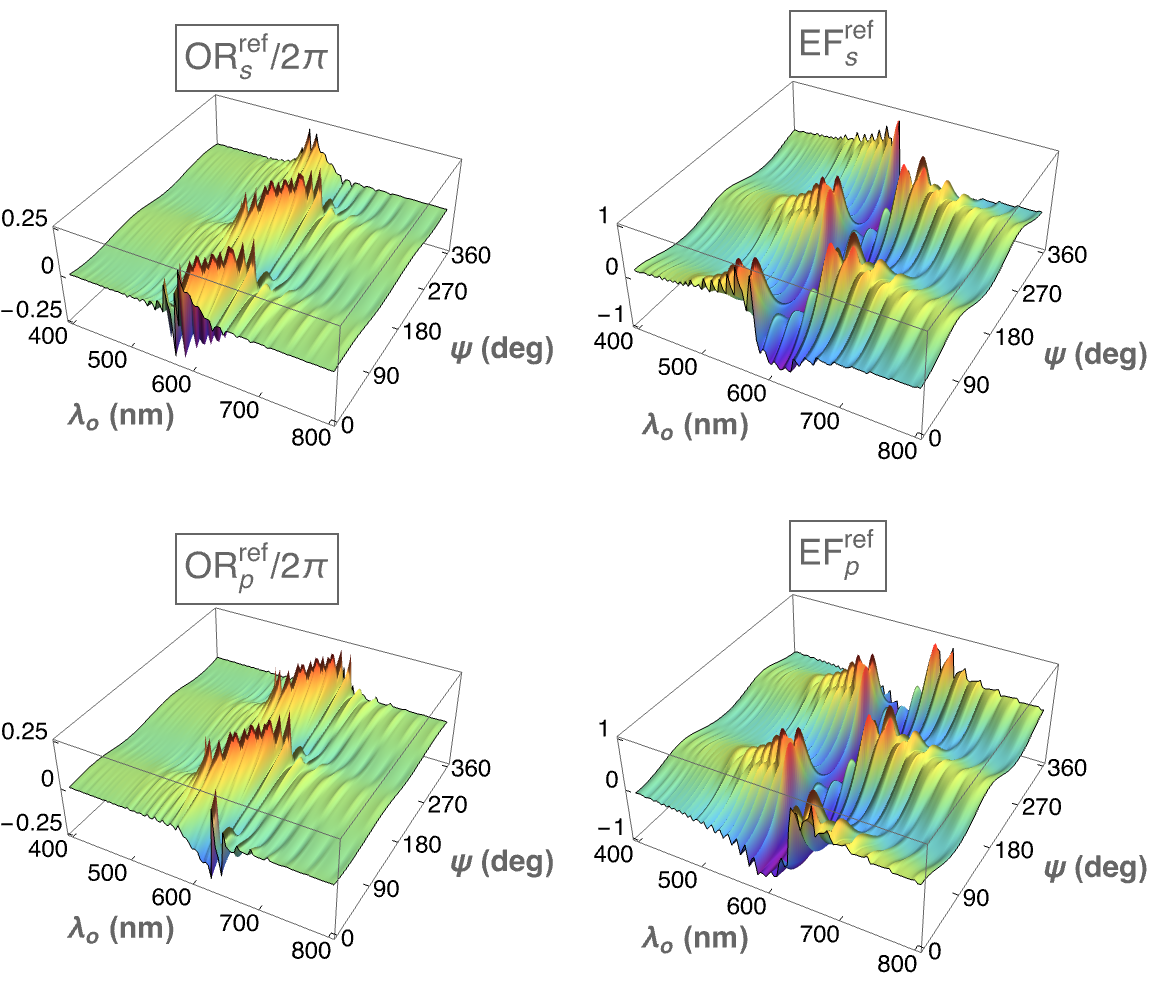} 
\hfill \vspace{0mm} 
 \caption{\label{OREFref-2} $h=1$, $\thetainc=0\deg$, and $\psi\in[0\deg,360\deg)$}
\end{subfigure} 
\\
\begin{subfigure}{0.5\textwidth}
\centering
\includegraphics[width=7cm]{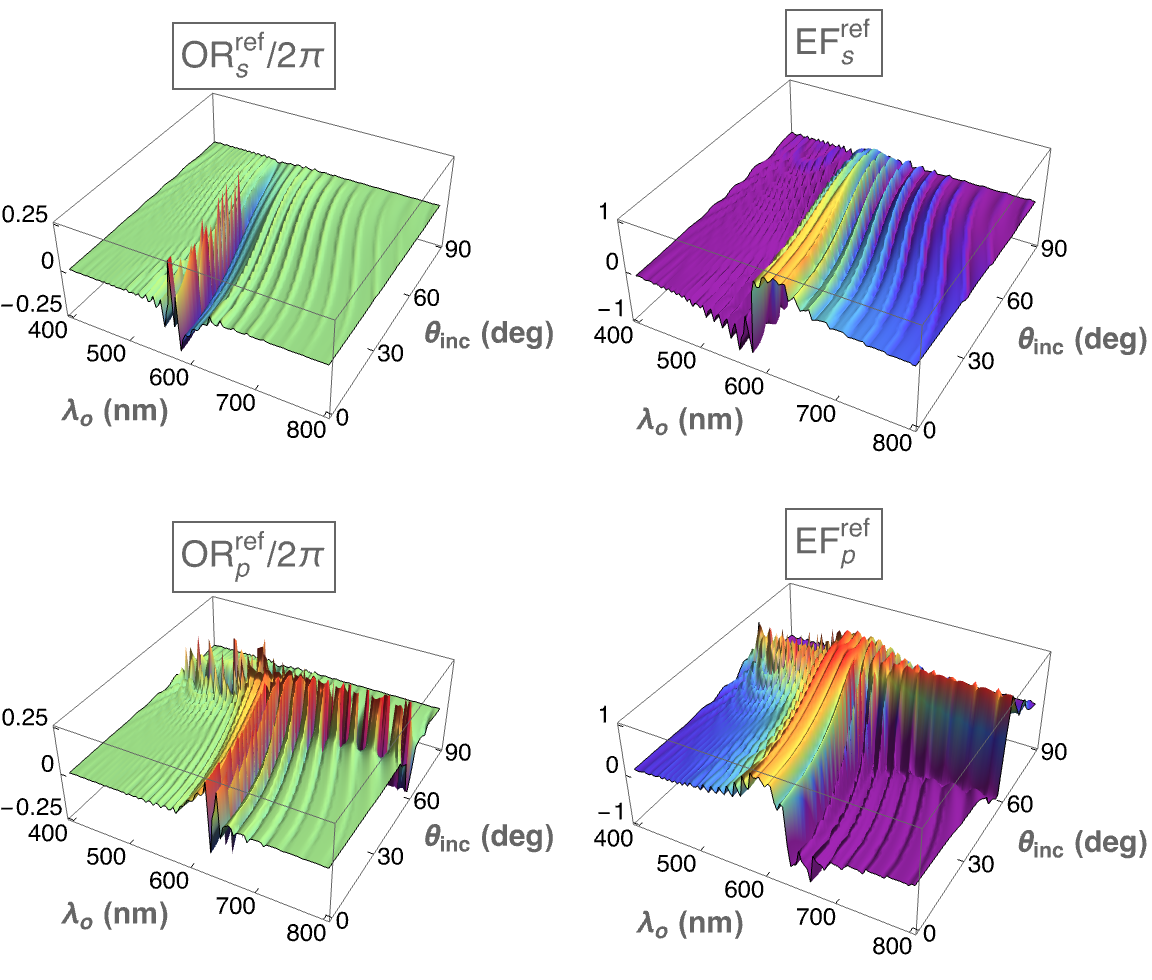} 
\hfill \vspace{0mm} 
 \caption{\label{OREFref-3} $h=-1$, $\thetainc\in[0\deg,90\deg)$, and $\psi=0\deg$}
\end{subfigure}  \hspace{-5mm} 
\begin{subfigure}{0.5\textwidth}
\centering
\includegraphics[width=7cm]{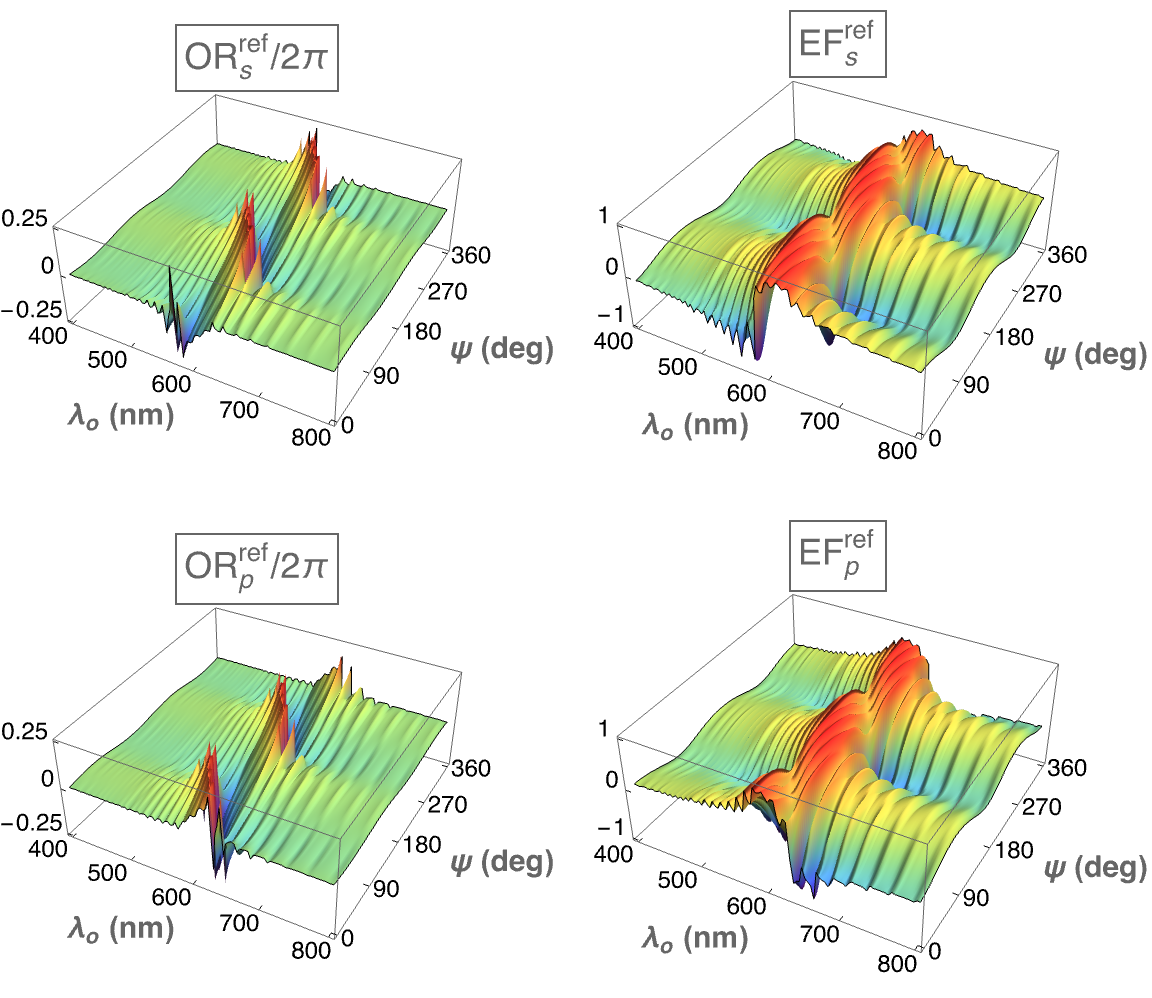} 
\hfill \vspace{0mm} 
 \caption{\label{OREFref-4} $h=-1$, $\thetainc=0\deg$, and $\psi\in[0\deg,360\deg)$}
\end{subfigure} 
	\caption{Spectral variations of the optical rotation $\ORref_\mu(\lambdao,\thetainc,\psi)$
	and  ellipticity function $\EFref_\mu(\lambdao,\thetainc,\psi)$, $\mu\in\lec s,p\ric$, of the reflected
	plane wave, when
	(a,b) $h=1$ and (c,d) $h=-1$. (a,c) $\thetainc\in [0\deg,90\deg)$ and $\psi= 0\deg$;  (b,d) $\thetainc=0\deg$ and $\psi\in[ 0\deg,360\deg)$.
	}
	\label{OREFref}
\end{figure}

\begin{figure}[ht]
\begin{subfigure}{0.5\textwidth}
\centering
\includegraphics[width=7cm]{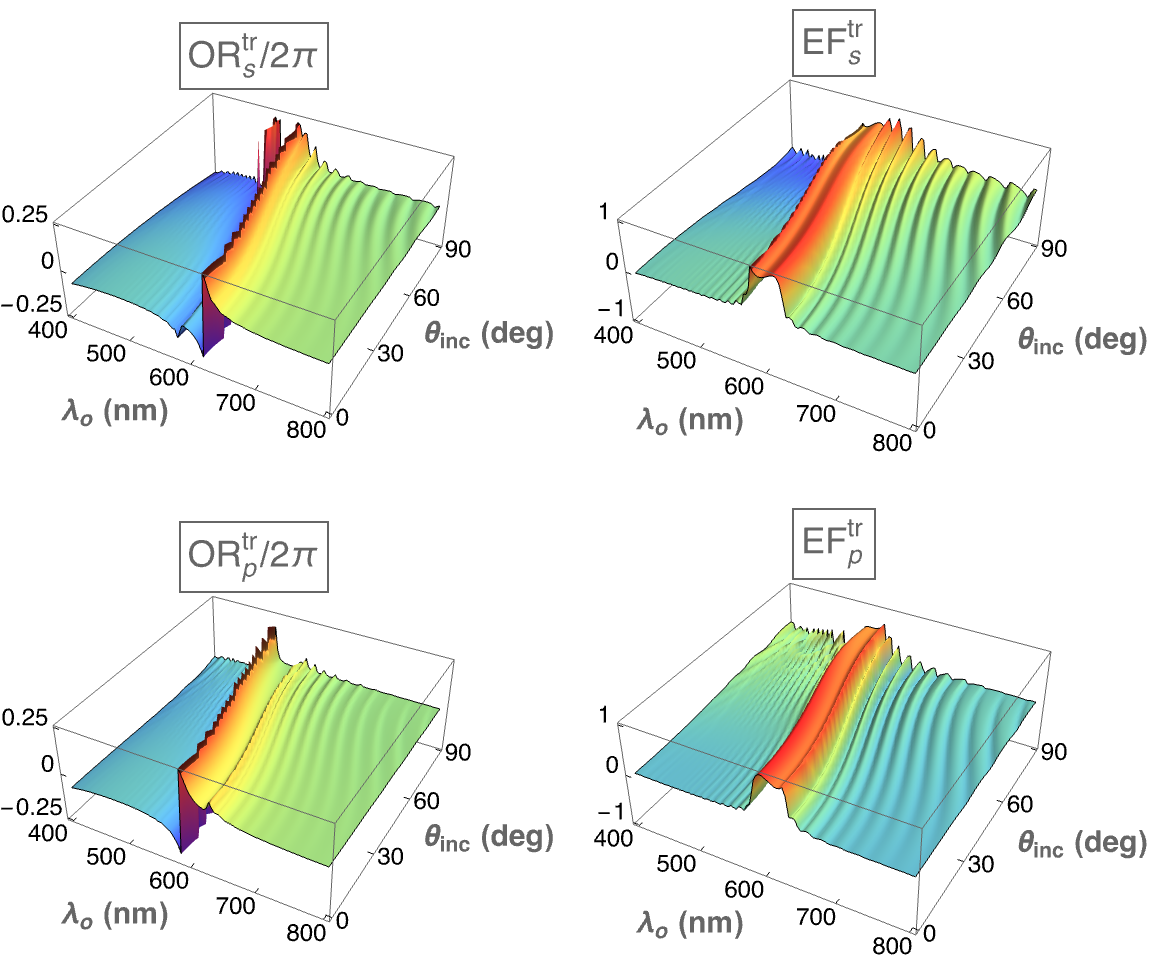} 
\hfill \vspace{0mm} 
 \caption{\label{OREFtrans-1} $h=1$, $\thetainc\in[0\deg,90\deg)$, and $\psi=0\deg$}
\end{subfigure}  \hspace{-5mm} 
\begin{subfigure}{0.5\textwidth}
\centering
\includegraphics[width=7cm]{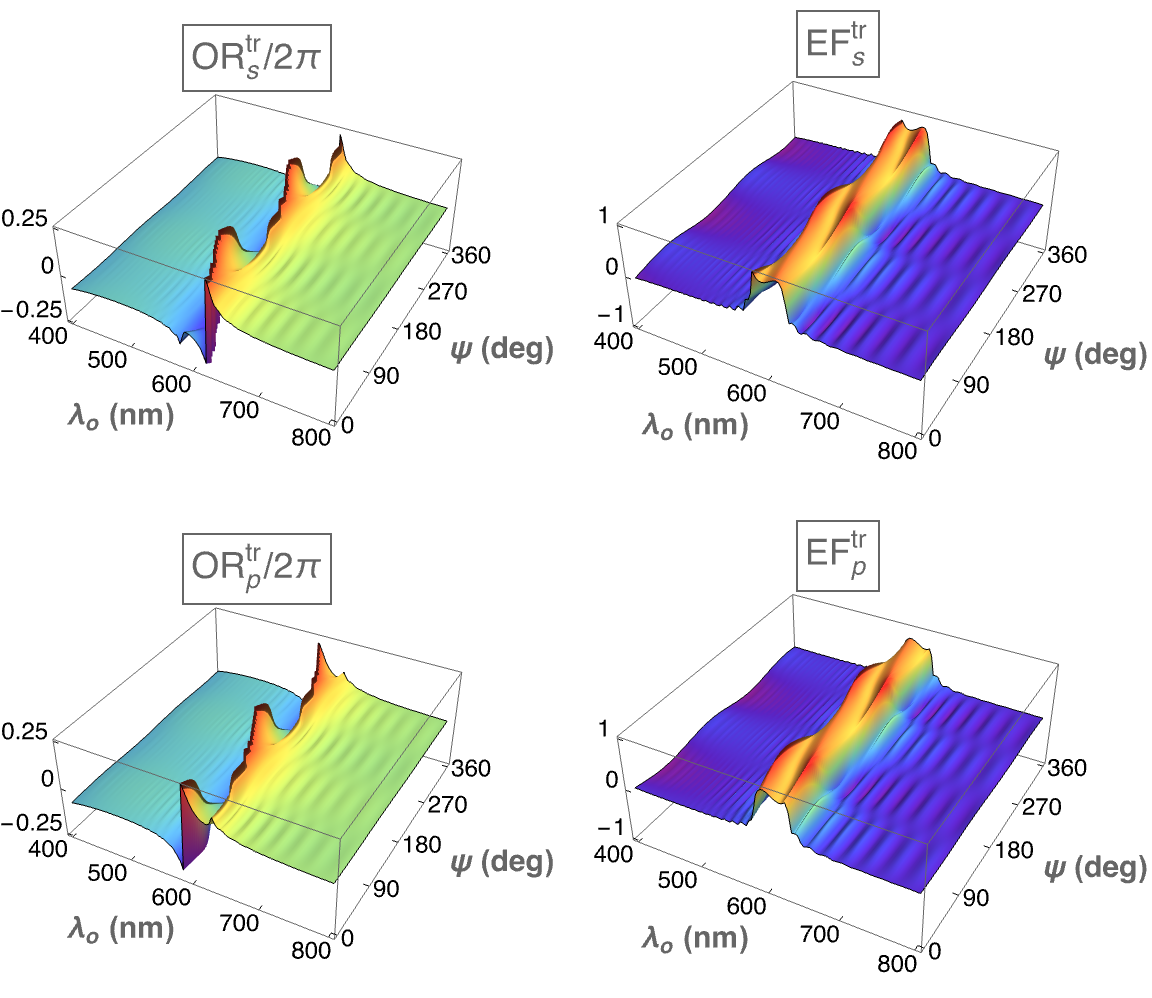} 
\hfill \vspace{0mm} 
 \caption{\label{OREFtrans-2} $h=1$, $\thetainc=0\deg$, and $\psi\in[0\deg,360\deg)$}
\end{subfigure} 
\\
\begin{subfigure}{0.5\textwidth}
\centering
\includegraphics[width=7cm]{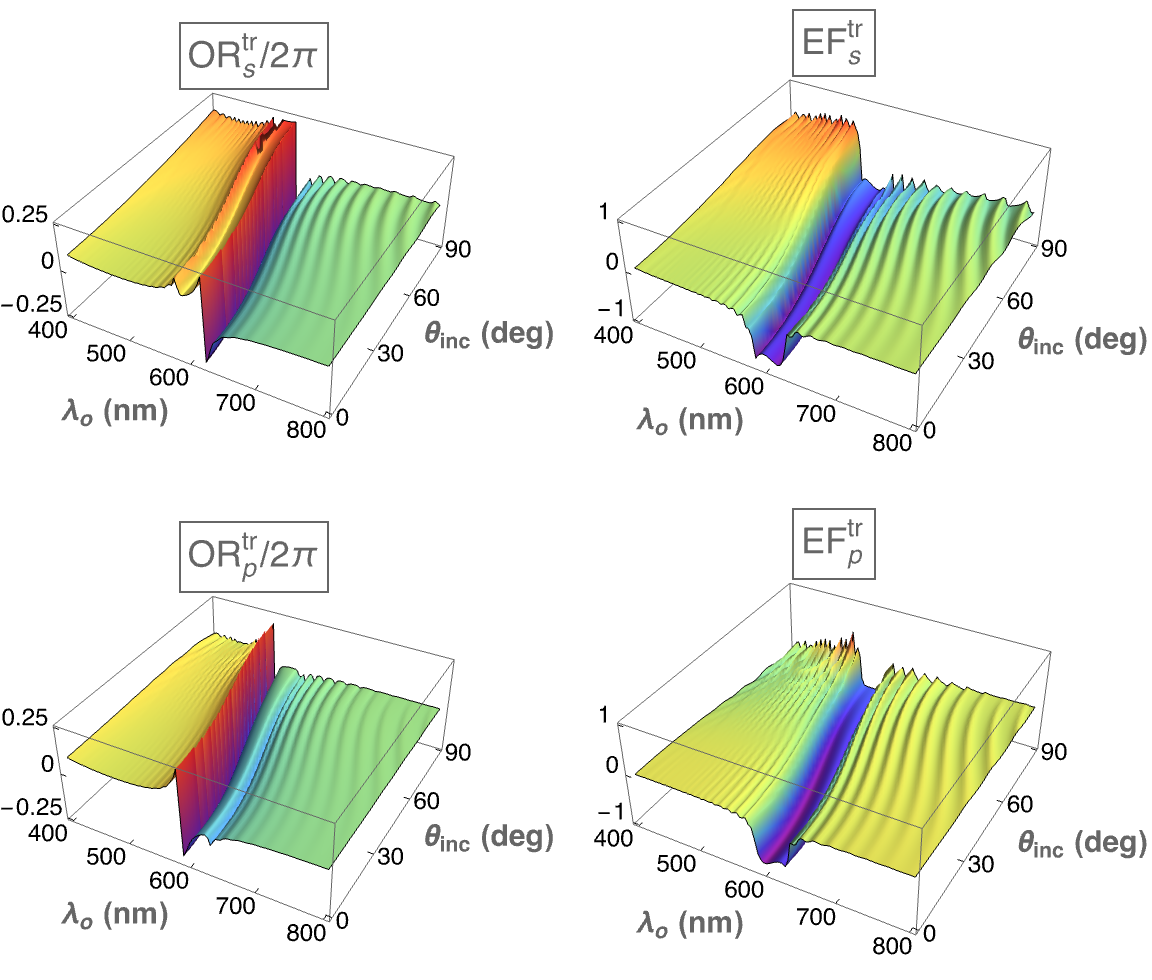} 
\hfill \vspace{0mm} 
 \caption{\label{OREFtrans-3} $h=-1$, $\thetainc\in[0\deg,90\deg)$, and $\psi=0\deg$}
\end{subfigure}  \hspace{-5mm} 
\begin{subfigure}{0.5\textwidth}
\centering
\includegraphics[width=7cm]{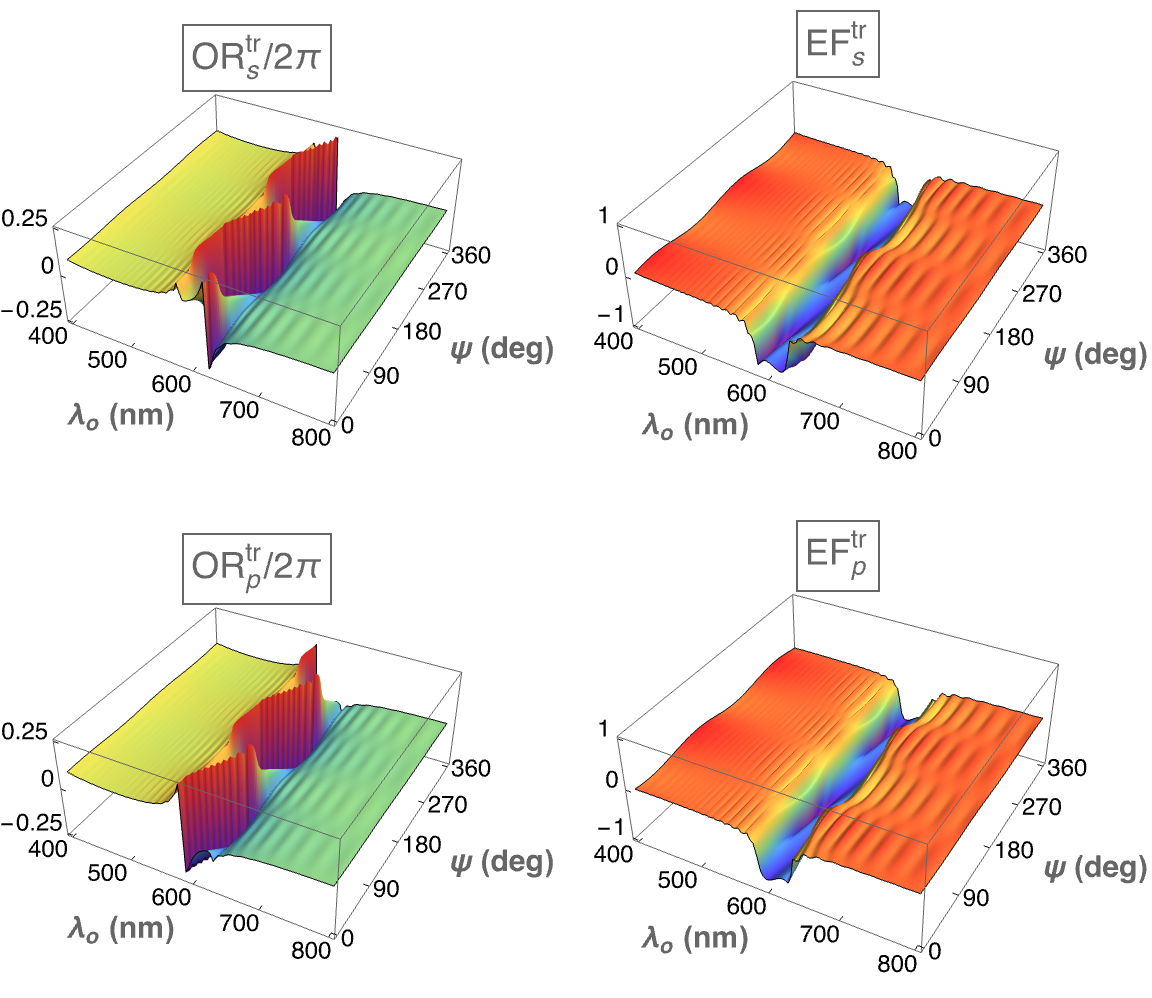} 
\hfill \vspace{0mm} 
 \caption{\label{OREFtrans-4} $h=-1$, $\thetainc=0\deg$, and $\psi\in[0\deg,360\deg)$}
\end{subfigure} 
	\caption{Spectral variations of the optical rotation $\ORtra_\mu(\lambdao,\thetainc,\psi)$
	and  ellipticity function $\EFtra_\mu(\lambdao,\thetainc,\psi)$, $\mu\in\lec s,p\ric$, of the transmitted
	plane wave, when
	(a,b) $h=1$ and (c,d) $h=-1$. (a,c) $\thetainc\in [0\deg,90\deg)$ and $\psi= 0\deg$;  (b,d) $\thetainc=0\deg$ and $\psi\in[ 0\deg,360\deg)$.
	}	
	\label{OREFtrans}
\end{figure}

Figure \ref{OREFref}   provides the spectral variations of $EF^{\rm ref}_{\rm s,p}$ and
$OR^{\rm ref}_{\rm s,p}$, and 
Fig.~\ref{OREFtrans}     the spectral variations of $EF^{\rm tra}_{\rm s,p}$ and
$OR^{\rm tra}_{\rm s,p}$. A feature representing the circular Bragg phenomenon is clearly evident in all 32 plots
in the two figures. For fixed $\psi$, the feature curves towards
 shorter wavelengths as $\thetainc$ increases. For normal
 incidence, the feature has two undulations with increasing $\psi$. Although measurements of optical rotation
 and ellipticity of the transmitted plane wave for normal incidence have been reported for over a century \cite{Reusch},
comprehensive experimental investigations for oblique incidence are very desirable in the near future.

\subsection{Geometric phases of reflected and transmitted plane waves}
The Stokes parameters of the incident plane wave are given by \cite{Jackson}
\begin{equation}
\label{Stokes-inc}
\left.\begin{array}{l}
\sinczero = \vert\aal\vert^2+\vert\aar\vert^2 =\vert\aas\vert^2+\vert\aap\vert^2
\\[5pt]
\sincone =2\,\Re\left(\aal\,\aar^\ast\right)=\vert\aap\vert^2-\vert\aas\vert^2
\\[5pt]
\sinctwo =2\,\Im\left(\aal\,\aar^\ast\right)=2\,\Re\left(\aas\,\aap^\ast\right)
\\[5pt]
\sincthree = \vert\aar\vert^2-\vert\aal\vert^2=2\,\Im\left(\aas\,\aap^\ast\right)
\end{array}
\right\}\,.
\end{equation} 
The   Poincar\'e spinor $\bphitinc$ of the incident plane wave can then be obtained from  Eqs.~(\ref{def-alphabeta}) and (\ref{def-PS})
in the Appendix.

After making the   changes $\lec\aal\to\bbl,\,\aar\to\bbr,\,\aas\to\bbs,\,\aap\to\bbp\ric$, Eqs.~(\ref{Stokes-inc}) can be used to determine the Stokes parameters $\srefzero$,  $\srefone$,  $\sreftwo$, and  $\srefthree$
of the reflected plane wave, and the   Poincar\'e spinor $\bphitref$ of the reflected plane wave 
can then be obtained from  Eqs.~(\ref{def-alphabeta}) and (\ref{def-PS}).
Calculation of the Poincar\'e spinor $\bphittra$ of the transmitted plane wave follows the same route.

Thereafter, the
 reflection-mode geometric phase $\bPhiref_{\ell}$ and the transmission-mode geometric phase $\bPhitra_{\ell}$,
 $\ell\in\left\{s,p,R,L\right\}$, can be calculated with respect to the incident plane wave using Eq.~(\ref{GP-def}) in available in the Appendix.
The subscript $\ell$ in both quantities indicates the polarization state of the incident plane wave:
 perpendicular ($s$), parallel ($p$), left-circular ($L$), or right-circular ($R$).

\begin{figure}[ht]
\begin{subfigure}{0.5\textwidth}
\centering
\includegraphics[width=7cm]{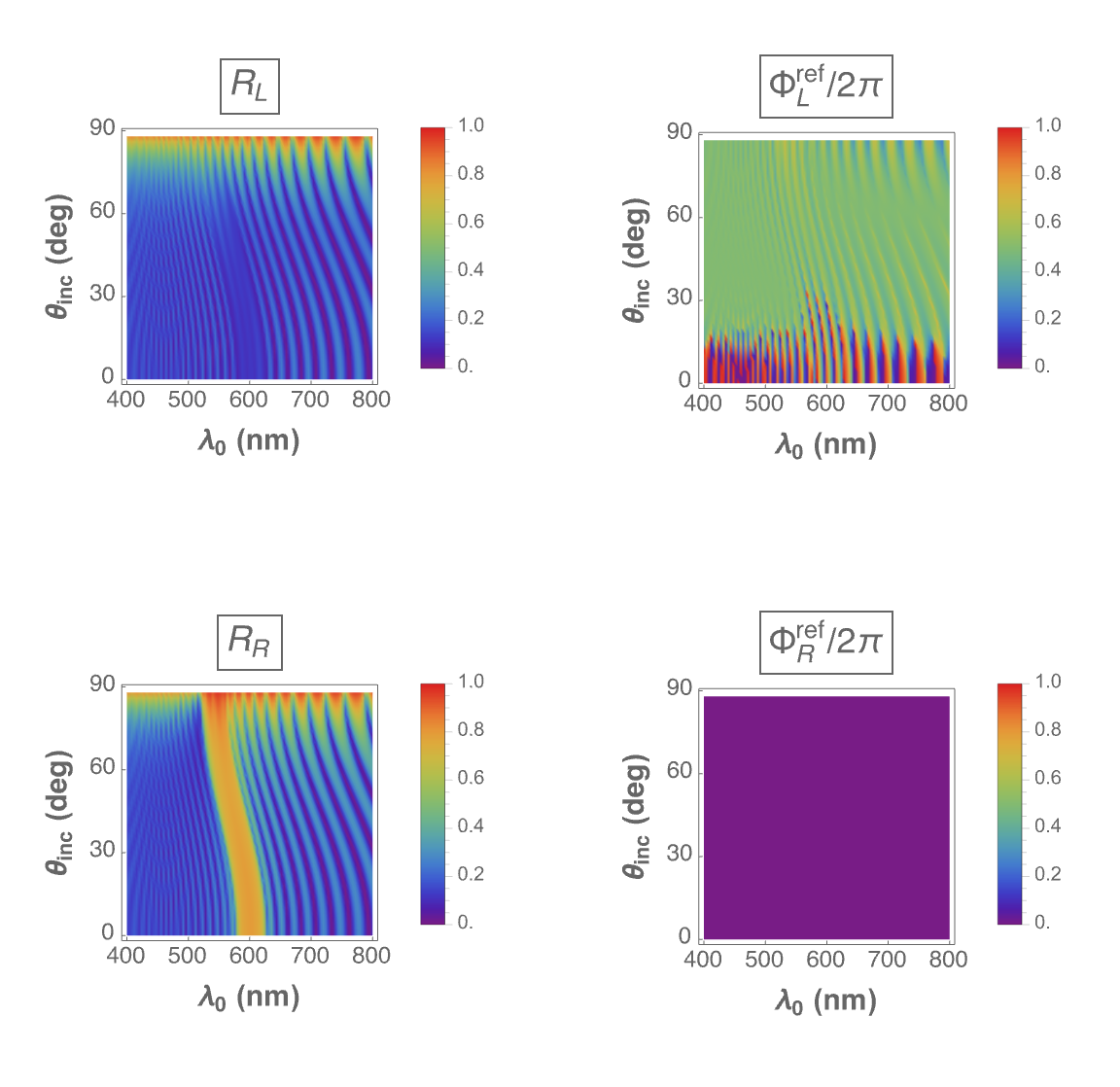} 
\hfill \vspace{0mm} 
 \caption{\label{CircGPref-1} $h=1$, $\thetainc\in[0\deg,90\deg)$, and $\psi=0\deg$}
\end{subfigure}  \hspace{-5mm} 
\begin{subfigure}{0.5\textwidth}
\centering
\includegraphics[width=7cm]{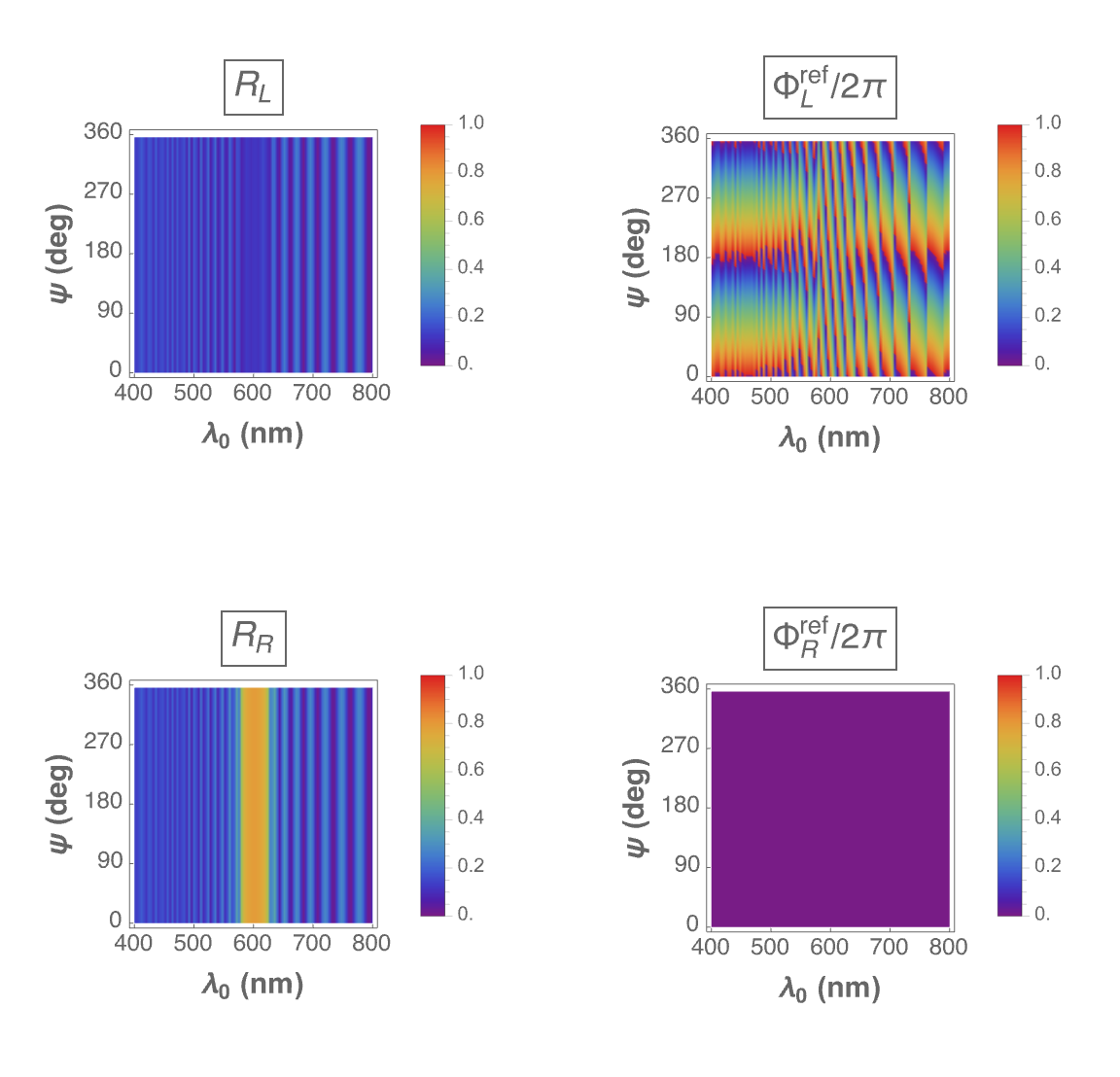} 
\hfill \vspace{0mm} 
 \caption{\label{CircGPref-2} $h=1$, $\thetainc=0\deg$, and $\psi\in[0\deg,360\deg)$}
\end{subfigure} 
\\
\begin{subfigure}{0.5\textwidth}
\centering
\includegraphics[width=7cm]{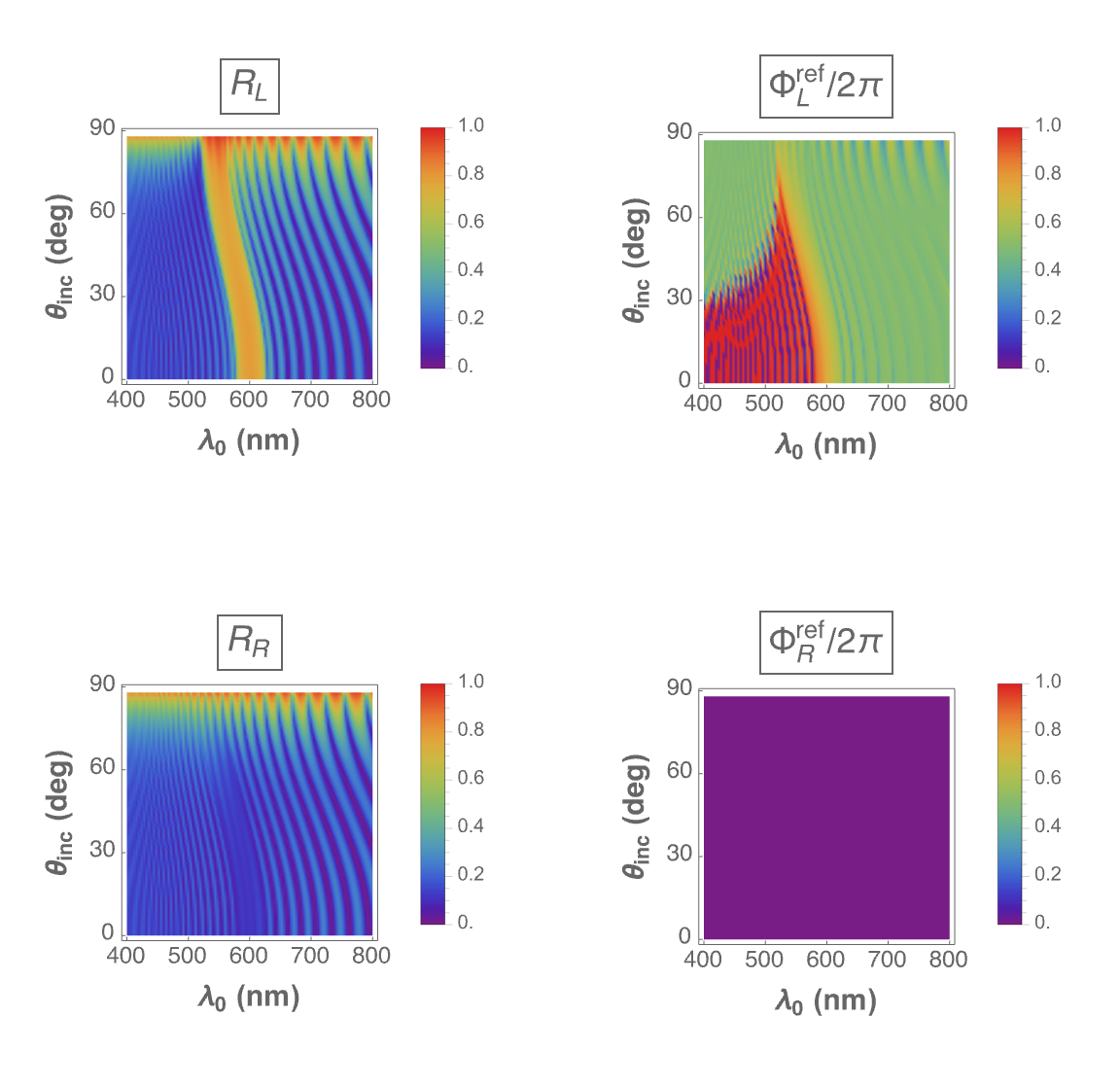} 
\hfill \vspace{0mm} 
 \caption{\label{CircGPref-3} $h=-1$, $\thetainc\in[0\deg,90\deg)$, and $\psi=0\deg$}
\end{subfigure}  \hspace{-5mm} 
\begin{subfigure}{0.5\textwidth}
\centering
\includegraphics[width=7cm]{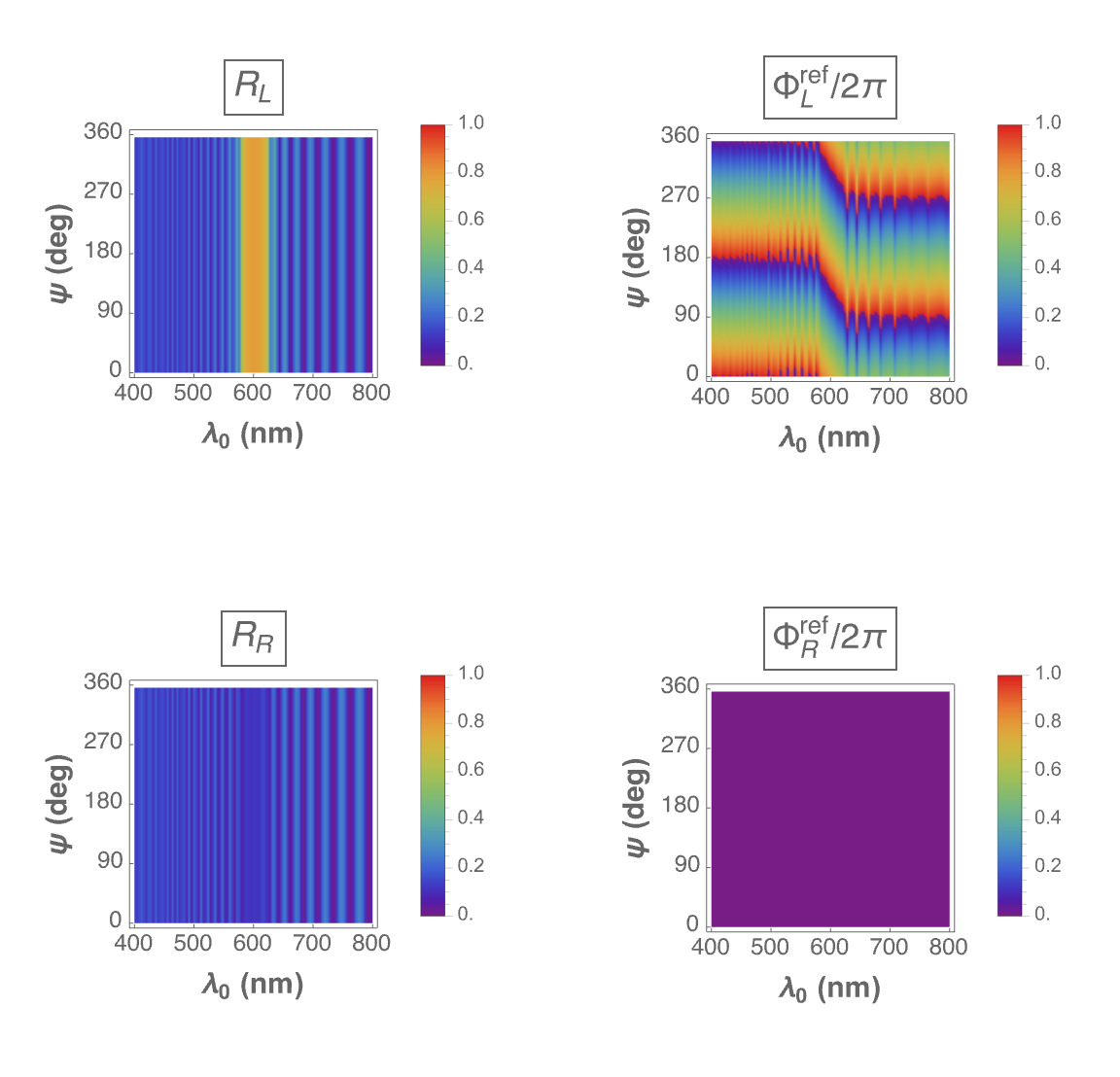} 
\hfill \vspace{0mm} 
 \caption{\label{CircGPref-4} $h=-1$, $\thetainc=0\deg$, and $\psi\in[0\deg,360\deg)$}
\end{subfigure} 
	\caption{Spectral variations of the total circular reflectance $R_{\mu}(\lambdao,\thetainc,\psi)$ and reflection-mode geometric phase $\bPhiref_{\mu}	(\lambdao,\thetainc,\psi)$,   $\mu\in\left\{L,R\right\}$, when
	(a,b) $h=1$ and (c,d) $h=-1$. (a,c) $\thetainc\in [0\deg,90\deg)$ and $\psi= 0\deg$;  (b,d) $\thetainc=0\deg$ and $\psi\in[ 0\deg,360\deg)$.
	}	
	\label{CircGPref}
\end{figure}

\begin{figure}[ht]
\begin{subfigure}{0.5\textwidth}
\centering
\includegraphics[width=7cm]{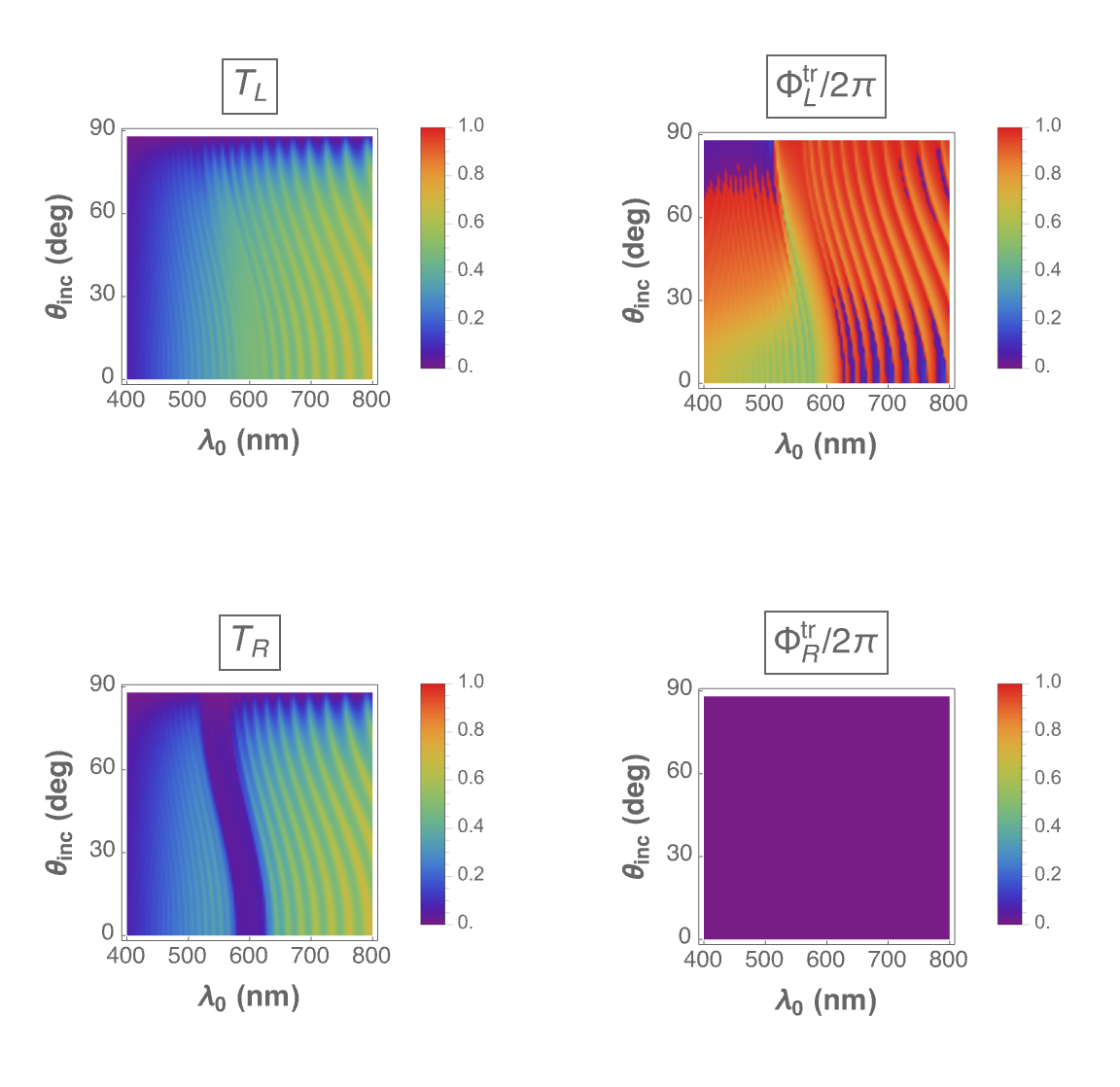} 
\hfill \vspace{0mm} 
 \caption{\label{CircGPtrans-1} $h=1$, $\thetainc\in[0\deg,90\deg)$, and $\psi=0\deg$}
\end{subfigure}  \hspace{-5mm} 
\begin{subfigure}{0.5\textwidth}
\centering
\includegraphics[width=7cm]{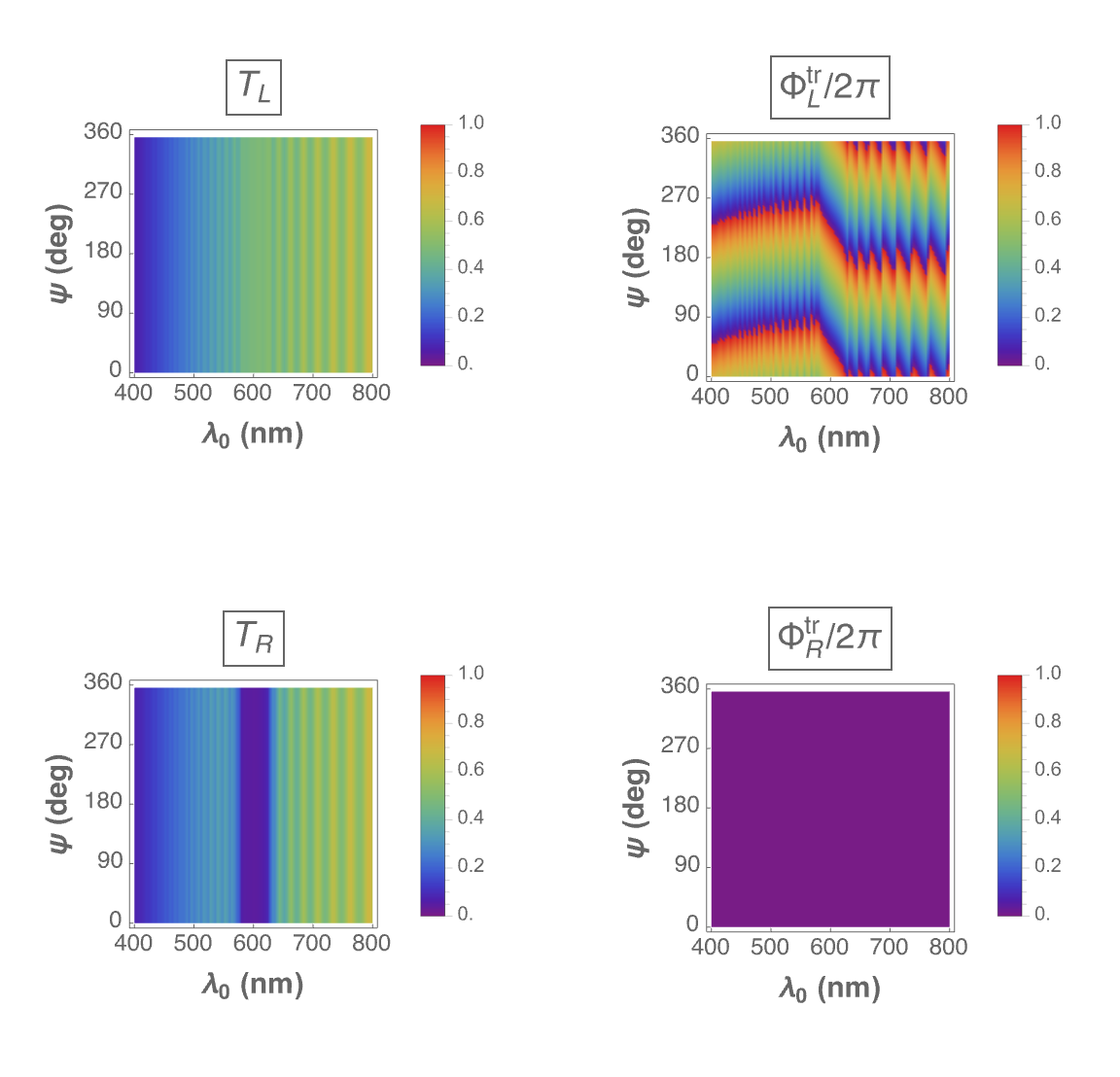} 
\hfill \vspace{0mm} 
 \caption{\label{CircGPtrans-2} $h=1$, $\thetainc=0\deg$, and $\psi\in[0\deg,360\deg)$}
\end{subfigure} 
\\
\begin{subfigure}{0.5\textwidth}
\centering
\includegraphics[width=7cm]{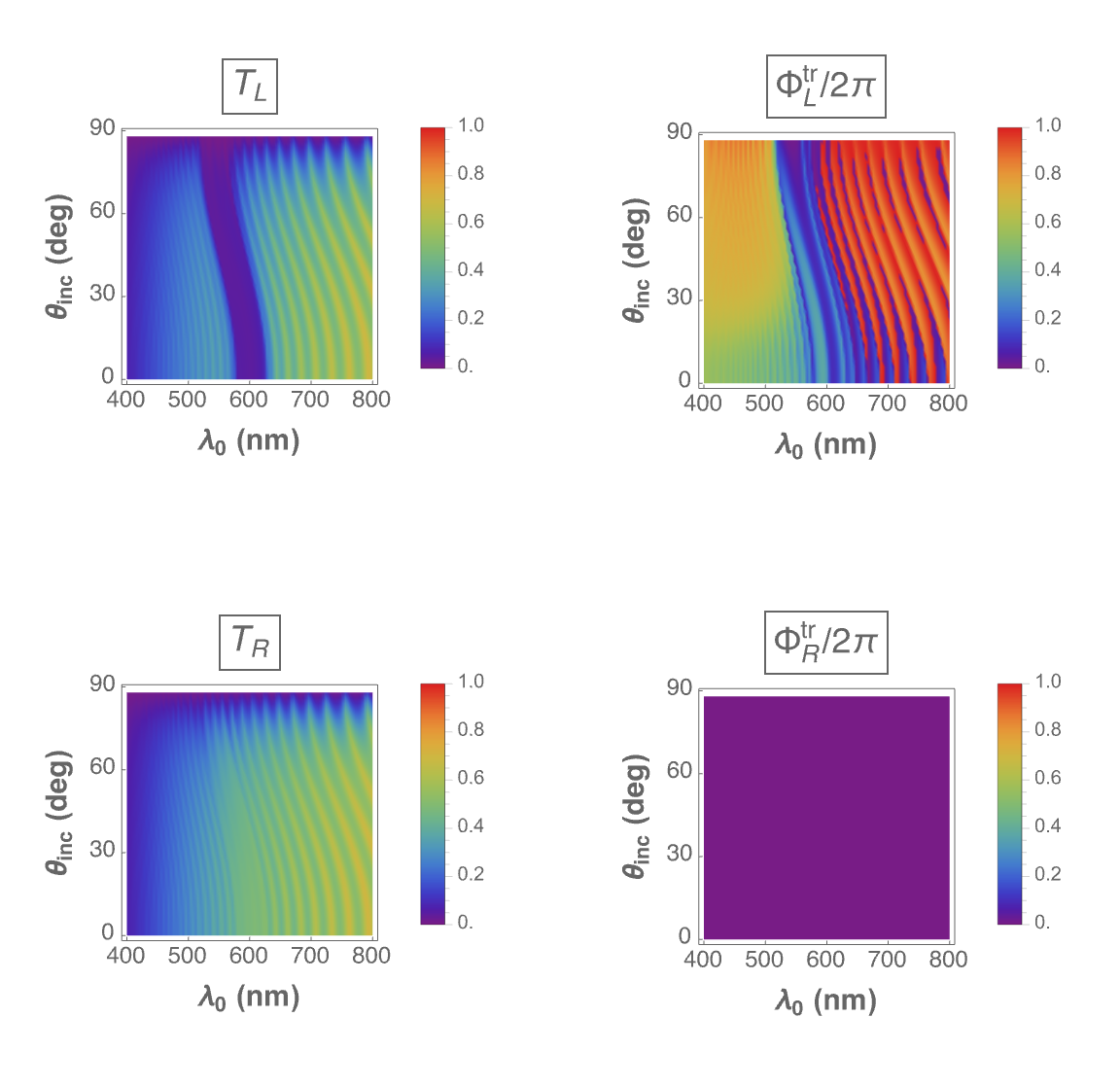} 
\hfill \vspace{0mm} 
 \caption{\label{CircGPtrans-3} $h=-1$, $\thetainc\in[0\deg,90\deg)$, and $\psi=0\deg$}
\end{subfigure}  \hspace{-5mm} 
\begin{subfigure}{0.5\textwidth}
\centering
\includegraphics[width=7cm]{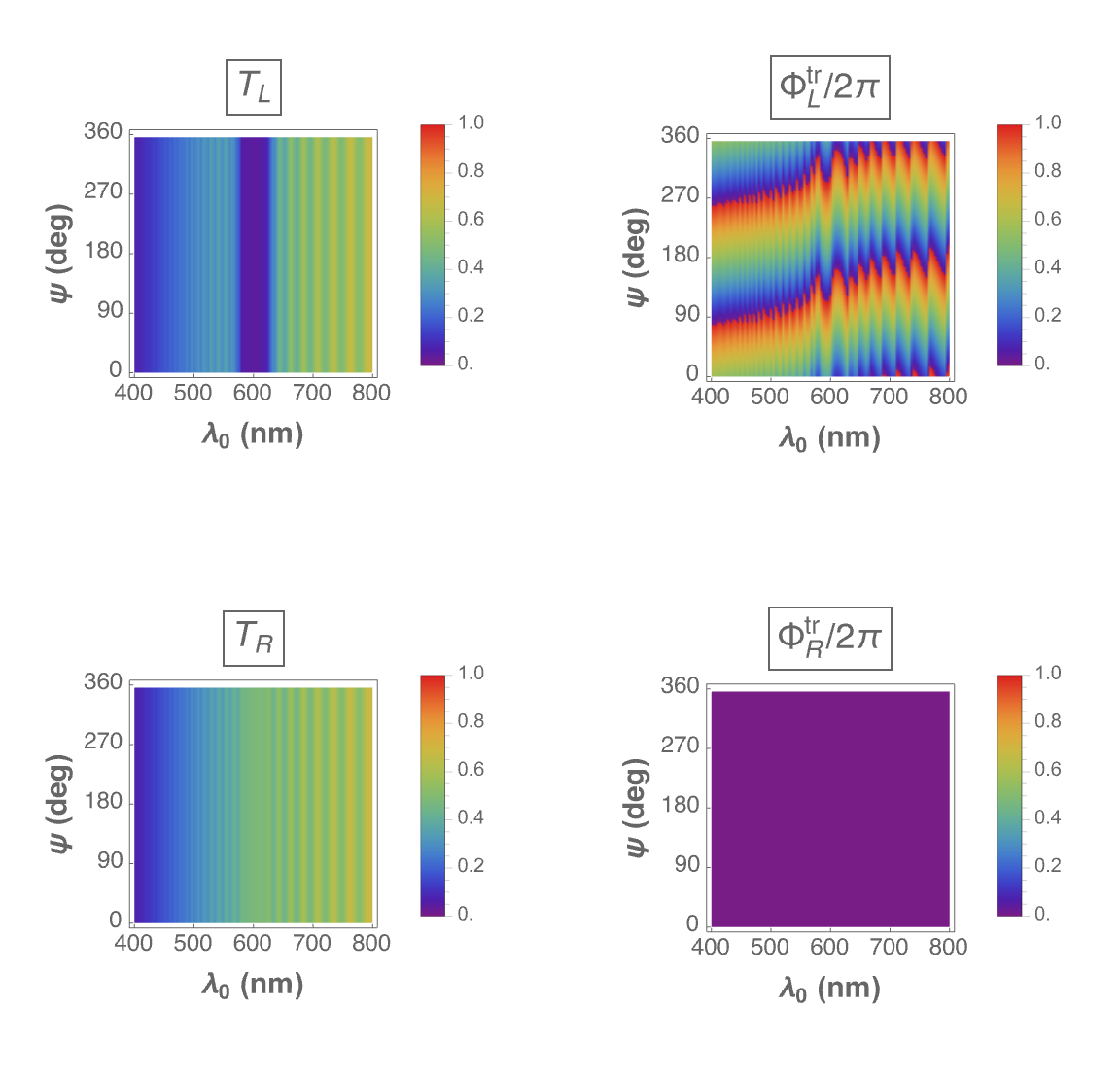} 
\hfill \vspace{0mm} 
 \caption{\label{CircGPtrans-4} $h=-1$, $\thetainc=0\deg$, and $\psi\in[0\deg,360\deg)$}
\end{subfigure} 
	\caption{Spectral variations of the total circular transmittance $T_{\mu}(\lambdao,\thetainc,\psi)$ and transmission-mode geometric phase $\bPhitra_{\mu}	(\lambdao,\thetainc,\psi)$,   $\mu\in\left\{L,R\right\}$, when
	(a,b) $h=1$ and (c,d) $h=-1$. (a,c) $\thetainc\in [0\deg,90\deg)$ and $\psi= 0\deg$;  (b,d) $\thetainc=0\deg$ and $\psi\in[ 0\deg,360\deg)$.
	}	
	\label{CircGPtrans}
\end{figure}

\begin{figure}[ht]
\begin{subfigure}{0.5\textwidth}
\centering
\includegraphics[width=7cm]{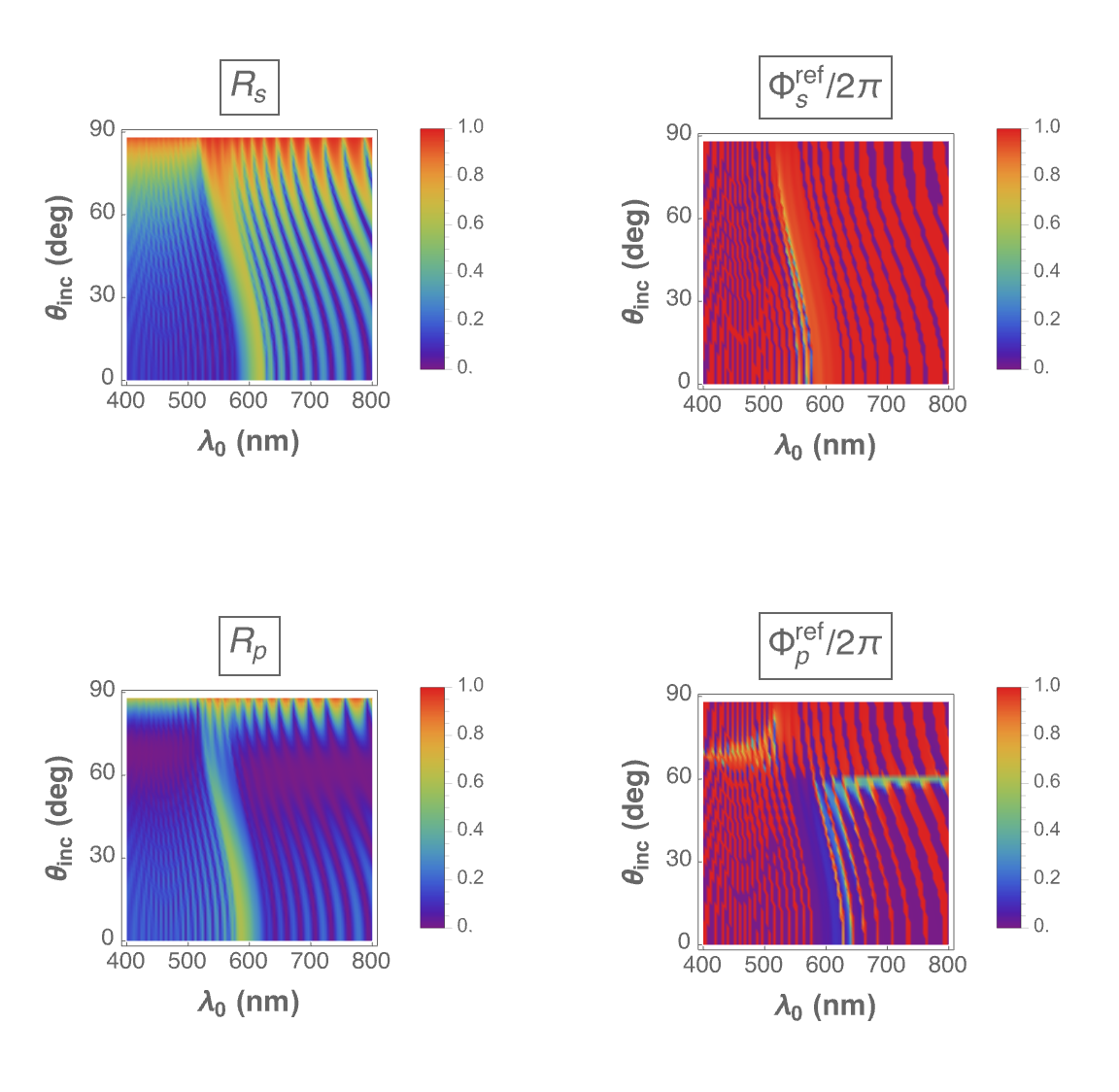} 
\hfill \vspace{0mm} 
 \caption{\label{LinGPref-1} $h=1$, $\thetainc\in[0\deg,90\deg)$, and $\psi=0\deg$}
\end{subfigure}  \hspace{-5mm} 
\begin{subfigure}{0.5\textwidth}
\centering
\includegraphics[width=7cm]{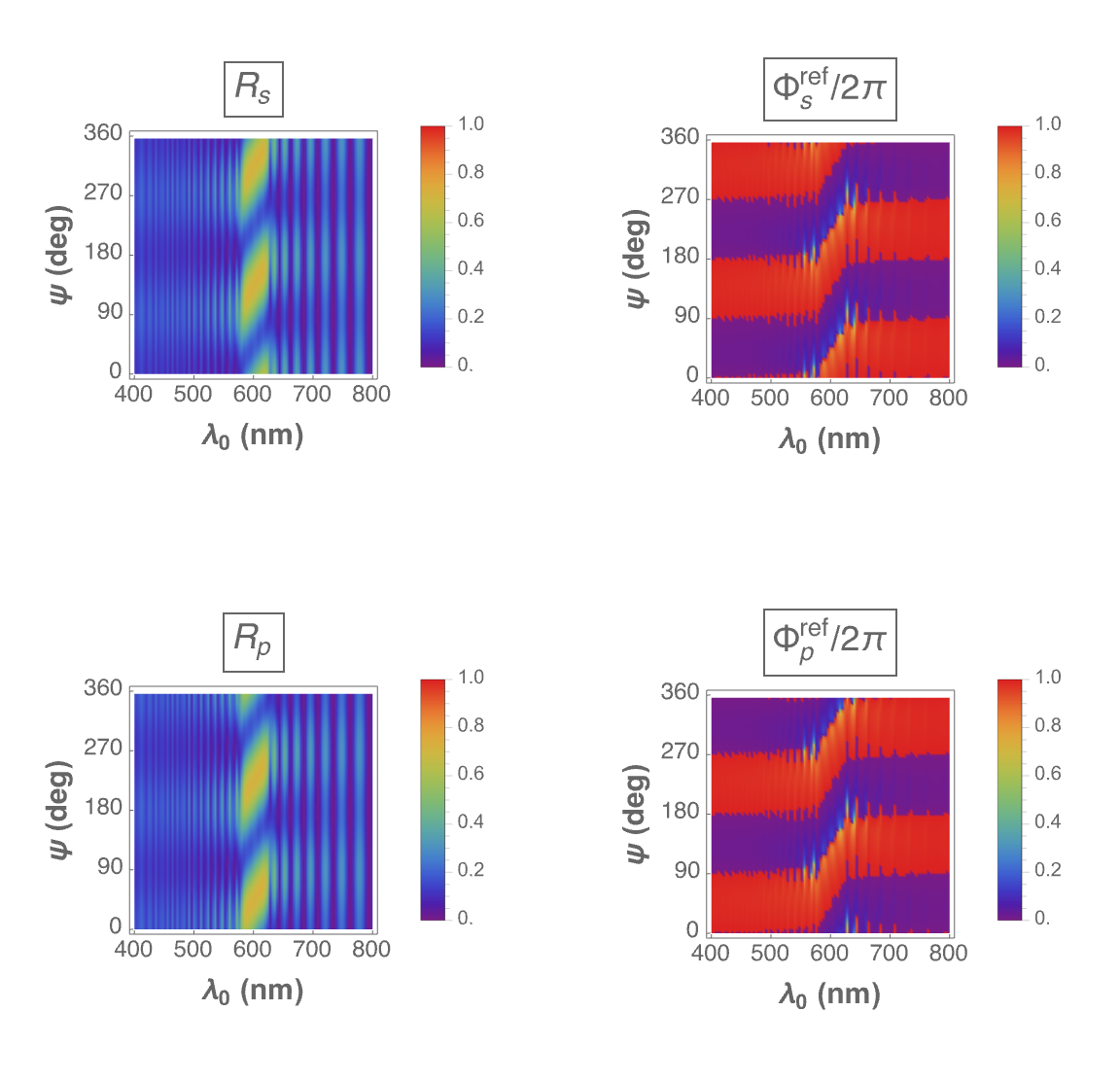} 
\hfill \vspace{0mm} 
 \caption{\label{LinGPref-2} $h=1$, $\thetainc=0\deg$, and $\psi\in[0\deg,360\deg)$}
\end{subfigure} 
\\
\begin{subfigure}{0.5\textwidth}
\centering
\includegraphics[width=7cm]{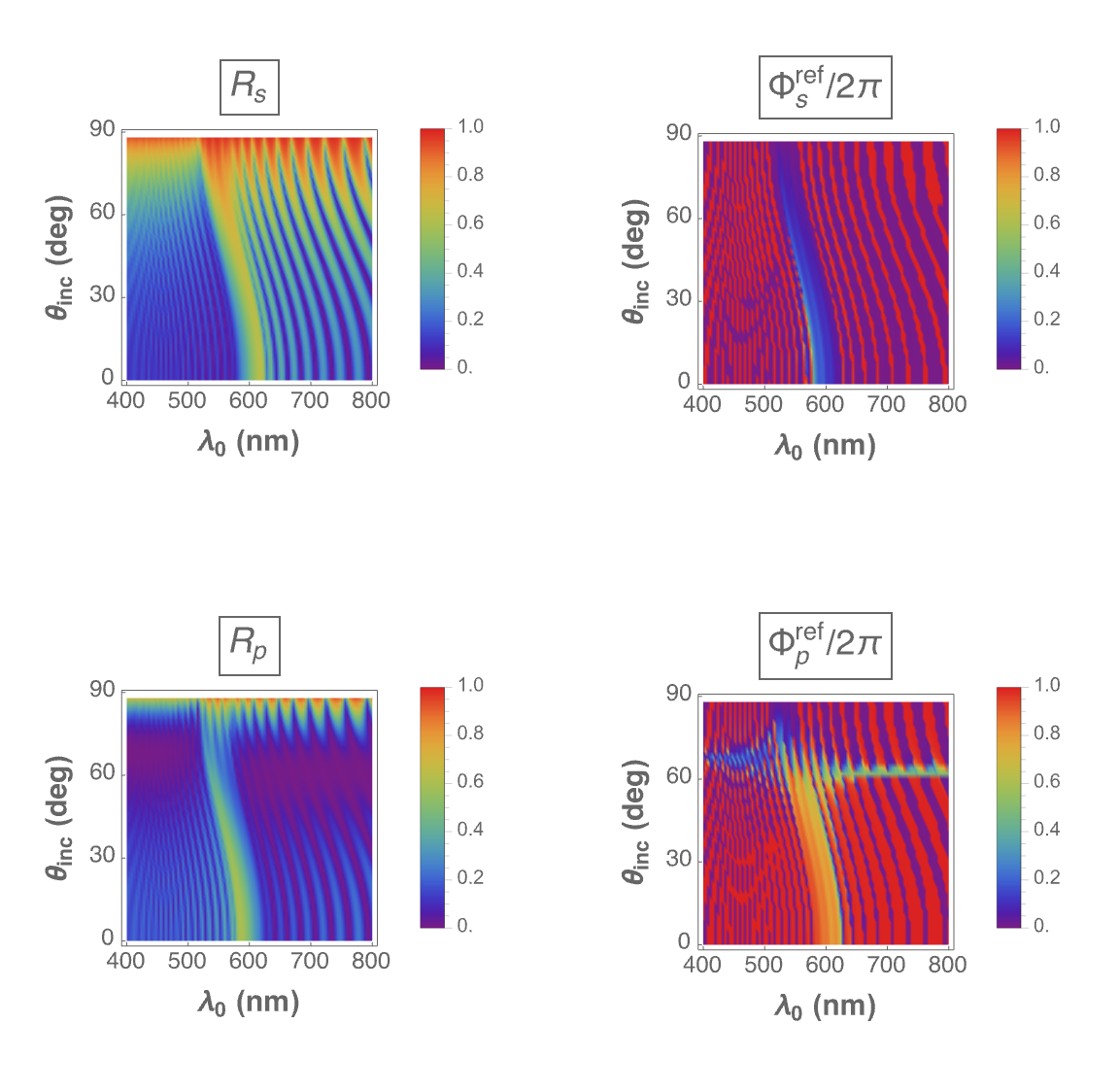} 
\hfill \vspace{0mm} 
 \caption{\label{LinGPref-3} $h=-1$, $\thetainc\in[0\deg,90\deg)$, and $\psi=0\deg$}
\end{subfigure}  \hspace{-5mm} 
\begin{subfigure}{0.5\textwidth}
\centering
\includegraphics[width=7cm]{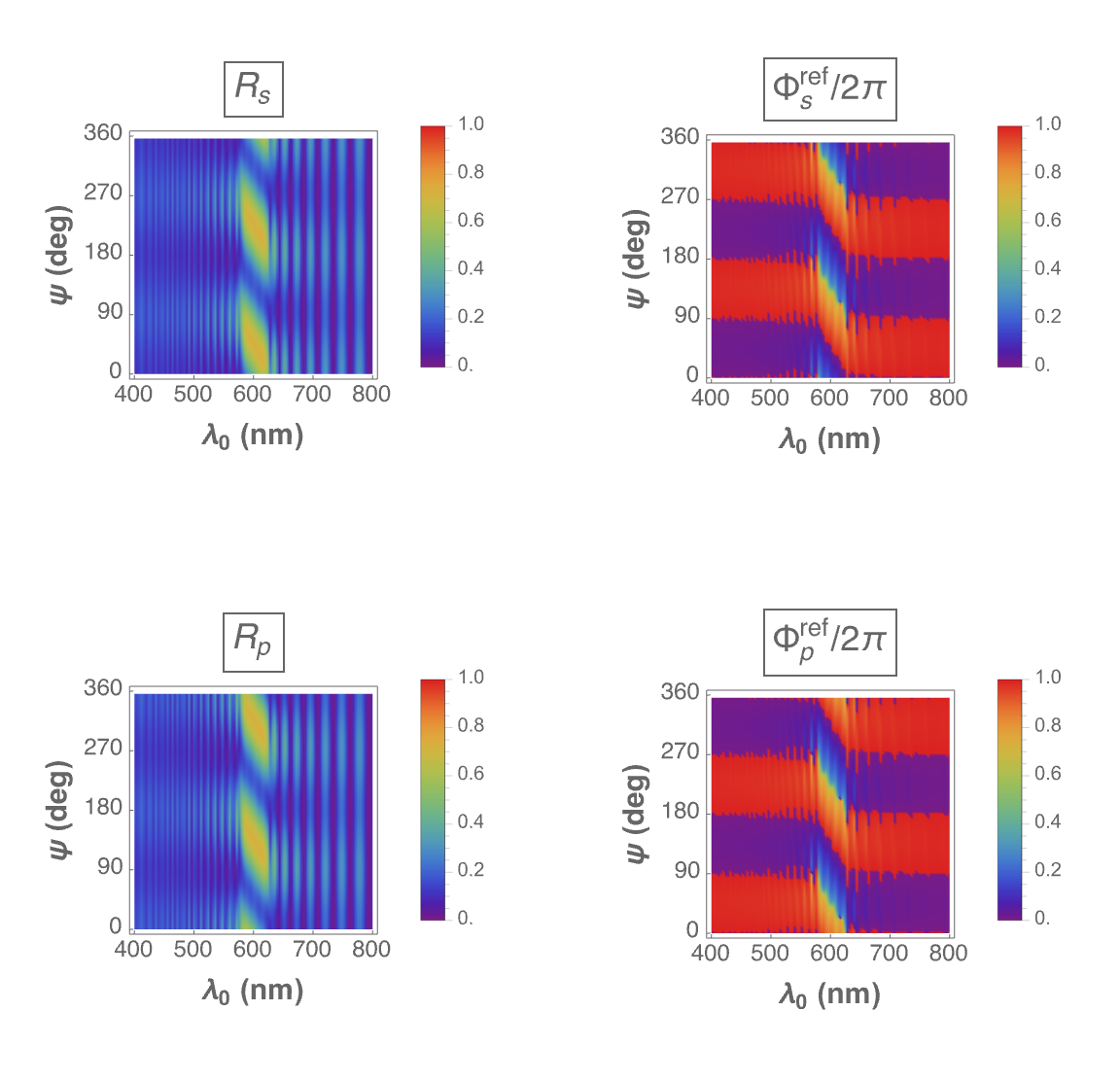} 
\hfill \vspace{0mm} 
 \caption{\label{LinGPref-4} $h=-1$, $\thetainc=0\deg$, and $\psi\in[0\deg,360\deg)$}
\end{subfigure} 
	\caption{Spectral variations of the total linear reflectance $R_{\mu}(\lambdao,\thetainc,\psi)$ and reflection-mode geometric phase $\bPhiref_{\mu}	(\lambdao,\thetainc,\psi)$,   $\mu\in\left\{s,p\right\}$, when
	(a,b) $h=1$ and (c,d) $h=-1$. (a,c) $\thetainc\in [0\deg,90\deg)$ and $\psi= 0\deg$;  (b,d) $\thetainc=0\deg$ and $\psi\in[ 0\deg,360\deg)$.
	}	
	\label{LinGPref}
\end{figure}

\begin{figure}[ht]
\begin{subfigure}{0.5\textwidth}
\centering
\includegraphics[width=7cm]{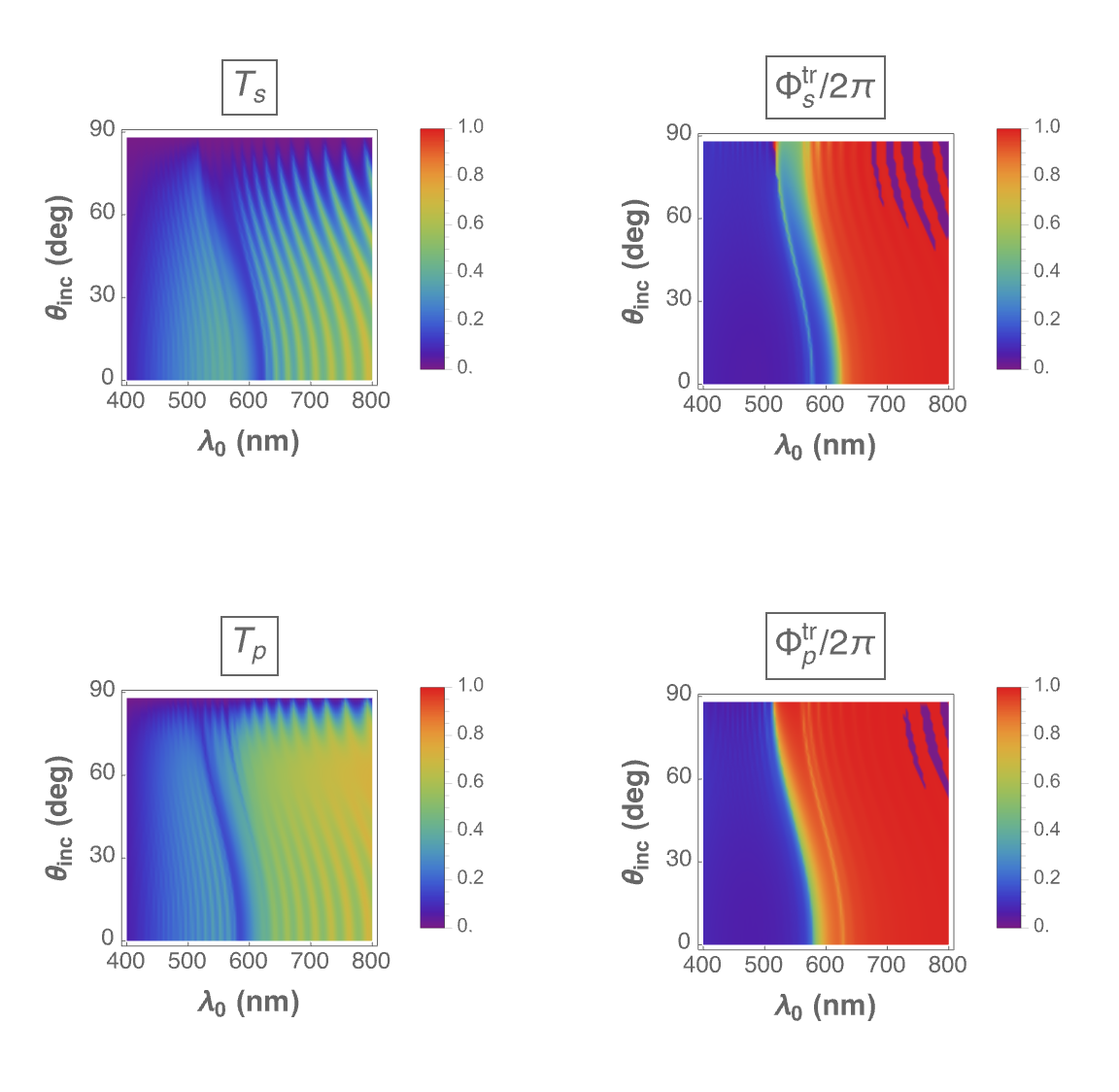} 
\hfill \vspace{0mm} 
 \caption{\label{LinGPtrans-1} $h=1$, $\thetainc\in[0\deg,90\deg)$, and $\psi=0\deg$}
\end{subfigure}  \hspace{-5mm} 
\begin{subfigure}{0.5\textwidth}
\centering
\includegraphics[width=7cm]{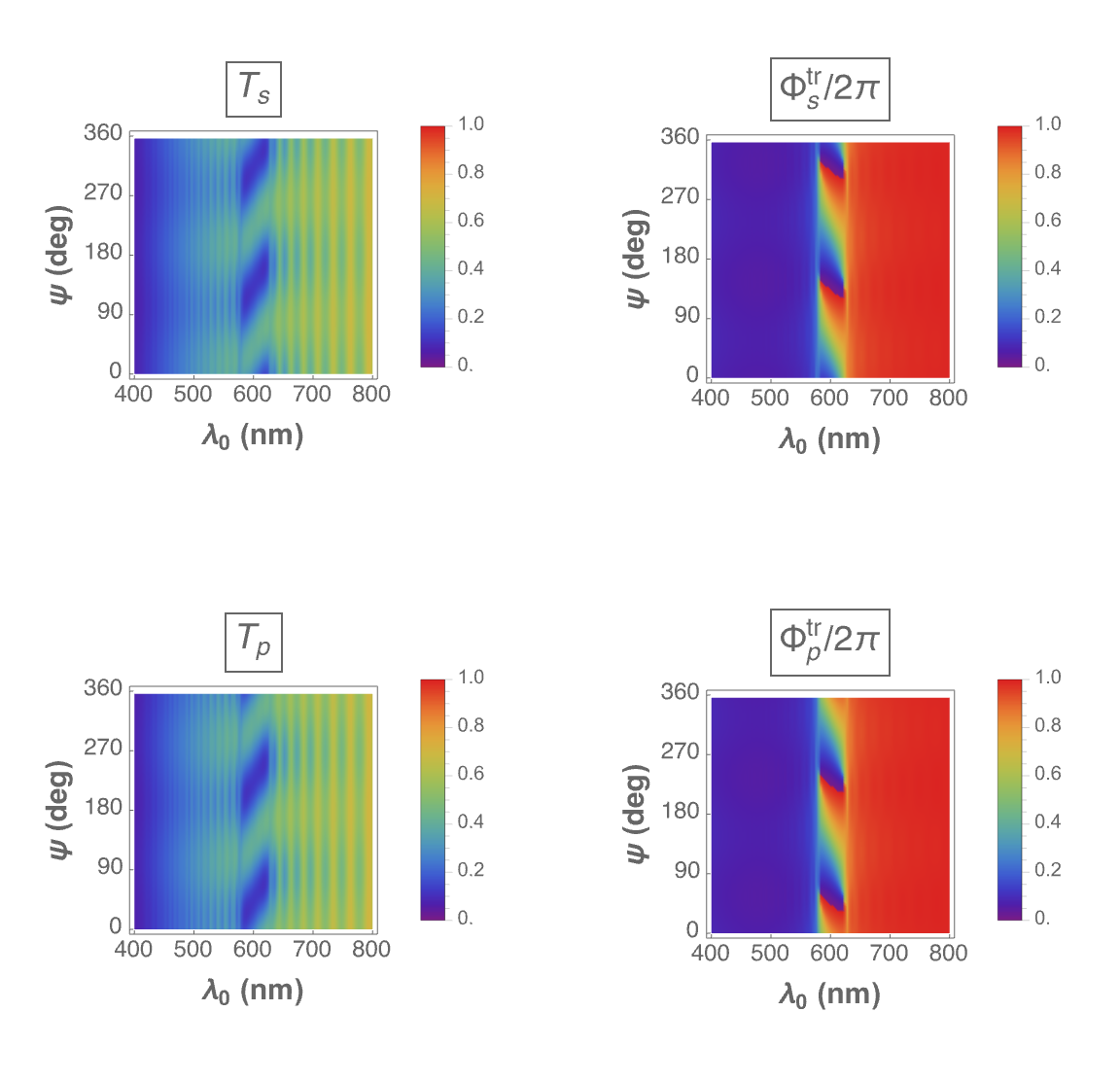} 
\hfill \vspace{0mm} 
 \caption{\label{LinGPtrans-2} $h=1$, $\thetainc=0\deg$, and $\psi\in[0\deg,360\deg)$}
\end{subfigure} 
\\
\begin{subfigure}{0.5\textwidth}
\centering
\includegraphics[width=7cm]{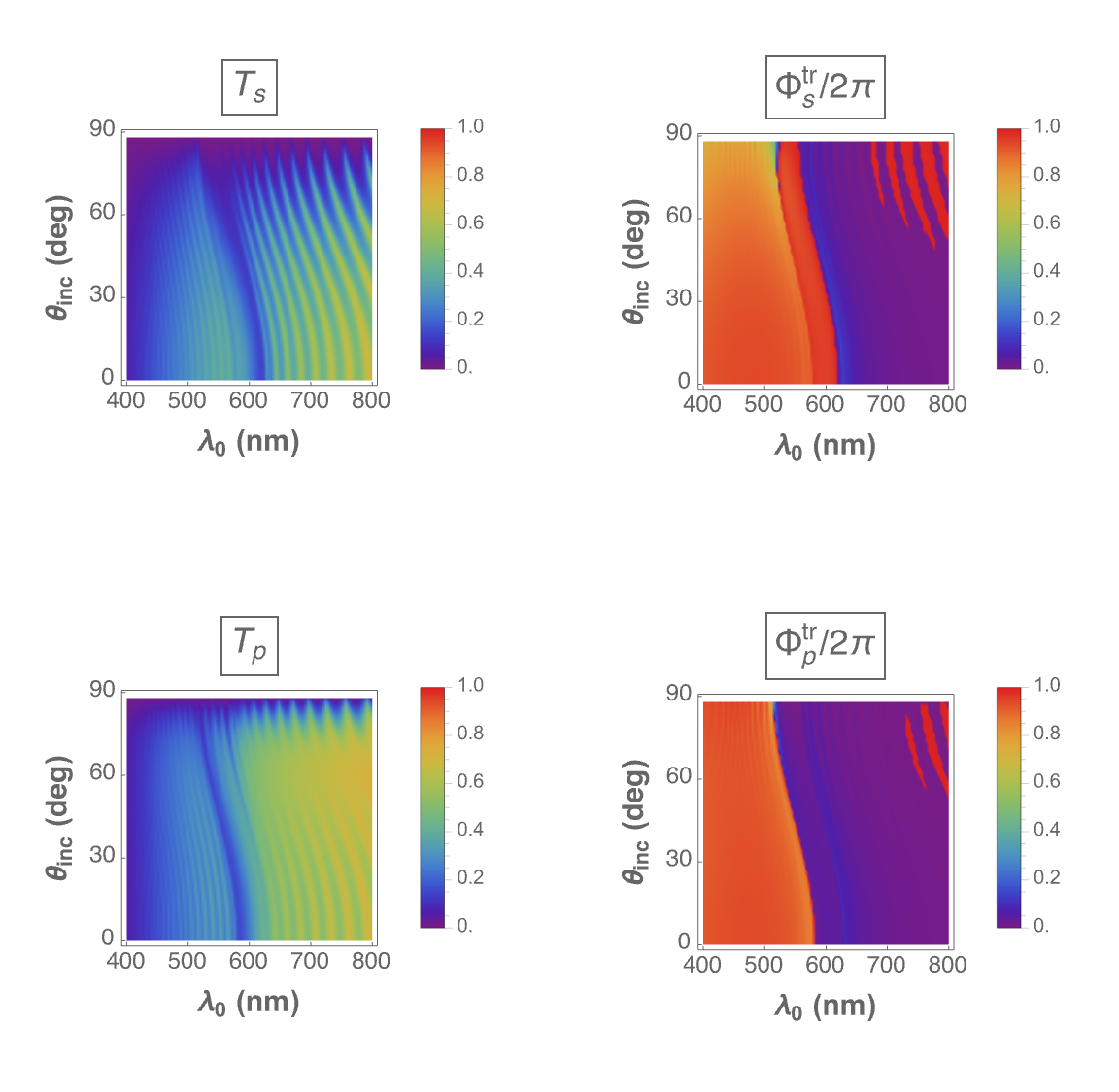} 
\hfill \vspace{0mm} 
 \caption{\label{LinGPtrans-3} $h=-1$, $\thetainc\in[0\deg,90\deg)$, and $\psi=0\deg$}
\end{subfigure}  \hspace{-5mm} 
\begin{subfigure}{0.5\textwidth}
\centering
\includegraphics[width=7cm]{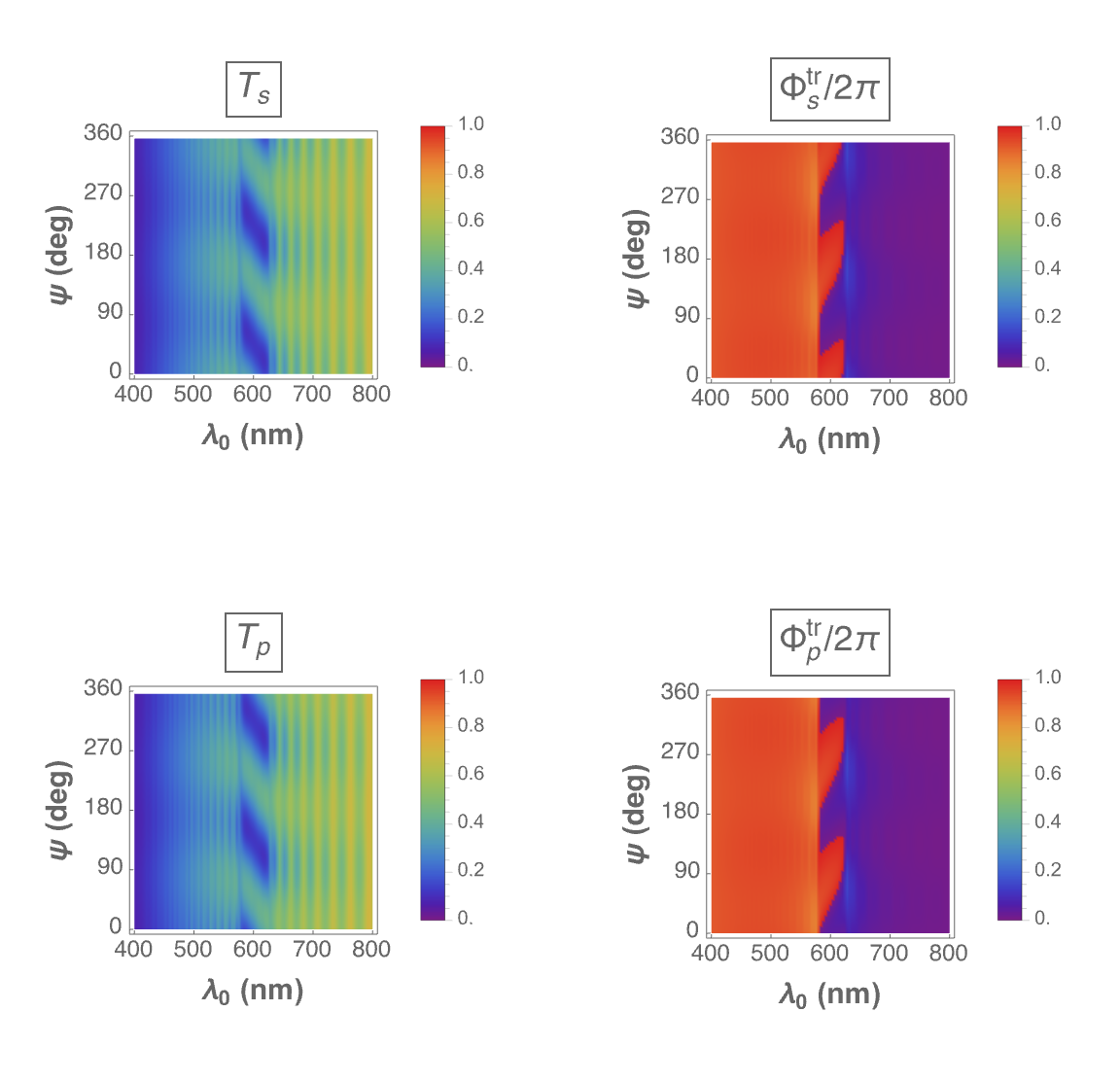} 
\hfill \vspace{0mm} 
 \caption{\label{LinGPtrans-4} $h=-1$, $\thetainc=0\deg$, and $\psi\in[0\deg,360\deg)$}
\end{subfigure} 
	\caption{Spectral variations of the total linear transmittance $T_{\mu}(\lambdao,\thetainc,\psi)$ and transmission-mode geometric phase $\bPhitra_{\mu}	(\lambdao,\thetainc,\psi)$,   $\mu\in\left\{s,p\right\}$, when
	(a,b) $h=1$ and (c,d) $h=-1$. (a,c) $\thetainc\in [0\deg,90\deg)$ and $\psi= 0\deg$;  (b,d) $\thetainc=0\deg$ and $\psi\in[ 0\deg,360\deg)$.
	}	
	\label{LinGPtrans}
\end{figure}

Note that $\bPhirefR=\bPhitraR\equiv0$ because of the structure of $\bphitinc$ for
an incident RCP plane wave \cite{Lakh2024josab,Lakh2024pra}. The other six geometric phases $\bPhiref_{\ell}$ and $\bPhitra_{\ell}$,
$\ell\in\left\{s,p,L\right\}$,
are generally non-zero in Figs.~\ref{CircGPref}--\ref{LinGPtrans}; furthermore,
their spectral dependencies have some resemblance to those of the corresponding total remittances defined in Eqs.~(\ref{def-Rcirc}),
(\ref{def-Tcirc}), (\ref{def-Rlin}), and (\ref{def-Tlin}).
Indeed, a feature representing the circular Bragg phenomenon is clearly evident in the plots of
$\bPhiref_{\ell}$ and $\bPhitra_{\ell}$,
$\ell\in\left\{s,p,L\right\}$.
The feature curves towards
 shorter wavelengths as $\thetainc$ increases while $\psi$ is fixed, and
 the feature has two undulations with increasing $\psi$ for normal
 incidence. 
 
 Although the geometric phase of the transmitted plane wave has been measured for normal
 incidence on a chiral sculptured thin film \cite{Das}, that was done only at a single value of
 $\lambdao$, that too in the long-wavelength neighborhood of the circular Bragg regime. Hopefully,
 experimental verification of the features evident in  Figs.~\ref{CircGPref}--\ref{LinGPtrans} will be carried
 out soon and the role of structural handedness clarified.

\section{Final remark}
Measurement of intensity-dependent observable quantities such as reflectances and transmittances
was instrumental in the identification of the circular Bragg phenomenon \cite{Mathieu,Fergason} and
continues to be undertaken \cite{Erten2015,McAtee2018,FialloPhD}. However,  measurement of phase-dependent
quantities, especially for oblique incidence, has been identified in this update as an arena for comprehensive research
in the near future.

\section*{Appendix: Poincar\'e spinor and geometric phase}
Any uniform plane wave propagating in free space can be represented as a point on the surface of the Poincar\'e sphere
$s_1^2+s_2^2+s_3^2=s_0^2$, where $s_0$, $s_{1}$, $s_2$, and $s_3$ are the four Stokes parameters. The plane wave's location
is identified by the longitude $\alpha\in[0,2\pi)$ and the latitude $\beta\in[-\pi/2,\pi/2]$ defined through the relations
\begin{equation}
\label{def-alphabeta}
\left.\begin{array}{l}
s_1=s_0\, \cos\beta\, \cos\alpha
\\[5pt]
s_2=s_0\, \cos\beta\, \sin\alpha
\\[5pt]
s_3=s_0\, \sin\beta 
\end{array}
\right\}\,.
\end{equation}
The  angles $\alpha$ and $\beta$ appear in the  Poincar\'e spinor
\begin{equation}
\label{def-PS}
\bphit=\les
\begin{array}{c}
\cos\left(\frac{\pi}{4}-\frac{\beta}{2}\right)
\\[5pt]
\sin\left(\frac{\pi}{4}-\frac{\beta}{2}\right)\exp\left(i\alpha\right)
\end{array}
\ris\,.
\end{equation}

With respect to a plane wave labeled ``1",
the geometric phase of a plane wave labeled ``2"  is defined
as the angle
\begin{equation}
\label{GP-def}
\Phi_{21}={\rm Arg}\lec {\les{\underline \phi}_1\ris}^\dag\cdot{\les{\underline \phi}_2\ris}\ric\,.
\end{equation}

 \section*{Acknowledgments}
This chapter is in appreciation of  the  national cricket team of U.S.A. that acquitted itself very well during the T20 World
Cup tournament held jointly in U.S.A. and West Indies in 2024. The author thanks the Charles Godfrey Binder Endowment at Penn State for supporting  his research from 2006 to 2024.

\end{document}